\pdfoutput=1
\documentclass[12pt]{article}
\usepackage[utf8]{inputenc}
\usepackage{tabularx}
\usepackage{longtable}

\usepackage[a4paper,width=160mm,top=25mm,bottom=35mm]{geometry}
\usepackage[small]{caption}
\usepackage{relsize}
\usepackage[table,svgnames,dvipsnames]{xcolor}
\usepackage{enumitem}
\usepackage[toc,page]{appendix}
\usepackage{listings}   
\usepackage[pdftex]{graphicx}
\usepackage{latexsym,amsmath,amsfonts,amssymb}
\usepackage[numbers,sort&compress]{natbib}
\usepackage{url}
\usepackage[colorlinks=true
  ,urlcolor=blue
  ,anchorcolor=blue
  ,citecolor=blue
  ,filecolor=blue
  ,menucolor=blue
  ,linktocpage=true
  ,pdfproducer=medialab
  ,pdfa=true
  ,linktoc=all
]{hyperref}  

\usepackage{amssymb} 
\usepackage{amsmath}                    			
\usepackage{amssymb}                    			
\usepackage{mathtools}                  			
\usepackage{latexsym}                   			
\usepackage{braket}                     			
\usepackage{graphicx}                               
\usepackage{subcaption}
\usepackage{wrapfig}                    			
\usepackage{sidecap}                    			
\usepackage{physics}                    			
\usepackage{color}                      			
\usepackage[dvipsnames]{xcolor}         		    
\usepackage{soul}                       			
\usepackage{array}                      			
\usepackage{enumitem}                   			
\usepackage{tensor}                     			
\usepackage{cancel}                     			
\usepackage{comment}                    			
\usepackage{empheq}                                 

\usepackage{tikz}
\usepackage{bbm}
\usepackage{float}
\usepackage{pdfscreen}                              
\usepackage[display]{texpower}                      
\usepackage{pifont}
\usepackage[thinlines]{easytable}
\usetikzlibrary{decorations.pathreplacing}
\usetikzlibrary{decorations.pathmorphing}
\usepackage{indentfirst}                            
\usepackage{csquotes}                               
\MakeOuterQuote{"}                                  

\numberwithin{equation}{section}

\usepackage{tocloft}

\newcommand{\h}{\hspace{0.5mm}}
\newcommand{\ie}{\textit{i.e.}}

\begin{document}
\begin{center}
 
\centerline{\Large {\bf Holographic Floquet states in low dimensions (II)}}

\vspace{8mm}
\hypersetup{linkcolor=black}
\renewcommand\thefootnote{\mbox{$\fnsymbol{footnote}$}}
Mart\'i Berenguer\footnote{marti.berenguer.mimo@usc.es}, Ana Garbayo\footnote{ana.garbayo.peon@usc.es},
Javier Mas\footnote{javier.mas@usc.es}
and  Alfonso V. Ramallo\footnote{alfonso.ramallo@usc.es}

\vspace{4mm}

{\small \sl Departamento de  F\'isica de Part\'iculas} \\
{\small \sl Universidade de Santiago de Compostela} \\
{\small \sl and} \\
{\small \sl Instituto Galego de F\'isica de Altas Enerx\'ias (IGFAE)} \\
{\small \sl E-15782 Santiago de Compostela, Spain} 
\vskip 0.2cm

\end{center}

\vspace{8mm}
\numberwithin{equation}{section}
\setcounter{footnote}{0}
\renewcommand\thefootnote{\mbox{\arabic{footnote}}}

\begin{abstract}

We continue the study in \cite{Garbayo_2020} of a  strongly coupled (2+1)-dimensional gauge theory subject to an external rotating electric field. The system is modelled holographically as a D3/D5 probe intersection. We add temperature to the D3 background and  analyze the phase diagram. Also here, the conductive phase extends down to vanishing external electric field at discrete values of the frequencies where {\em vector meson Floquet  condensates} form. For all temperatures, at given intercalated frequencies, we find  new dual states that we name {\em Floquet suppression points} where the  vacuum polarization vanishes even in the presence of an electric field. From the data we infer that these  states exist both in the conductive and insulating phases.   In the massless limit we find a linear and instantaneous conductivity law, recovering known general results in 2+1 dimensions. We also examine the photovoltaic AC and DC  current as the response to an oscillating probe electric field and see that rising the temperature suppresses  the photovoltaic Hall current. All the results  obtained  carry over qualitatively unaltered to the case of D3/D7. 

\end{abstract}

\clearpage

\hypersetup{linkcolor=black}
\tableofcontents
\hypersetup{linkcolor=blue}

\section{Introduction}

Periodically driven systems form a separate chapter in the book of non-equilibrium dynamics. Much progress has been achieved both at theoretical and experimental levels in the path to control their effective long time dynamics \cite{Bukov_2014,Holthaus_2015,Eckardt_2016,Weinberg_2016,Oka_2018,Giovannini_2019,Rudner_2020}. This has opened the way to engineer  Hamiltonians that  embody non trivial phenomena and new  phases of quantum materials.  Of particular interest is the case  of  solutions where  energy injection and dissipation  balance, thereby reaching a Floquet type of {\em non-equilibrium steady state} (NESS). In a series of previous papers, the existence of a Floquet NESS has been studied in the context of the AdS/CFT correspondence both in the case of a D3/D7 system \cite{Hashimoto_2016,Kinoshita_2017}, and of a D3/D5 system \cite{Garbayo_2020}. In these intersections the higher dimensional flavor brane is treated as a probe in the backgrond of a $AdS_5\times {\mathbb S}^5$ geometry. The flavor degrees of freedom experience the rotating external field, while they are coupled to the non abelian "gluon" vacuum acting as a bath. The crucial ingredient that allows for the system to be solved numericallly is the fact that, in the rotating frame, the action becomes time independent and, therefore, all differential equations turn out to be of the ordinary type. This is a remarkable ansatz where the technique developed in \cite{Karch_2007} can still be applied to obtain the fully non-linear one point functions just from demanding reality of the action. In the literature, this extremely useful IR fixing mechanism has been applied to several static configurations. For time dependent sources very little is known. In spatial dimension $d_s=2$, and for massless charge carriers, the current response is linear and instantaneous $J(t) = \sigma E(t)$ \cite{Karch_2010}. Our results are consistent with this observation in the limit of small mass flavours.

In this paper we deal with the D3/D5 system, in which the D3- and D5-branes share two spatial directions. The field theory dual to this brane setup is well-known and consists of a supersymmetric theory with flavor hypermultiplets living in a two-dimensional defect coupled to an ambient ${\cal N}=4$ four-dimensional Yang-Mills theory \cite{DeWolfe_2001, KarchKatz_2002, Erdmenger_2002}. In this work, which is a continuation of \cite{Garbayo_2020}, we add a background temperature to the adjoint degrees of freedom which, therefore, are now deconfined. The nonzero temperature breaks supersymmetry and adds charged carriers to the ones previously formed by Schwinger pair production. These are naturally melted mesons that are present at finite temperature for low enough quark mass.  Therefore, the phenomenology is expected to yield  a continuous deformation of the case at zero temperature.

It is known that  the fluctuations of the flavor brane degrees of freedom feel another temperature, $T_{\rm eff}$, through the so-called {\it open-string metric} (OSM) \cite{Seiberg_1999,Gibbons_2000}. That this is a {\em bona fide} temperature has been the subject of careful studies that examined, for example, the universality of the fluctuation dissipation relations \cite{Sonner_2012}. In most known examples, $T_\text{eff} > T$, where $T$ is the temperature of the bulk plasma. While exceptions to this inequality have been found \cite{Nakamura_2013}, our analysis across parameter space reveals no such anomalies in the D3/D5 model.

The plan of the paper is the following. Section \ref{sec:D3D5temp} sets up the holographic model, introduces the relevant field theory quantities, and discusses the classification of brane embeddings. We also analyze the behavior of the effective temperature. In Section \ref{sec:phasespace}, we examine the phase structure of the system in detail. Section \ref{sec:conductivitiesD3D5} is devoted to transport phenomena: we study both the rotating current induced by the periodic driving and the photovoltaic current, which characterizes the response to a secondary probe field on top of the rotating electric field. In Section \ref{sec:mesons}, we analyze the mesonic spectrum via linearized Minkowski embeddings. In Section \ref{sec:noneqthermo} we comment on the difficulties of interpreting the on-shell action thermodynamically. Our findings and open questions are summarized in Section \ref{sec:Conclusions}.

Finally, the paper includes several appendices with technical details and supplementary results. In Appendix \ref{app:coordinates} we summarize the different coordinate systems we use. Appendix \ref{app:analyticsols} presents analytic solutions for zero and small quark mass. Appendix \ref{app:holoreno} contains the holographic renormalization and dictionary. Appendix \ref{app:opticalcond} elaborates on the calculation of optical conductivities, including the analytic massless case. For completeness, Appendix \ref{app:D3D7} presents the corresponding phase structure for the D3/D7 system.

\section{D3/D5 system at finite temperature}\label{sec:D3D5temp}

In this section we will set up the stage and review the main results obtained in this study for the D3/D5 systems. We chose to parametrize the D5 brane embeddings  with an angular coordinate function $\psi(u)\in [0,1]$ where $\psi(u) =0$ corresponds to massless embeddings and the boundary of AdS sits at $u=\infty$. The details on the coordinate systems we use are given in Appendix \ref{app:coordinates}.

In these coordinates, the AdS$_5\times S^5$ metric is given by
\begin{equation}
\begin{aligned}
    ds^2=\frac{u^2}{L^2}&\left(-\frac{g(u)^2}{h(u)}dt^2+h(u)\left(dx^2+dy^2+dz^2\right)\right)+\frac{L^2}{u^2}du^2\\
    &\phantom{\left(-\frac{g(u)^2}{h(u)}dt^2+h(u)\left(dx^2\right.\right.}+L^2\left(\frac{d\psi^2}{1-\psi^2}+(1-\psi^2)d\Omega_2^2+\psi^2 d\Omega_2^2\right)~,
    \label{eq:AdS5xS5FloquetII}
\end{aligned}
\end{equation}
with the functions $g(u)$ and $h(u)$ given by
\begin{equation}
    g(u)=1-\frac{u_h^4}{u^4}~~,\hspace{5mm}h(u)=1+\frac{u_h^4}{u^4}~,
    \label{eq:ghfunctions}
\end{equation}
The black hole horizon is located at $u_h=r_h/\sqrt{2}$, with the Hawking temperature given by
\begin{equation}
    T_H=\frac{r_h}{\pi L^2}=\frac{\sqrt{2}u_h}{\pi L^2}~.
    \label{eq:TBH}
\end{equation}
In the following, we set $L=1$. We will parametrize background temperatures in terms of $r_h/m$.

We want to study the response of the system to an external circularly polarized electric field,
\begin{equation}
    \begin{pmatrix} \mathcal{E}_x(t) \\ \mathcal{E}_y(t) \end{pmatrix}=\begin{pmatrix} \cos\Omega t & -\sin\Omega t\\ \sin\Omega t & \cos\Omega t \end{pmatrix}\begin{pmatrix} E_x \\ E_y\end{pmatrix}\equiv \mathcal{O}(t)\vec{E}~,
    \label{eq:rotatingE}
\end{equation}
with $\vec E=(E_x,E_y)$ denoting the electric field at $t=0$. It is convenient to introduce the complexified electric field
\begin{equation}
    \mathcal{E}_x+i\mathcal{E}_y=Ee^{i\Omega t}~,
    \label{eq:complexE}
\end{equation}
with $E=E_x+iE_y$. This electric field can be derived from a vector potential
\begin{equation}
    A_x+iA_y=Ae^{i\Omega t}~,
    \label{eq:complexA}
\end{equation}
with $A=iE/\Omega$. Holographically, this corresponds to turning on a worldvolume gauge field $A(t,u)$ that approaches \eqref{eq:complexA} at the boundary of AdS$_4$. We introduce it as
\begin{equation}
    2\pi\alpha'A(t,u)=A_x(t,u)~dx+A_y(t,u)~dy~,
\end{equation}
with field strength given by
\begin{equation}
    2\pi\alpha'F=\dot{A}_x~dt\wedge dx+A_x'~du\wedge dx+\dot{A}_y~dt\wedge dy+A_y'~du\wedge dy~.
\end{equation}
The complexifications above motivate to do the same at the level of the bulk fields,
\begin{equation}
\begin{aligned}
    2\pi\alpha'\h A(t,u)=A_x(t,u)+i A_y(t,u)&\equiv c(t,u)e^{i\Omega t}\\
    &\equiv b(t,u)e^{i\left(\Omega t+\chi(t,u)\right)}~,
    \label{eq:defctu}
\end{aligned}
\end{equation}
where in the second line we have defined two new real variables, $b(t,u)$ and $\chi(t,u)$, representing the magnitude and the phase of $c(t,u)$, respectively.

The D5-branes will be located at $z=0$. For the embedding function we are going to consider a general dependence on the radial coordinate, and time, $\psi=\psi(t,u)$. However, as we will see shortly, all time-dependence can be eliminated and the resulting problem becomes one-dimensional, with dependence only on the holographic coordinate. The induced metric on the D5-branes is
\begin{equation}
\begin{aligned}
    ds^2=&-\left(u^2\frac{g(u)^2}{h(u)}-\frac{\dot{\psi}^2}{1-\psi^2}\right)dt^2+u^2h(u)\left(dx^2+dy^2\right)+\left(\frac{1}{u^2}+\frac{\psi'^2}{1-\psi^2}\right)du^2\\    
    &+\frac{2\dot{\psi}\psi'}{1-\psi^2}dt\hspace{0.5mm}du+(1-\psi^2)d\Omega_2^2~.
\end{aligned}
\end{equation}

In terms of these variables, the DBI action for the $N_f$ probe D5-branes is given by
\begin{equation}
\begin{aligned}
    S_{D5}=&-\mathcal{N}\int dt~du\sqrt{1-\psi^2}\Bigg[h\left(1-\psi^2+u^2\psi'^2\right)\left(u^4g^2-\dot{b}^2-\left(\Omega+\dot{\chi}\right)^2b^2\right)\Bigg.\\
    &\phantom{-N_f T_{D5}}-u^2h~\dot{\psi}^2\left(h+b'^2+b^2\chi'^2\right)+2u^2h~\dot{\psi}~\psi'\left(\dot{b}~b'+\left(\Omega+\dot{\chi}\right)b^2\chi'^2\right)\\
    &\phantom{-N_f T_{D5}}\Bigg.+\left(1-\psi^2\right)\left[u^4g^2\left(b'^2+b^2\chi'^2\right)-b^2\left(\dot{b}\chi'-\left(\Omega+\dot{\chi}\right)b'\right)^2\right]\Bigg]^{1/2}~,
    \label{eq:actionD5tdep}
\end{aligned}
\end{equation}
with $\mathcal{N}=4\pi N_fT_{\text{D5}} V_{\mathbb{R}^{2,1}}$. The factors $V_{\mathbb{R}^{2,1}}$ and $4\pi$ arise from the integration along the $(x,y)$-directions and the two-sphere, respectively. It turns out that a consistent ansatz corresponds to taking the fields $\psi(t,u)$, $b(t,u)$ and $\chi(t,u)$ to be time-independent:
\begin{equation}
    \psi=\psi(u)~,\quad\quad b=b(u)~,\quad\quad\chi=\chi(u)~.
\end{equation} 
With this truncation, the action \eqref{eq:actionD5tdep} becomes
\begin{align}
\begin{split}
    S_{D5}=-\mathcal{N}\int du \sqrt{1-\psi^2}\Bigg[\left(u^4g^2-\Omega^2b^2\right)\left[\left(1-\psi^2\right)\left(h+b'^2\right)+u^2h\psi'^2\right]\Bigg.\\
    \Bigg.+u^4g^2b^2\left(1-\psi^2\right)\chi'^2\Bigg]^{1/2}~,
    \label{eq:actionD5}
\end{split}
\end{align}
where we have re-defined the overall constant to $\mathcal{N}=4\pi N_fT_{\text{D5}}$. On both sides we have implicitly divided by the volume of $\mathbb{R}^{2,1}$. The prefactor $\mathcal{N}$ may be written in terms of field theory quantities as
\begin{equation}
    \mathcal{N}=4\pi N_fT_{\text{D5}}=\frac{N_fN_c\sqrt{\lambda}}{2\pi^3}~.
    \label{eq:ND3D5}
\end{equation}

From this action it is obvious that all time-dependence has been removed, and the system has effectively become one-dimensional, with only dependence on the holographic coordinate, $u$. The fact that the electric field is time-dependent and rotating is seen through the $\Omega$-dependence in the action\footnote{This is a notorious difference with respect to the case of a constant electric field, that is at the heart of many differences among the two cases. In the constant field case, $A_x = -Et +...$ \cite{Karch_2007}, the action depends explicitly on $E$, whereas here the electric field will be extracted from the asymptotic behavior of the $b$ and $\chi$ fields. See \eqref{eq:UVexpansions}.}. This is the action we will use to derive the equations of motion for the fields $\psi$, $b$ and $\chi$. These equations are long and unilluminating, so they won't be reproduced here.

By imposing that the equations must be satisfied order by order in $u$ near the UV boundary, we can find the asymptotic behavior of the fields. By expanding $\psi(u)=\sum_{n=0}^\infty \psi_n u^{-n}$, and similarly for $b$ and $\chi$, we find the following behavior:
\begin{align}
    \psi(u)&=\frac{m}{u}+\frac{C}{u^2}+...~,\\
    b(u)e^{i\chi(u)}&=\frac{i E}{\Omega}+\frac{J}{u}+...~,
    \label{eq:UVexpansions}
\end{align}
where we have denoted $\psi_1$ and $\psi_2$ as $m$ and $C$, respectively, and in the second line we have
\begin{equation}
\begin{aligned}
    \frac{iE}{\Omega} &= \lim_{u\to\infty}b(u)e^{i\chi(u)} = b_0\cos\chi_0+i\h b_0\sin\chi_0~,\\
    J & = -\lim_{u\to\infty} u^2 \left[b'(u)+ib(u)\chi'(u)\right]e^{i\chi(u)} =  e^{i\chi_0}\left(b_1+ib_0\chi_1\right)~,
    \label{eq:UVexpansionsEJ}
\end{aligned}
\end{equation}
with $E$ the complex electric field, defined in \eqref{eq:complexE}-\eqref{eq:complexA}. The same complexified notation is used for $J$, defined as $J=J_x+iJ_y$. From here we can read off the one-point functions for the quark condensate $\langle O_m\rangle$ and the electric current $\mathcal{J}_\text{YM}$ in the boundary theory. The precise relation between $E$, $J$, $m$ and $C$ and the electric field ${\cal E}_\text{YM}(t)$, the electric current ${\cal J}_\text{YM}(t)$, quark mass $m_q$ and quark condensate $\langle O_m\rangle $ in the boundary theory is derived in Appendix \ref{app:holoreno}:
\begin{equation}
\begin{aligned}
    \mathcal{E}_\text{YM}(t) &= \frac{\sqrt{\lambda}}{2\pi}E e^{i\Omega t}~, & \qquad \mathcal{J}_\text{YM}(t) &= \frac{N_fN_c}{\pi^2}Je^{i\Omega t}~, \\
    m_q~ &= \frac{\sqrt{\lambda}}{2\pi}m~, & \qquad \langle O_m\rangle &= -\frac{N_fN_c}{\pi^2}C~,
    \label{eq:dictionaryD3D5}
\end{aligned}
\end{equation}
where $\lambda=g^2_\text{YM}N_c$ is the 't Hooft coupling of the $\mathcal{N}=4$ theory.

It is worth noting that the action \eqref{eq:actionD5} depends only on the derivative of $\chi(u)$, namely $\chi'(u)$, and not on $\chi(u)$ itself. Therefore, there exists a conserved quantity defined as
\begin{equation}
    q\equiv\Omega \frac{\partial \mathcal{L}}{\partial \chi'}\sim J_xE_x+J_yE_y=\vec{J}\cdot \vec{E}~,
    \label{eq:qconserved}
\end{equation}
with $\mathcal{L}$ the DBI Lagrangian. With $\sim$ we denote the UV expansion of this quantity, using \eqref{eq:UVexpansions}. Therefore we see that $q$ acquires the meaning of a Joule heating. Only black hole embeddings have a non-zero value for this quantity. The stationarity of the background metric upon this energy injection can only be understood as a transient effect due to the imbalance $N_f/N_c \sim 0$ that is present in the probe limit. 
In the presence of a black hole in the bulk, the long time effect of a non-negligible backreaction  would be a slow increase in the horizon radius.

We can use the conserved quantity to obtain $\chi'$ as
\begin{equation}
    \chi'^2=q^2\frac{\left(u^4g^2-\Omega^2b^2\right)}{u^4g^2b^2\left(1-\psi^2\right)\left(u^4g^2\Omega^2b^2\left(1-\psi^2\right)^2-q^2\right)}\left(\left(1-\psi^2\right)b'^2+h\left(1-\psi^2+u^2\psi'^2\right)\right)~.
\end{equation}
Using this, we can Legendre transform the action to get rid of $\chi'$, and write an action in terms of $q$,
\begin{equation}
\begin{aligned}
    \bar{S}_{\text{D5}}&=S_{\text{D5}}-\int du~ \chi'\frac{\partial\mathcal{L}}{\partial \chi'}\\
    &=-\mathcal{N}\int du \frac{\sqrt{b'^2+h\left(1+\frac{u^2\psi'^2}{1-\psi^2}\right)}}{u^2g\Omega\abs{b}}\sqrt{\left[u^4g^2-\Omega^2b^2\right]\left[u^4g^2\Omega^2b^2\left(1-\psi^2\right)^2-q^2\right]}~.
    \label{eq:actionLTtransf}
\end{aligned}
\end{equation}

\begin{figure}
    \centering
    \includegraphics[width=0.34\textwidth]{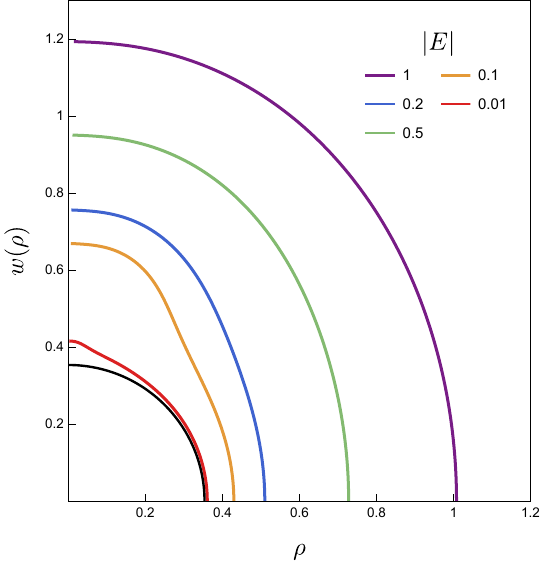}
    \hspace{0.02\textwidth} 
    \includegraphics[width=0.45\textwidth]{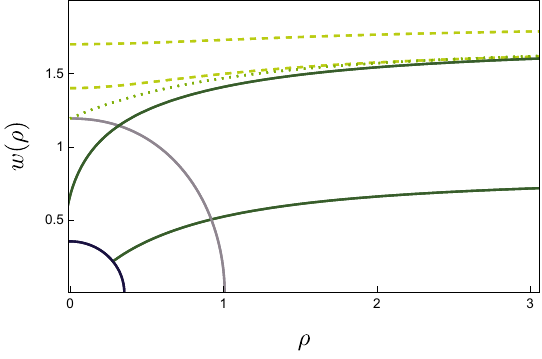}
    \caption{Left: shape of the singular shell for different applied electric fields. The axes are $\rho =u\sqrt{1-\psi^2}$ and $w=u\psi$ with $\psi=\sin\theta$. The curves are the points $\rho^2+w^2=u_c(E,\psi_0)$ with $\psi_0=\sin\theta_0$ the embedding angle at the singular shell. Outside the limit $E\to 0$ the shell shape is non spherical. Hence, unlike the case of a constant electric field, in the rotating situation we have a non-trivial dependence of $u_c$ on the mass $m$ of the D5-brane at fixed $|E|$. Right: profiles of different embeddings for $\Omega=|E|=1$. The dashed lines represent black hole embeddings. They can be either regular (thermal), ending on the horizon, or singular, ending in a conical singularity. The dotted line represents a critical embedding. The solid lines represent Minkowski embeddings. The black hole and pseudo-horizon are shown by the black (inner) and gray (outer) lines.}
    \label{fig:rcEmbeddings}
\end{figure}

As has become the usual case when there is an electric field switched on \cite{Karch_2007}, demanding reality of the Legendre-transformed action \eqref{eq:actionLTtransf} imposes that the two terms under the second square root must change sign (and therefore, vanish) at the same point $u=u_c$, which locates the singular shell. At once, this condition fixes both  the value of $u_c$ and that of the conserved quantity $q$
\begin{equation}
    \Omega\h b_0=\frac{u_c^4-u_h^4}{u_c^2}~, \qquad q=\frac{(u_c^4-u_h^4)^2}{u_c^4}(1-\psi_0^2)~,
    \label{eq:b0andq}
\end{equation}
where $b_0=b(u_c)$, $\psi_0=\psi(u_c)$. As $\Omega b_0\geq 0$, we see that  $u_c\geq u_h$, and the singular shell sits always at a larger radius than the black hole horizon. Moreover, here it is the IR data, $b_0$, what controls the position of the singular shell, $u_c(u_h, b_0)$. This is another important difference with respect to the case of constant $E$  \cite{Karch_2007}, where the position of the singular shell depends exclusively on the UV value $\abs{E}$. Therefore, in this case the shape of the critical surface at constant $|E|$ is non-spherical, as can be seen in the left plot of Fig. \ref{fig:rcEmbeddings}.

An important remark for later use is the fact that both the worldvolume electric field and the black hole horizon add up their effects of bending the brane in the IR towards the origin. This will mean that as we increase the temperature we will find black hole embeddings with milder electric fields.

The following  is a  scaling symmetry of the lagrangian  $\mathcal{L} \to \alpha^2\mathcal{L}$ and the boundary conditions
\begin{equation}
\begin{aligned}
    t &\to t/\alpha~, & u &\to \alpha\h u~, & w & \to \alpha\h w~, & b & \to \alpha\h b~, & \chi & \to\chi~,\\
    \Omega &\to \alpha\h \Omega~, & E &\to \alpha^2\h E~, & J & \to \alpha^2\h J~, & \psi & \to \psi~, & m & \to\alpha\h m~,\\
    C &\to \alpha^2C, & q &\to \alpha^4q~, & T & \to \alpha\h T~, 
    \label{eq:scaling}
\end{aligned}
\end{equation}
By choosing $\alpha=1/m$ in \eqref{eq:scaling} we can take $m=1$ and deal with the remaining quantities in units of (the  appropriate powers of) $m$.

\subsection{Types of embeddings}

There are three types of embeddings in place now. First of all, we find the \textit{Minkowski} embeddings, which do not intersect the singular shell. They end up closing  smoothly at a value of $u=u_0>u_c$, where $\psi=1$. With \textit{black hole} (BH) embeddings we will generically denote solutions that intersect the singular shell. This accounts for the fact that for a worldvolume observer the singular shell acts as an event  horizon, inducing thermal effects through Schwinger pair production. For this reason, we will term interchangeably singular shell and effective horizon. Finally, the \textit{conical} embeddings will denote the solutions ending outside but just above the singular shell. Notice that this differs from the usual definition of critical embeddings when the electric field is not present, where they are defined as those ending just above the black hole horizon. We depict the different embeddings in the right plot of Fig. \ref{fig:rcEmbeddings}.

\subsubsection{Black hole embeddings}
Black hole embeddings  can be further subdivided into two classes,  \textit{thermal} and \textit{conical}.  The first ones hit the bulk black hole horizon, while the second ones close up at $\psi=1$ with a conical singularity, most likely a reflection of the sink of energy pumped by the electric field in a conducting, albeit non-dissipative system. These conical black hole embeddings are the remnant of the ones that were studied in \cite{Garbayo_2020} at zero temperature. For a constant electric field they were analyzed in \cite{Mas_2009,Kim_2011}. The two lower embeddings in the right plot of Fig. \ref{fig:rcEmbeddings} are examples of each class.

As a technical  remark, notice that, in order to solve numerically for the  black hole embeddings,  boundary conditions must be placed at the singular shell. This can be done by expanding the fields as
\begin{equation}
\begin{aligned}
    \psi(u) & = \psi_0+\psi_1(u-u_c)+...\\
    b(u) & = b_0+b_1(u-u_c)+...\\
    \chi(u) & = \chi_0 +\chi_1(u-u_c)+...~,
    \label{eq:coefsBHemb}
\end{aligned}
\end{equation}
with $\psi_0=\psi(u_c)$, and similarly for $b_0$ and $\chi_0$. These parameters can be chosen freely and give rise to different solutions with different UV asymptotics. However, the equations of motion derived from the action \eqref{eq:actionD5} are naïvely divergent at $u=u_c$. Therefore, the three derivatives $\psi_1$, $b_1$ and $\chi_1$ are not free but commanded by regularity. The precise expressions are lengthy and provide no further insight, so they are omitted here.

In contrast, we could choose boundary conditions at the background horizon for the embedding function $\psi$ and the module $b$, but not for the phase  $\chi$, since this function diverges logarithmically as $u\to u_h$. Starting from the effective horizon, then, one can integrate either outwards or inwards and this is how the thermal and/or conical embeddings are distinguished.

\subsubsection{Minkowski and critical embeddings}
Minkowski embeddings do not intersect the singular shell. They close off smoothly at $u=u_0>u_c$, where $\psi=1$. The values $b_0=b(u_0)$ and $\chi_0=\chi(u_0)$ are free parameters and, again, regularity of the equations at $u=u_c$ determines the fields to behave as
\begin{equation}
\begin{aligned}
    \psi(u)&=1-\frac{3\left(u_0^4+u_h^4\right)\left(u_0^8+u_h^8-u_0^4\left(2u_h^4+\Omega^2b_0^2\right)\right)}{u_0\left(u_0^4-u_h^4\right)\left(3\left(u_0^8+u_h^8\right)+u_0^4\left(2u_h^4-\Omega^2b_0^2\right)\right)}(u-u_0)+...\\
    b(u)&=b_0-\frac{u_0b_0\Omega^2\left(u_0^4+u_h^4\right)^2}{\left(u_0^4-u_h^4\right)\left(3\left(u_0^8+u_h^8\right)+u_0^4\left(2u_h^4-\Omega^2b_0^2\right)\right)}(u-u_0)+...
\end{aligned}
\end{equation}
We haven't written an expansion for $\chi$ because the equations of motion impose all the terms $\chi_n$, with $n\ge 1$, to vanish. Therefore, for Minkowski embeddings, $\chi(u)=\chi_0$ is a constant along the holographic direction, only determining the initial direction at $t=0$ of the electric field in the boundary theory. For this reason, as mentioned earlier, the Joule heating $q$ vanishes for Minkowski embeddings (recall that $q\propto \chi'$).

Critical embeddings can be understood as Minkowski embeddings, in the limit where $\Omega b_0\to\frac{u_c^4-u_h^4}{u_c^2}$, as dictated by the position of the singular shell, Eq. \eqref{eq:b0andq}. This leads to the following expansion for the critical embeddings:
\begin{equation}
\begin{aligned}
    \psi(u)&=1-\frac{4\left(u_c^4+u_h^4\right)+\Omega^2u_c^2}{2u_c^2\left(u_c^4+u_h^4\right)}(u-u_0)^2+...\\
    b(u)&=\frac{u_c^4-u_h^4}{\Omega u_c^2}-\frac{\Omega}{2u_c}(u-u_0)+...
\end{aligned}
\end{equation}

\subsection{Effective metric and effective temperature}\label{subsec:Teff}

As mentioned earlier, the critical radius $u_c$ signals the position of an event horizon in the induced open string metric, which governs the dynamics of the worldsheet fluctuations. The effective temperature $T_\text{eff}$ is defined as the Hawking temperature associated with the black hole horizon of this metric. We now extend the results to the present case of a rotating electric field.

Let $h_{ab}$ be the induced six-dimensional metric and $\mathcal{F}_{ab}$ the worldvolume gauge field. The effective open string metric $\gamma_{ab}$ is defined as
\begin{equation}
    \gamma_{ab}=h_{ab}+(2\pi\alpha')^2\mathcal{F}_{ac}\mathcal{F}_{bd}~h^{cd}~.
\end{equation}
In order to write the form of this metric for our ansatz when the embedding is parameterized as a function $\psi=\psi(u)$, let us define the function $F(u)$ as
\begin{equation}
    F(u)\equiv\frac{1}{h}\left(u^2g^2-\frac{\Omega^2\abs{c}^2}{u^2}\right)~,
\end{equation}
and the complex one-forms $e_{\pm}$ as
\begin{equation}
    e_{\pm}=e^{\mp i\Omega t}(dx\pm idy)~.
\end{equation}
Then, we have
\begin{align}
\begin{split}
    \gamma_{ab}d\xi^a d\xi^b=-F(u) dt^2+\left[\frac{1}{u^2}+\frac{\psi'^2}{1-\psi^2}+\frac{\abs{c'}^2}{u^2h}\right]du^2-\frac{2\h \Omega}{u^2h}\Im(c\bar{c}')dt\h du\\
    +\frac{1}{4}\left[\frac{1}{u^2}+\frac{\psi'^2}{1-\psi^2}\right]^{-1}(c'e_-+\bar{c}'e_+)^2+\frac{\Omega^2h}{4u^2g^2}(c\h e_-+\bar{c}\h e_+)^2+\frac{h^2F(u)}{g^2}e_+e_-~,
    \label{eq:eff_metric_non_diagonal}
\end{split}
\end{align}
where $c(u)=b(u)\h e^{i\chi(u)}$ is the complexified field potential in the rotating frame. We can diagonalize the $(t,u)$ part of the metric with the change
\begin{equation}
    t=\tau-h(u)~,\qquad \text{where}\quad h'(u)=\frac{\gamma_{ut}}{\gamma_{tt}}~.
\end{equation}
In the new variables, the $(\tau,u)$ part of the metric reads
\begin{equation}
\begin{aligned}
    \gamma_{ab}d\xi^a d\xi^b\big\rvert_{\tau,u}&=\gamma_{tt}d\tau^2+\left(\gamma_{uu}-\frac{\gamma_{tu}^2}{\gamma_{tt}}\right)du^2\\
    &=-F(u)d\tau^2+\left[\frac{1}{u^2}+\frac{\psi'^2}{1-\psi^2}+\frac{\abs{c'}^2}{u^2h}+\frac{\Omega^2}{u^2h^2F(u)}\Im(c\h \bar{c}')\right]du^2
\end{aligned}
\end{equation}

\begin{figure}
    \centering
    \begin{subfigure}[t]{0.48\textwidth}
        \centering
        \includegraphics[width=\textwidth]{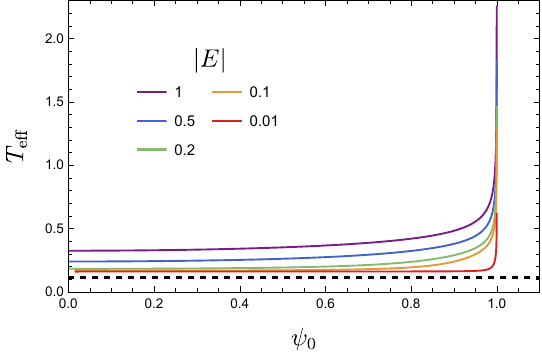}
        \label{sfig:TPlot}
    \end{subfigure}
    \hspace{0.02\textwidth} 
    \begin{subfigure}[t]{0.48\textwidth}
        \centering
        \vspace{-5.05cm} 
        \includegraphics[width=\textwidth]{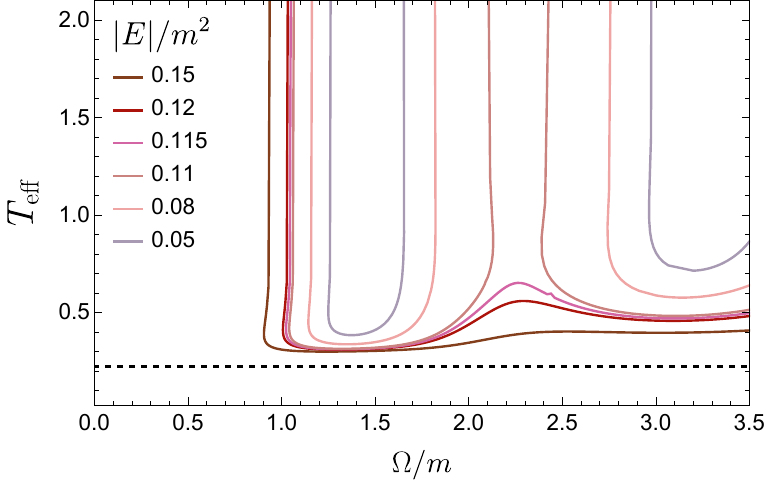}
        \label{sfig:EffectiveTlobes}
    \end{subfigure}
    \caption{Left: effective temperature for different BH embeddings, labeled by the insertion angle at the effective horizon, $\psi_0=\psi(u_c)$. Right: effective temperature as we vary $\Omega/m$ at fixed $|E|/m^2$. In both cases, the divergences in $T_{\text{eff}}$ arise as the curves come close to the critical embeddings. In both plots, the lowest dashed line in black signals the background temperature.}
    \label{fig:Teff}
\end{figure}

This metric has an event horizon at the singular shell, $u=u_c$, where the function $F(u)$ vanishes. Using the field expansions around $u=u_h$ defined in \eqref{eq:coefsBHemb} the Euclidean metric can be expanded near the horizon. Enforcing the correct periodicity of the time coordinate to eliminate the conical singularity at the origin leads to the effective temperature of the OSM,
\begin{equation}
    T_{\text{eff}}=\frac{2u_ch(u_c)-\Omega b_1}{2\pi b_0\chi_1}~,
    \label{eq:TeffD5}
\end{equation}
where $b_0$, $b_1$ and $\chi_1$ are the coefficients defined in \eqref{eq:coefsBHemb} for black hole embeddings. As in the vast majority of the situations encountered in the literature \cite{Kim_2011,Kundu_2013,Nakamura_2013,Kundu_2019}, here we also find that $T_{\text{eff}} > T_H$, as far as we have been able to  scan, as shown in Fig. \ref{fig:Teff}.

\section{Phase space}\label{sec:phasespace}

As just discussed, the different types of IR boundary conditions correspond to different behaviors for the probe branes. This behavior has to have an effect on the field theory phase space. The standard lore in flavor branes is that Minkowski (black hole) embeddings are dual to insulating (conducting) phases of the dual field theory. In the case of a rotating electric field, we must be more careful. Actually, two types of currents emerge. Black hole embeddings carry dissipating currents because of the presence of fundamental carriers. The external driving has to supply energy in order to maintain the stationary rotating current. 

For Minkowski embeddings, $J$ is a polarization current. In analogy with the case of the D3/D7 intersection \cite{Kinoshita_2017}, we interpret this polarization as a coherent alignement of the vector meson vacuum fluctuations parallel to the electric field. The polarization current is the time derivative of the polarization and rotates at right angles with the electric field, signalling zero Joule heating, hence not dissipating any energy. This conservative aspect allows for the  possibility to have non-zero persistent current even in the limit of vanishing driving field. This effect happens at discrete driving frequencies. As we will see, there exists a dual possibility, in which the polarizability is dynamically suppressed, even at finite driving field.

Using the different IR boundary conditions, we can integrate the equations of motion up to the UV boundary. For black hole embeddings, this is done by specifying the values $\psi_0$, $b_0$, $\chi_0$ at the singular shell $u=u_c$, with the derivatives of the fields dictated by regularity. For Minkowski embeddings, the IR boundary conditions are imposed at the point where the brane ends, $u=u_0$. In both cases, the solution is unique under these boundary conditions. From the asymptotic behavior of the fields, Eq. \eqref{eq:UVexpansions}, we can read off the values of $m$, $C$, $E$, $J$. Due to the scaling symmetry \eqref{eq:scaling}, these quantities are not all independent. Therefore, we will work with the dimensionless quantities $T/m$, $r_h/m$, $\Omega/m$, $E/m^2$, $J/m^2$ and ${\cal C}/m^2$.

\begin{figure}
    \centering
    \begin{subfigure}[t]{0.48\textwidth}
        \centering
        \includegraphics[width=\textwidth]{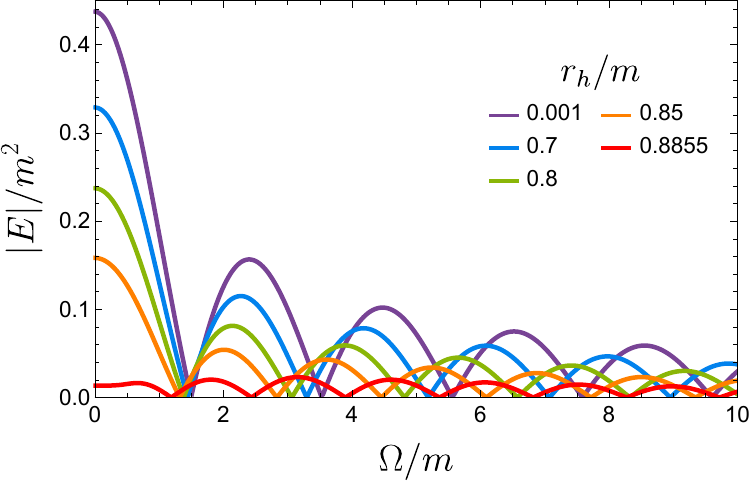}
    \end{subfigure}
    \hspace{0.02\textwidth}
    \begin{subfigure}[t]{0.48\textwidth}
        \centering
        \includegraphics[width=\textwidth]{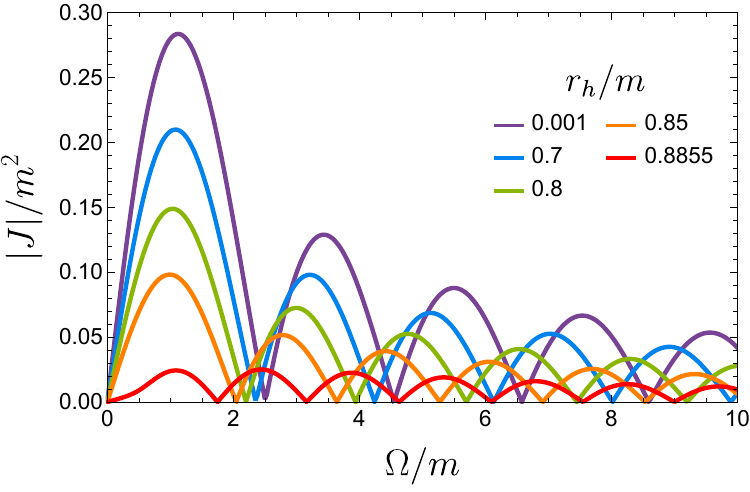}
    \end{subfigure}
    \caption{Electric field and current of the critical embeddings versus driving frequency, for different values of $r_h/m$. The frequencies for which $|E|/m^2$ vanishes are the critical frequencies $\Omega_c/m$ of the vector meson Floquet condensates. The regions of black hole embeddings between lobes, above such critical points, are termed \textit{wedges} in the text. In the right plot we show the value of the current along the curve of critical embeddings. We observe points with $|J|/m^2=0$ somewhere close to the maxima of the lobes in the left plot. We will term these points \textit{Floquet suppression points}, and will study them in detail in Section \ref{subsec:suprpoints}.}
    \label{fig:LobesEJD3D5}
\end{figure}

In Fig. \ref{fig:LobesEJD3D5} we show the phase diagram in the $(E/m^2,\Omega/m)$ parameter space. Different lines correspond to different temperatures, parametrized by $r_h/m$. The limit $r_h/m\to 0$ (purple line) recovers the results of \cite{Garbayo_2020}. For each temperature, the solid line represents the electric field and frequency of the critical embeddings. Roughly speaking, points above such curves correspond to black hole embeddings, whereas those below are Minkowski\footnote{However, close to the critical embeddings, the boundary quantities are usually multivalued (see \cite{Mateos_2006,Mateos_2007} and Fig. \ref{fig:JCvsE} below) and we can find both types of embeddings in the near vicinity.}.

This lobed structure was also present at zero temperature \cite{Garbayo_2020}, and in the D3/D7 model \cite{Kinoshita_2017}. We notice that $E/m^2$ has a series of maxima whose height decreases as $\Omega/m$ increases. Of particular interest are the points in which, for particular values of the driving frequency, the critical electric field vanishes. We denote these frequencies as $\Omega_c$. For these frequencies, however, the electric current is non-zero, as we see in the right plot of Fig. \ref{fig:LobesEJD3D5}. This means that, for these particular frequencies, the insulator-conductor transition can be triggered with vanishing external electric field. Physically, for these frequencies the driving field enters in resonance with the vector meson excitations of the gauge theory. We will refer to these states as \textit{vector meson Floquet condensates}, following \cite{Hashimoto_2016,Kinoshita_2017}. Actually, these $E=0$ Floquet states exist also for Minkowski embeddings in a finite range of frequencies, $\Omega_c\le\Omega\le \Omega_{meson}$, where $\Omega_{meson}$ denotes the mass of a vector meson in the supersymmetric defect theory. These masses were computed in \cite{Arean_2006} and are given by
\begin{equation}
    \Omega_{meson}/m=2\sqrt{\left(n+\frac{1}{2}\right)\left(n+\frac{3}{2}\right)}~,\qquad\qquad n=0,1,2,...
    \label{eq:mesonfreqs}
\end{equation}
For the zero-temperature D3/D5 intersection, these frequencies can be obtained analytically by considering the linearized fluctuations of the brane, and imposing that the critical electric field vanishes. This was done in \cite{Garbayo_2020}, and in Section \ref{sec:mesons} we extend it to non-zero temperature, which has to be solved numerically. For comparison, at zero temperature the critical frequencies and the meson frequencies are given by
\begin{equation}
\begin{aligned}
    \Omega_c/m & = 1.497,~3.531,~5.568,~7.585,~...\\
    \Omega_{meson}/m & = 1.732,~3.873,~5.916,~7.937,~...
\end{aligned}
\end{equation}

From Fig. \ref{fig:LobesEJD3D5} we can observe that the effect of increasing the temperature is a depletion of the  height of the lobes with the rising of $r_h/m$, until they fully disappear beyond some temperature. Let us pause to describe the origin of this damping effect.  We choose to measure  dimensionful quantities in units of the quark mass. In particular, the curves above are drawn each one for a fixed value of $r_h/m$. Remember that both the electric field and the temperature tend to bend the probe brane towards the origin in the IR. Let us fix a mass  $m=1$ for concreteness. Then, for a small value of $r_h$, we can still switch on and fine tune the electric field to make the embedding bend  enough so as to touch the critical surface. As $r_h$ grows, this supplemental field needed becomes less and less, which accounts for the drop in the lobe structure to be seen on the  plots in  Fig. \ref{fig:LobesEJD3D5}. Finally, there is a maximum value for $r_h/m=0.8897$ beyond which all the embeddings are of black hole type for any value of $|E|$ and $\Omega$.

Fig. \ref{fig:JCvsE} unfolds the fine structure in the vicinity of the first vector meson Floquet condensate. The left (right) plots show the values of the current $|J|/m^2$ (condensate $C/m^2$) as a function of the applied electric field, $|E|/m^2$, for different values of the frequency $\Omega/m$, close to the first resonant frequency $\Omega_c$. From top to bottom, the temperature increases parametrized by $r_h/m$. The upper case, with $r_h/m=0.5$ is almost indistinguishable from the case with $r_h/m=0$ studied in \cite{Kinoshita_2017,Garbayo_2020}. Points on the continuous (dashed) lines are for  black hole (Minkowski) embeddings. From the lower left corner, all curves start at the  Minkowski solution with $|J|=|E|=0$ (no singular shell). For $\Omega<\Omega_c$, the behavior is as shown in the green curves with $\Omega/m=1.2$. As the electric field increases, a non-dissipative polarization current builds up along the dashed portion of the green curves. At some point, the curve becomes multivalued, the prelude of a presumably discontinuous phase transition to a black hole configuration (a point on the continuous curve segment upwards) where a dissipative conduction current is allowed. The nature and exact occurrence of this transition is beyond the reach of equilibrium thermodynamics where free energy evaluation is enough. We will comment more about these issues in Section \ref{sec:noneqthermo}.

\begin{figure}
    \centering
    \begin{subfigure}[t]{0.48\textwidth}
        \centering
        \includegraphics[width=\textwidth]{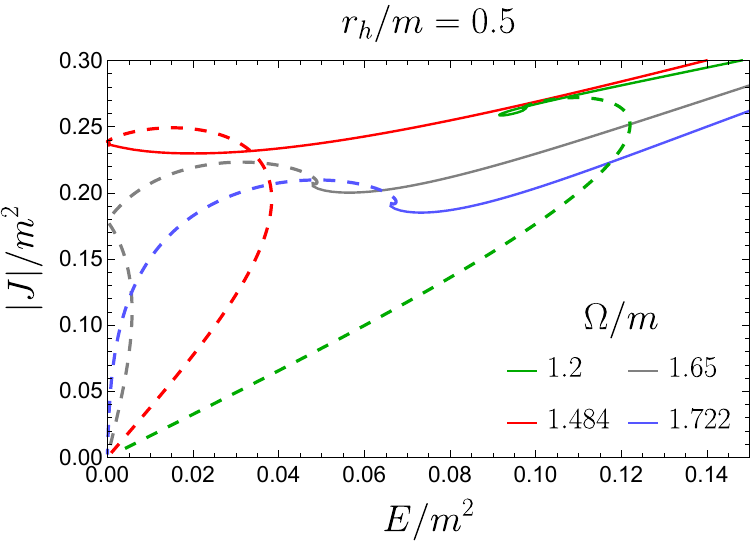}
        \label{sfig:JErh0p5}
    \end{subfigure}
    \hspace{0.02\textwidth}
    \begin{subfigure}[t]{0.48\textwidth}
        \centering
        \includegraphics[width=\textwidth]{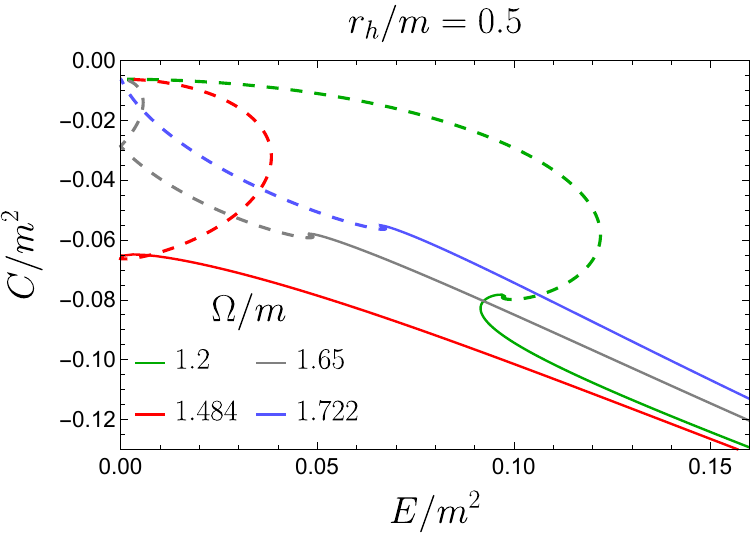}
        \label{sfig:cErh0p5}
    \end{subfigure}
    
    \centering
    \begin{subfigure}[t]{0.48\textwidth}
        \centering
        \includegraphics[width=\textwidth]{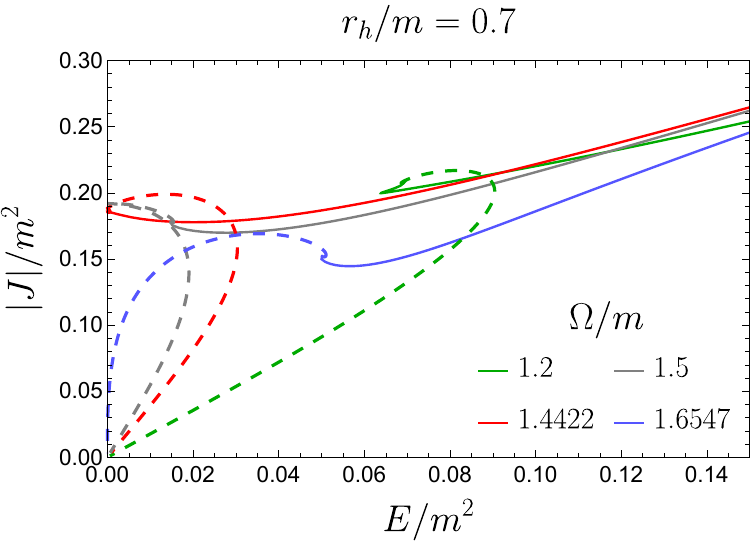}
        \label{sfig:JErh0p7}
    \end{subfigure}
    \hspace{0.02\textwidth}
    \begin{subfigure}[t]{0.48\textwidth}
        \centering
        \includegraphics[width=\textwidth]{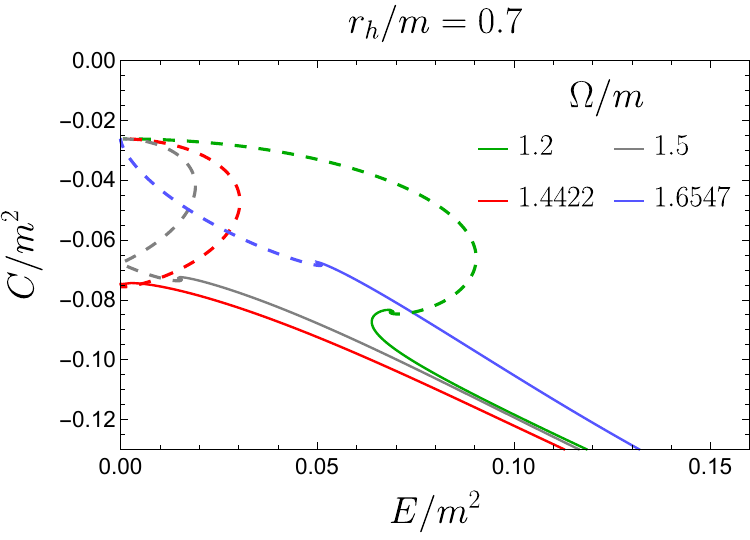}
        \label{sfig:cErh0p7}
    \end{subfigure}

    \centering
    \begin{subfigure}[t]{0.48\textwidth}
        \centering
        \includegraphics[width=\textwidth]{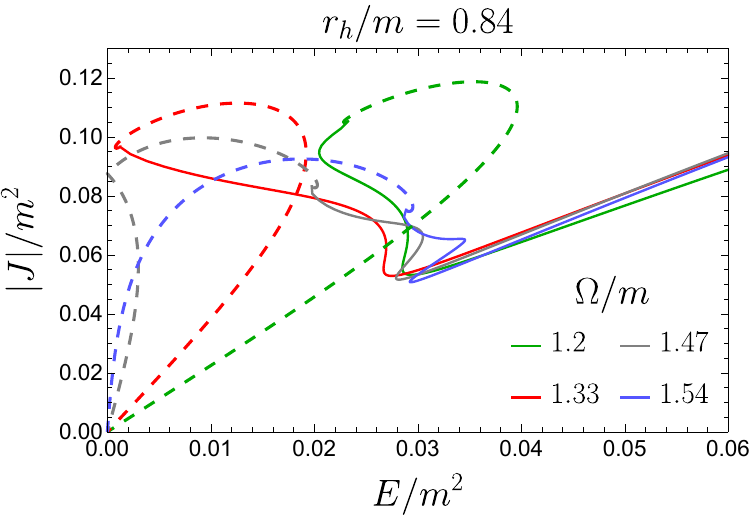}
        \label{sfig:JErh0p84}
    \end{subfigure}
    \hspace{0.02\textwidth}
    \begin{subfigure}[t]{0.48\textwidth}
        \centering
        \includegraphics[width=\textwidth]{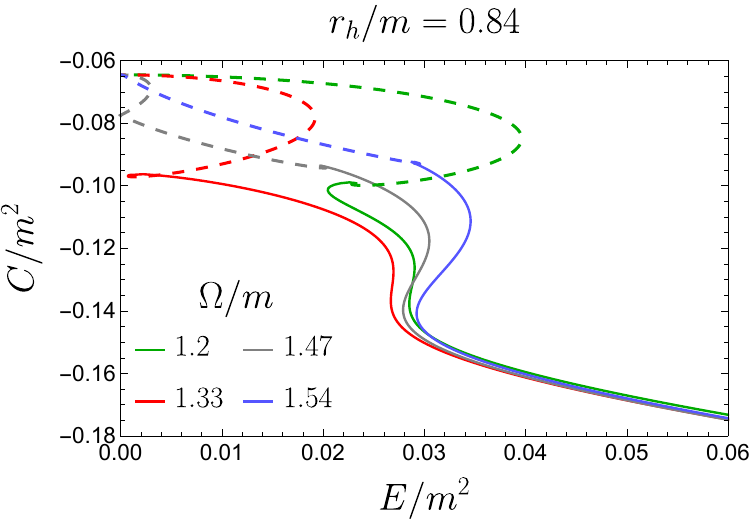}
        \label{sfig:cErh0p84}
    \end{subfigure}
    \caption{We show the electric current $|J|/m^2$ (left) and condensate $C/m^2$ (right) versus electric field for $r_h/m=0.5$, $0.7$ and $0.84$, from top to bottom. The dashed curves represent the insulator (Minkowski) phase and the solid curves the conductive (BH) phase (beware the opposite code with respect to \cite{Garbayo_2020}). In all figures, we show curves for a driving frequency $\Omega<\Omega_c$ (green), $\Omega=\Omega_c$ (red), $\Omega_c<\Omega<\Omega_{meson}$ (gray) and $\Omega=\Omega_{meson}$ (blue). We show that $E/m^2$ can only vanish for $J/m^2\neq0$ for Minkowski embeddings with $\Omega_c\leq \Omega\leq \Omega_{meson}$.}
    \label{fig:JCvsE}
\end{figure}

When increasing the frequency $\Omega$, all the curves get displaced towards the left and, at the value $ \Omega_c$, which depends on $r_h/m$, they contact the axis $E/m^2=0$ at a non-zero value for both $|J|/m^2$ and $C/m^2$ (red curves). Precise computation reveals the contact point to correspond to a critical embedding, as it is evident from the change dashed $\to$ solid at the contact point. This is the first zero in Fig. \ref{fig:LobesEJD3D5}. Further increase in $\Omega$ makes this contact point slide down the vertical axis, now inside the Minkowski branch. Eventually, it reaches zero, merging again with the trivial Minkowski embedding with  $|E|=|J| = 0$. All the embeddings having $|J|/m^2\neq 0$ with $|E|/m^2=0$ build the manifold of \textit{vector meson  Floquet condensates} \cite{Kinoshita_2017}, which exist for $\Omega_c\le\Omega\le \Omega_{meson}$, as already mentioned before.

\begin{figure}
    \centering
    \begin{subfigure}[t]{0.48\textwidth}
        \centering
        \includegraphics[width=\textwidth]{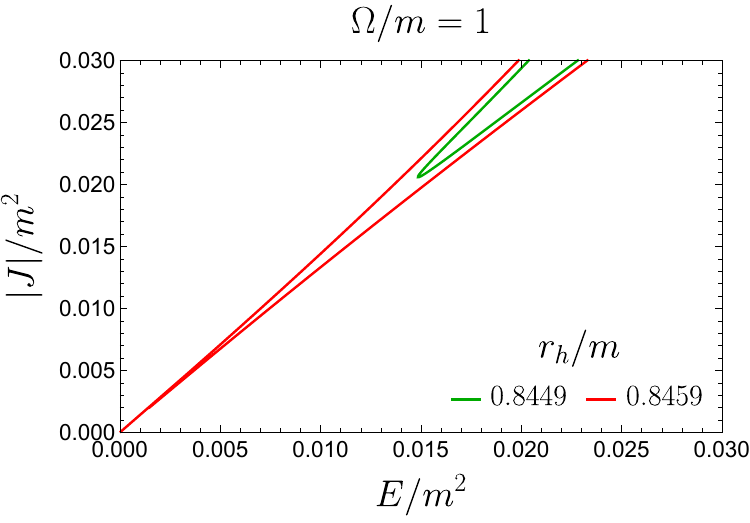}
    \end{subfigure}
    \hspace{0.02\textwidth}
    \begin{subfigure}[t]{0.48\textwidth}
        \centering
        \includegraphics[width=\textwidth]{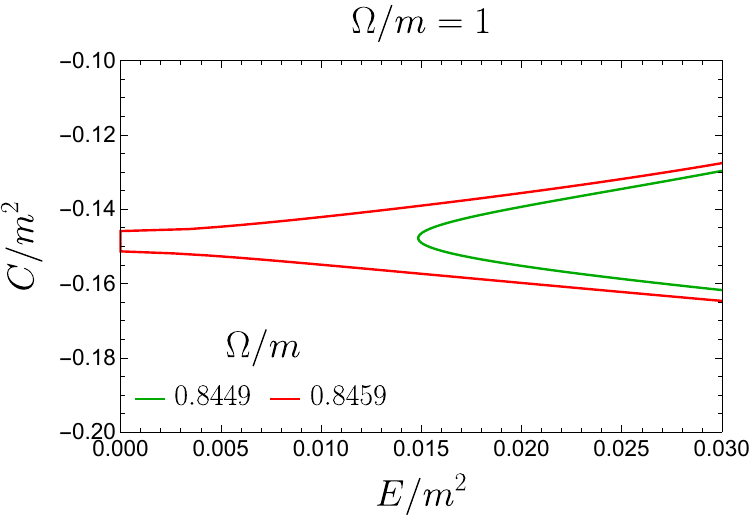}
    \end{subfigure}
    
    \centering
    \begin{subfigure}[t]{0.48\textwidth}
        \centering
        \includegraphics[width=\textwidth]{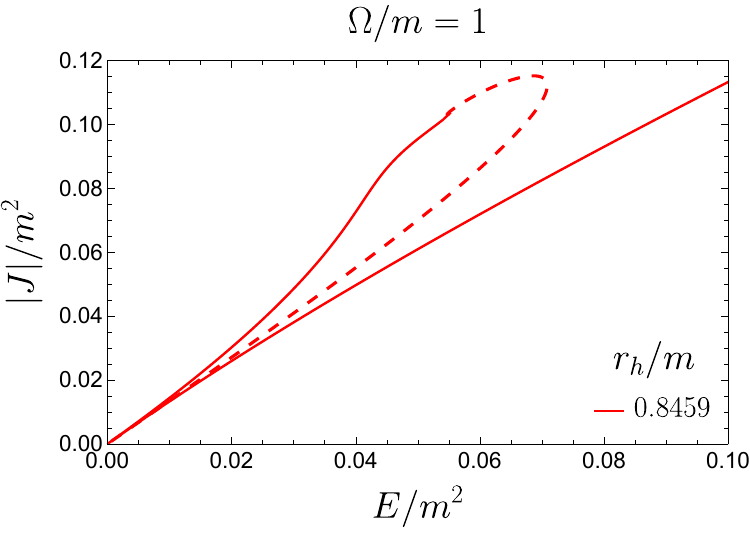}
    \end{subfigure}
    \hspace{0.02\textwidth}
    \begin{subfigure}[t]{0.48\textwidth}
        \centering
        \includegraphics[width=\textwidth]{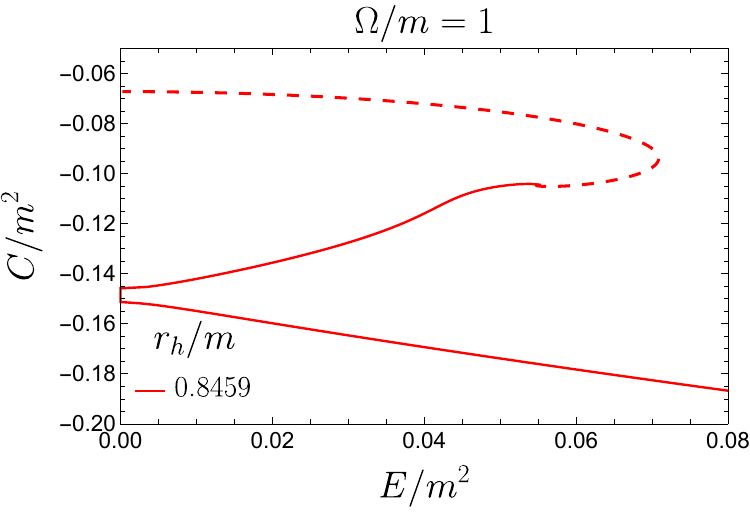}
    \end{subfigure}
    \caption{Electric current (left) and condensate (right) versus $|E|/m^2$ for $\Omega/m=1$. The continuous lines represent BH embeddings, while the dashed ones are the Minkowski embeddings. The upper plots show that for $r_h/m$ around 0.845, we start getting $E/m^2=0$ for BH embeddings. The lower plots are the full curves for the $r_h/m=0.8459$ case. Notice that the electric field increases again, to recover the limit $|J| = \sigma |E|$.}
    \label{fig:JFullCurves}
\end{figure}

When $r_h/m\simeq 0.84$ we start seeing an interference between the two horizons as they come close together. The effect is a multivaluedness in the curve of black hole embeddings (solid lines) that precludes the monotonic growth of $|J|/m^2$ and $|C|/m^2$ with $|E|/m^2$ that is seen at lower temperatures. In all cases, in the large-field regime, the Ohmic behavior $|J| = \sigma |E|$, with constant $\sigma$, is reached. We prove this in Appendix \ref{app:analyticsols}, where we find analytic solutions for the fields in this limit. However, in the small interval  $r_h/m \in (0.84, 0.8897)$, this regime is not approached monotonically, and we find a multivaluedness of $|J|/m^2$ and $C/m^2$ as functions of $|E|/m^2$. This looks similar to the multivaluedness encountered in \cite{Ishii_2018} within the superconducting phase. We, however, encounter this multivaluedness in the conducting phase (the normal phase there). In Fig. \ref{fig:JFullCurves}, we see the interference effect between the two nearby horizons is so effective that a trivial configuration with  $|J|=|E|=0$  is again attained, but now inside the branch of black hole embeddings.

\begin{figure}[H]
    \centering
    \begin{subfigure}[t]{0.48\textwidth}
        \centering
        \includegraphics[width=\textwidth]{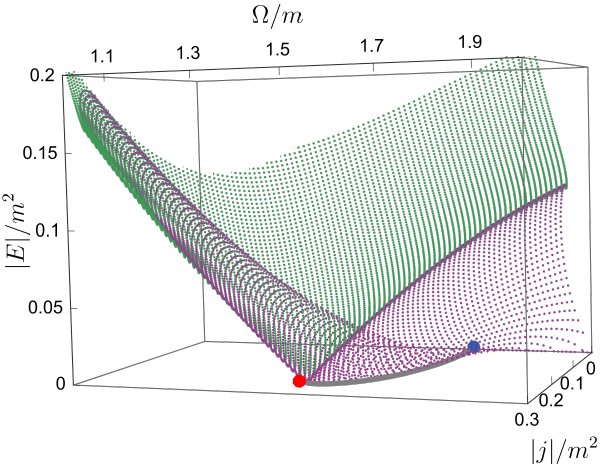}
    \end{subfigure}
    \hspace{0.02\textwidth}
    \begin{subfigure}[t]{0.48\textwidth}
        \centering
        \includegraphics[width=\textwidth]{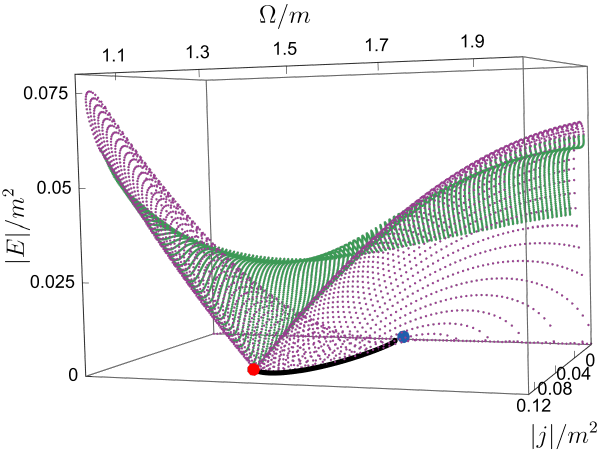}
    \end{subfigure}
    
    \centering
    \begin{subfigure}[t]{0.48\textwidth}
        \centering
        \includegraphics[width=\textwidth]{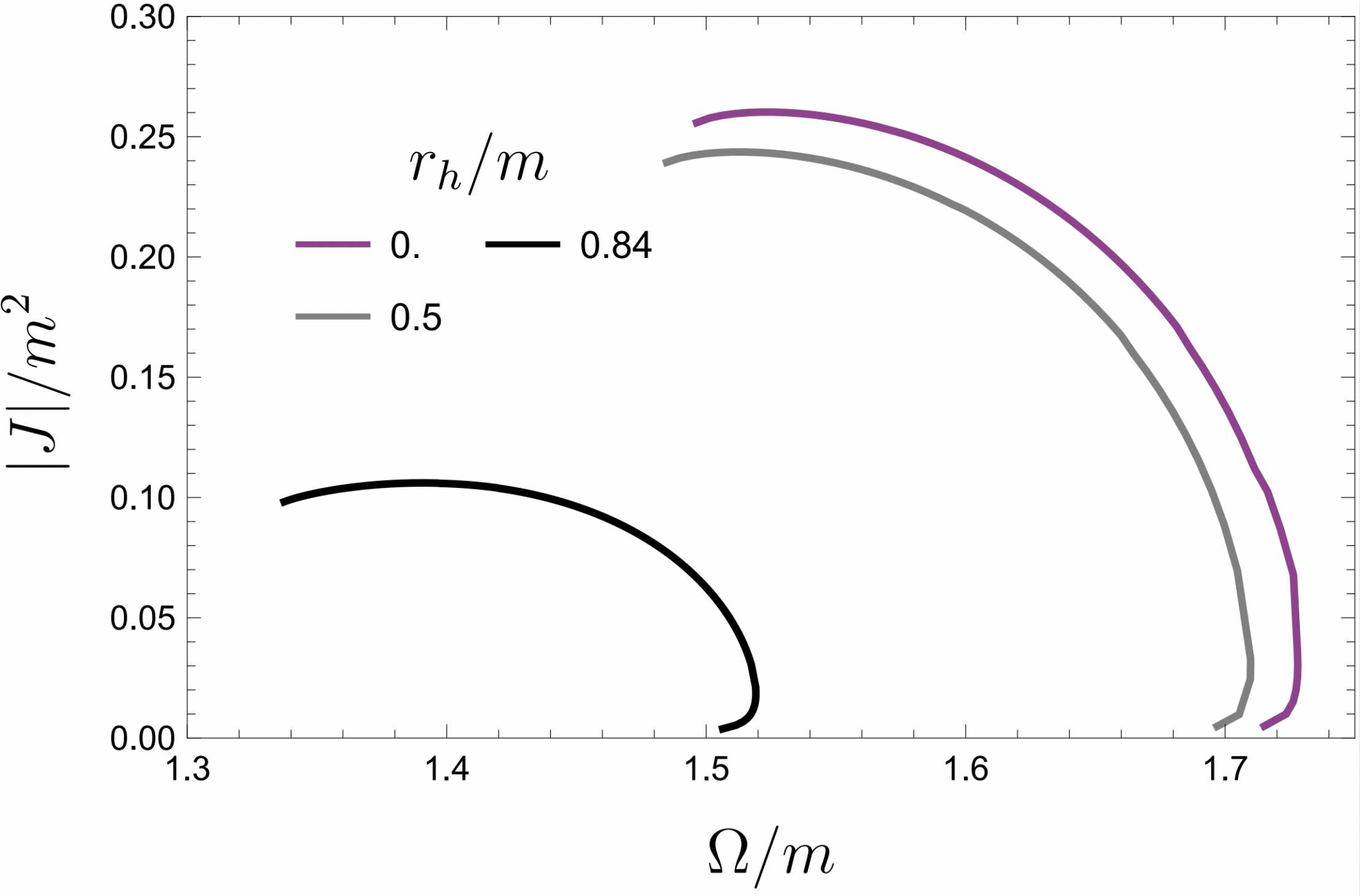}
    \end{subfigure}
    \hspace{0.02\textwidth}
    \begin{subfigure}[t]{0.48\textwidth}
        \centering
        \includegraphics[width=\textwidth]{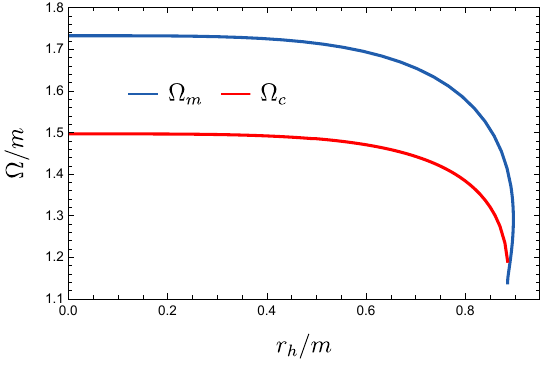}
    \end{subfigure}
    \caption{\textit{Top:} 3D development of the lobed curves in Fig. \ref{fig:LobesEJD3D5} in the vicinity of the first critical point, with an extra $\Omega$ axis, for $r_h/m=0.5$ (left) and $r_h/m=0.84$ (right). Purple (green) surfaces belong to the Minkowski (BH) phase.  Beware the difference in vertical scales. The gray thick lines represent the $|E|=0$ vector meson Floquet condensates, from $\Omega_c$ (red dot) to $\Omega_{meson}$ (blue dot). \textit{Lower left:} vertical view of the upper left plot, where the gray lines in that plot have been graphed for three different temperatures. \textit{Lower right:}  movement of the two extreme points in the curve as a function of $r_h$. They are the critical embedding and the first meson mass frequency respectively. We see that the effect of the background temperature is  mild (less than 10 \%) until the value $r_h/m \sim 0.8$ is reached.}
    \label{fig:3D}
\end{figure}

In Fig. \ref{fig:3D} we zoom again in the region near the first resonance. We have promoted the frequencies in each of the plots in Fig. \ref{fig:JCvsE} to a third $\Omega$ axis, where the associated curves of \ref{fig:JCvsE} are sections of a 3D surface. The surface "touches" the bottom plane $|E|/m^2=0$ in a curve which is the full manifold of vector meson Floquet condensates. This curve interpolates between two endpoints. On one end, $|J|/m^2\neq0$ (red dot), corresponding to $\Omega_c$. On the other end (blue dot), $|J|/m^2\neq0$ and we find the frequency corresponding to the mass of the first vector meson condensate, \ie, the fluctuations in the probe brane worldvolume gauge field \cite{Arean_2006}. As mentioned before, at non-zero temperature these masses  \noindent have to be found numerically, which we do in Section \ref{sec:mesons} and check they correspond to this point (see Fig. \ref{fig:mesons} for the first three meson masses). Turning on $r_h$ causes an overall shift of these curves ($\Omega_c$ and $\Omega_{meson}$) towards lower values of $\Omega$, which can be inferred from the movement of the extreme points as shown on the lower right plot in Fig. \ref{fig:3D}.

On general grounds, the influence of the temperature on the results at $r_h/m=0$ is small until we approach the maximum temperature $r_h/m = 0.8897$, when the lobes in Fig. \ref{fig:LobesEJD3D5} are very small. Regarding the upper plots in Fig. \ref{fig:3D}, notice how the manifold of black hole embeddings folds down for the right plot $r_h/m=0.85$, in contrast with the monotonic growth on the left one, at $r_h/m=0.5$. This is precisely the multivaluedness remarked for higher temperatures inside the branch of BH embeddings, also observed in Fig. \ref{fig:JCvsE}.

To finish this section, let us comment on the possibility of accurately locating the first-order phase transitions where the phase space curves become multivalued. The usual prescription of comparing the free energies is valid in thermodynamical equilibrium. The usual holographic prescription that proposes the euclidean gravitational action for such a construct is not working properly in the present context of a non-equilibrium steady state (see Section \ref{sec:noneqthermo} for details, and also \cite{Ishii_2018} for similar concerns and \cite{Kundu_2019} for a review on the topic). A more sophisticated approach using techniques tailored for non-equilibrium open systems as applied to the holographic context is an interesting project to carry out also here. Eventually, an exact dynamical simulation with a slowly varying $|E|/m^2$ should be the right thing to do.

\subsection{Floquet supression points}\label{subsec:suprpoints}

In Fig. \ref{fig:LobesEJD3D5}, on the right plot, we already mentioned the presence of points within the line of critical embeddings where the current vanishes $\abs{J}=0$, even in the presence of a non-zero electric field. They roughly coincide with the points where the electric field becomes maximal within the same family. We will term these points \textit{Floquet suppression points} and the corresponding states \textit{Floquet-suppressed states}. As we will show, the existence of these points extends to the Minkowski embeddings and, in a sense to be explained in the next section, also to the black hole embeddings. 
Focusing on critical and Minkowksi embeddings, we have already mentioned that the current has its origin in the polarizability, $\tilde\pi$, of the vacuum, $\mathcal{P} = \tilde\pi \mathcal{E}$, with $\mathcal{P}= \langle \bar{\psi} (\gamma_x +i\gamma_y) \psi \rangle$ \cite{Kinoshita_2017}. Hence $\mathcal{J} = \dot{\mathcal{P}} =i\Omega \tilde\pi \mathcal{E}$ and a vanishing value of $\mathcal{J} = 0$ implies $\tilde\pi = 0$, i.e.  the polarizability is \textit{dynamically} suppressed.

This suppression of the vacuum polarizability for certain frequencies is similar to  well known (and searched for) dynamical effects in other examples of Floquet engineering. For example, in periodically driven lattices, hopping between neighbouring sites, although present in the bare hamiltonian, can be completely suppressed by tuning the ratio of frequency to amplitude, leading to induced dynamical localization (see \cite{Bukov_2014,Oka_2018,Giovannini_2019} for references).

\begin{figure}[H]
    \centering
    \begin{subfigure}[t]{0.48\textwidth}
        \centering
        \includegraphics[width=\textwidth]{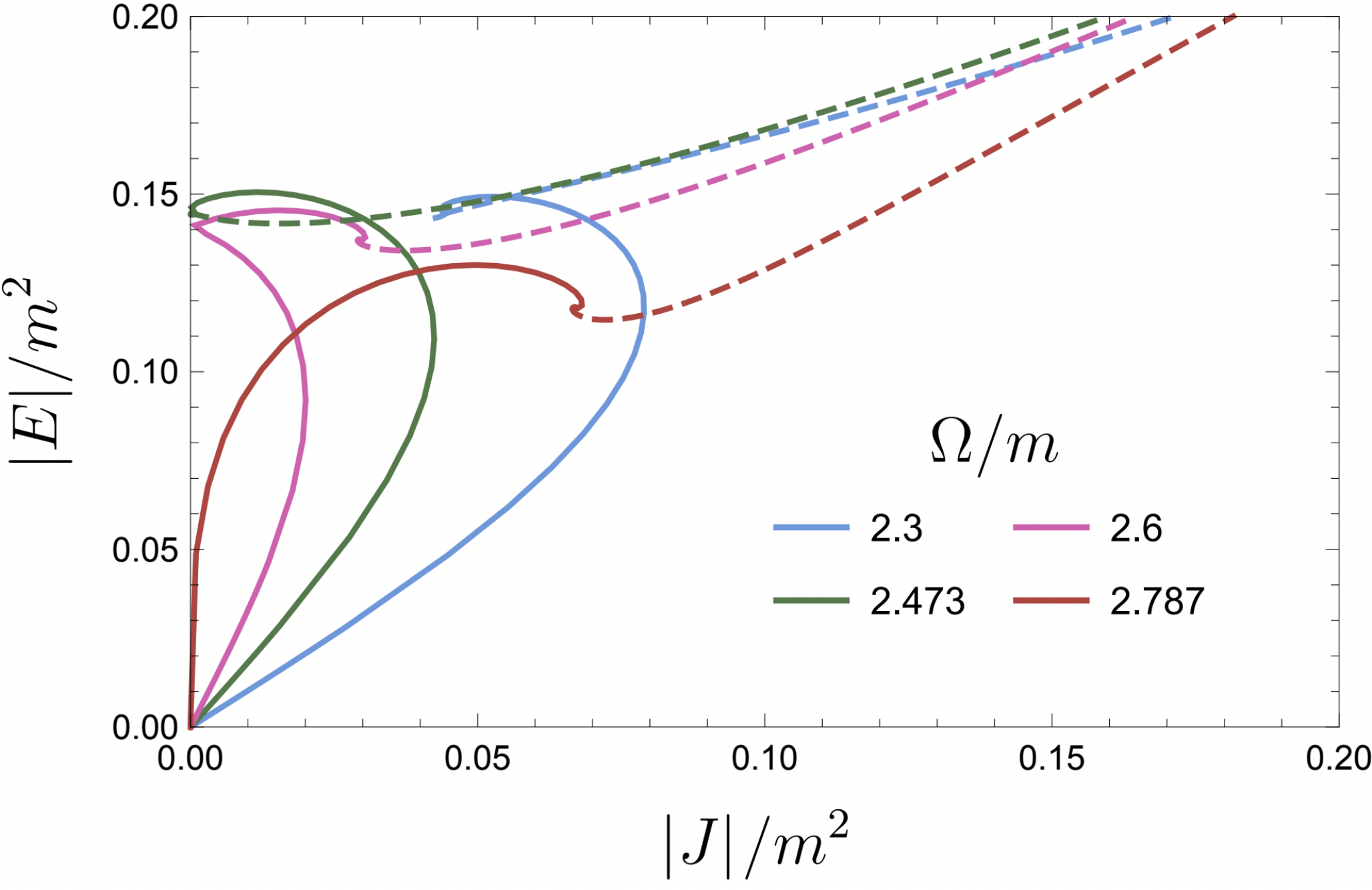}
    \end{subfigure}
    \hspace{0.02\textwidth}
    \begin{subfigure}[t]{0.48\textwidth}
        \centering
        \includegraphics[width=\textwidth]{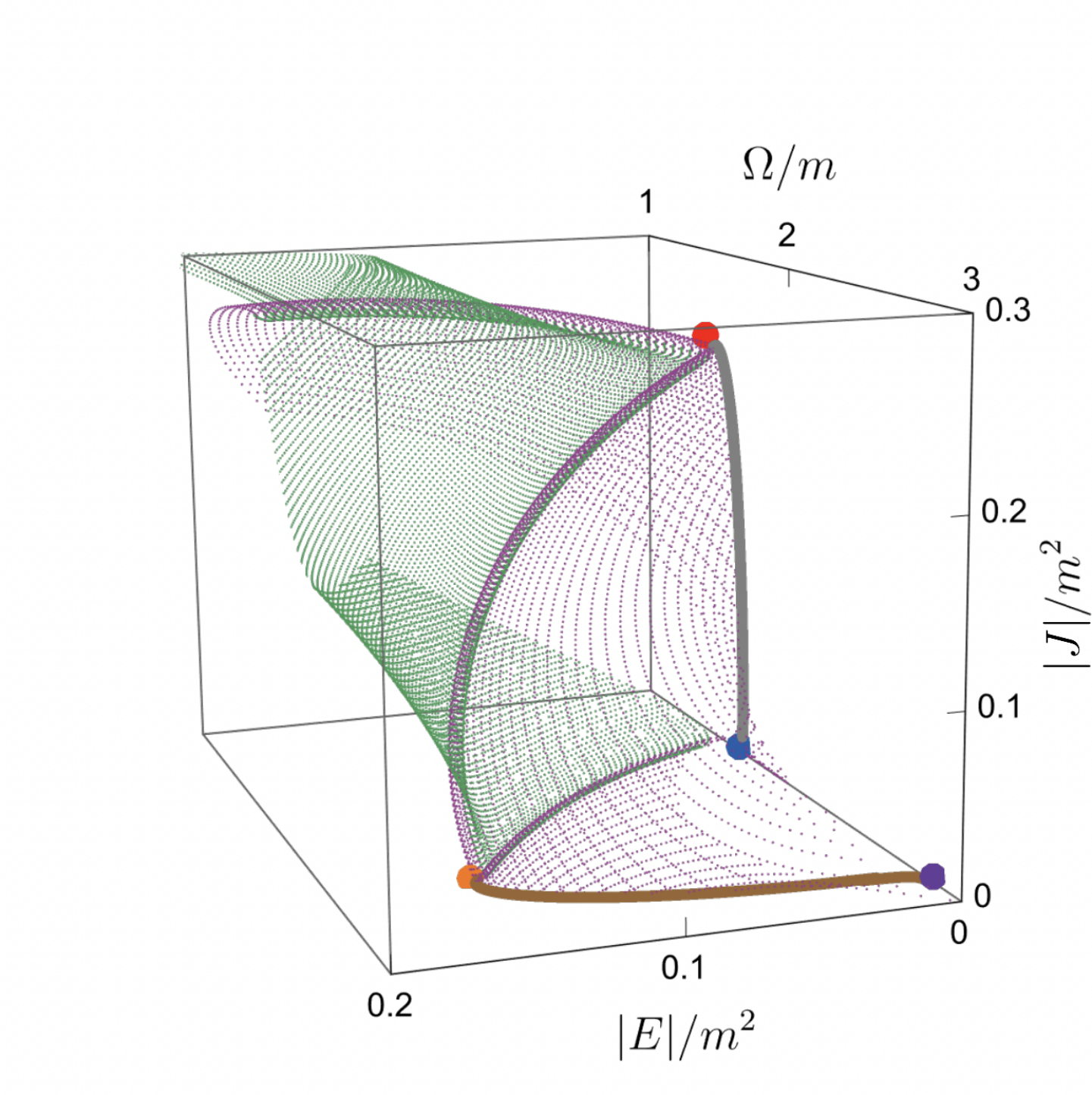}
    \end{subfigure}
    
    \centering
    \begin{subfigure}[t]{0.48\textwidth}
        \centering
        \includegraphics[width=\textwidth]{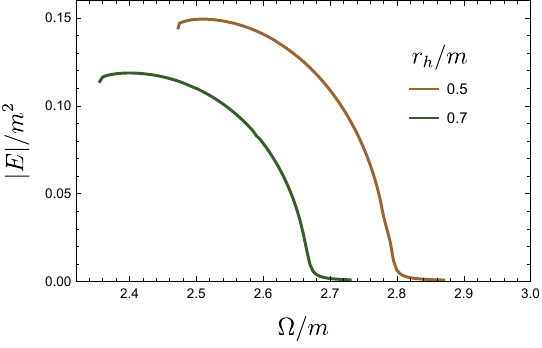}
    \end{subfigure}
    \hspace{0.02\textwidth}
    \begin{subfigure}[t]{0.48\textwidth}
        \centering
        \includegraphics[width=\textwidth]{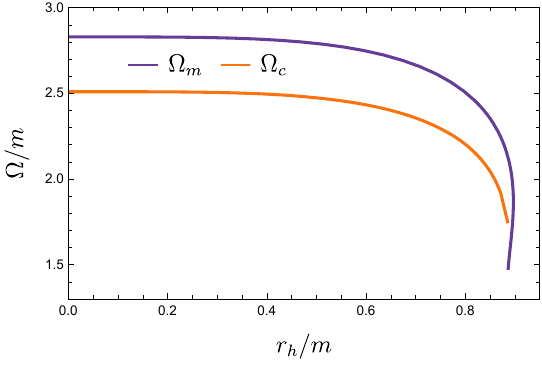}
    \end{subfigure}
    \caption{\textit{Top:} On the left, plots of $|J|/m^2$ vs. $|E|/m^2$, where we exchanged the axes to make apparent the striking similarity with the plots in Fig. \ref{fig:JCvsE}. On the right, 3D plot where the range of $\Omega$ has been extended to cover the first vector meson Floquet condensate (grey curve at $|E|=0$, now vertical), as well as the first Floquet suppressed condensate (brown curve at $|J|=0$, horizontal). \textit{Bottom:} The left plot shows view of the $|J|=0$ plane of the upper plots. The suppressing effect of the temperature is apparent. The right plot is the downshift in $\Omega$ of the two extreme points (orange and violet) in the plot above this, as a function of $r_h/m$.}
    \label{fig:3Dj}
\end{figure}

Fig. \ref{fig:JCvsE} was built by scanning frequencies $\Omega \in (1.2, 1.7)$, \ie, around the point of the first vector meson Floquet condensate. In Fig. \ref{fig:3Dj} we show, on the upper left plot, the similar curves setting instead $\Omega \in (2.3,2.7)$, that is, in a small interval around the first Floquet suppression point. The result exhibits a remarkable similarity, but with $|J|/m^2$ and $|E|/m^2$ axes exchanged. Indeed the symmetry is not exact, as can be seen by comparing the curves in the lower left plots in Figs. \ref{fig:3D}  and \ref{fig:3Dj}. These are, respectively, the curves of vector meson Floquet condensates and Floquet suppression points. In the same way as  \noindent for the vector meson condensates, the frequencies at the endpoints corresponding to the Minkowski branch can be obtained by studying the linearized fluctuations of the worldvolume gauge field, subject to the boundary condition $|J|=0$, in this case. At zero temperature this calculation can be performed analytically giving (see \cite{Garbayo_2020} eqs. (C.2) and (C.3))
\begin{equation}
    \abs{J} = 0 \quad \to \quad\Omega_k/m  = 2\sqrt{k(k+1)}=2.828,~4.899,~...
\end{equation}
This is the "dual" result of \eqref{eq:mesonfreqs}.

Upon rising the temperature, $r_h>0$, these quantities get shifted downwards, as shown on the lower right plots in Figs. \ref{fig:3D} and \ref{fig:3Dj}. The (almost) symmetry between vector meson Floquet condensates and  Floquet suppression points is highlighted on the top right plot of Fig. \ref{fig:3Dj}, where both manifolds have been included within the same 3D development.

It is worth mentioning that the existence of these Floquet suppression points is not restricted to the D3/D5 system, hence is not apparently linked to the dimensionality. For completeness, we devote Appendix \ref{app:D3D7} to the twin version of this section in the context of a D3/D7 scenario. Apart from the discrepancy in the precise numerical values, the global picture is the same. For example, in Fig. \ref{fig:LobesEJD3D7} we reproduce the lobe structure for the D3/D7 system, which is analogous to the one found in Fig. \ref{fig:LobesEJD3D5} for D3/D5. Also the $(|J|/m^2,|E|/m^2)$
curves in Fig. \ref{fig:JvsED3D7} are very similar counterparts of the ones in Figs. \ref{fig:JCvsE} and \ref{fig:3Dj}.

\section{Conductivities}\label{sec:conductivitiesD3D5}

In this section we analyze the response of the system under the externally applied electric fields, focusing on two disctinct but related notions of conductivity. The first, which we refer to as \textit{non-linear conductivity}, characterizes the non-linear relation between the induced current and the applied electric field. It was mentioned there that, in the massless case, the conductivity is a constant for any time-dependent electric field \cite{Karch_2010}. Our results below agree with this behavior.

The second part of our analysis involves optical conductivities, where we introduce a small linearly polarized probe electric field on top of the existing rotating background field. From the response of the system to this small perturbation we are able to extract AC and DC conductivities, following the proposal of \cite{Oka_2010} and mimicking the strategy in \cite{Hashimoto_2016,Garbayo_2020}.

\subsection{Non-linear conductivity}\label{subsec:nonlinearcond}

The relation between the current vector and the electric field vector defines a \textit{rotating current (RC) conductivity}\footnote{Notice  that  in the rotating frame we write $\sigma_{RC}$ as we are dealing here with a single  Fourier component  $\Omega$  of the rotational time dependence. In general, in the lab frame, we would write instead $\mathcal{J}(t)=\int d\tau \sigma_{RC}(\tau) \mathcal{E}(t-\tau)$. Also the use of complex instead of vector notation is implicit.}

\begin{equation}
    \begin{pmatrix} J_x\\ J_y \end{pmatrix}=\begin{pmatrix} \sigma_{xx} & \sigma_{xy} \\ -\sigma_{xy} & \sigma_{xx} \end{pmatrix}\begin{pmatrix} E_x \\ E_y\end{pmatrix}~,
    \label{eq:sigmaRC}
\end{equation}
where the form of the matrix is dictated by rotating symmetry in the $(x,y)$-plane. Inverting these relation yields
\begin{equation}
    \sigma_{xx}=\frac{E_xJ_x+E_yJ_y}{E_x^2+E_y^2}=\frac{\abs{J}}{\abs{E}}\cos\delta~,\qquad\qquad \sigma_{xy}=\frac{E_yJ_x-E_xJ_y}{E_x^2+E_y^2}=\frac{\abs{J}}{\abs{E}}\sin\delta~,
\end{equation}
where $\delta$ is the angle formed by the vectors $(J_x,J_y)$ and $(E_x,E_y)$ in the $(x,y)$-plane. Therefore, in complex notation we can write it as
\begin{equation}
    J=\sigma_{RC}E~,
\end{equation}
where $\sigma_{RC}=\sigma_{xx}-i\sigma_{xy}$ is a complex number which is, itself, a non-linear function of $|E|$ (by rotational symmetry) and $\Omega$.

\begin{figure}
    \centering
    \begin{subfigure}[t]{0.48\textwidth}
        \centering
        \includegraphics[width=\textwidth]{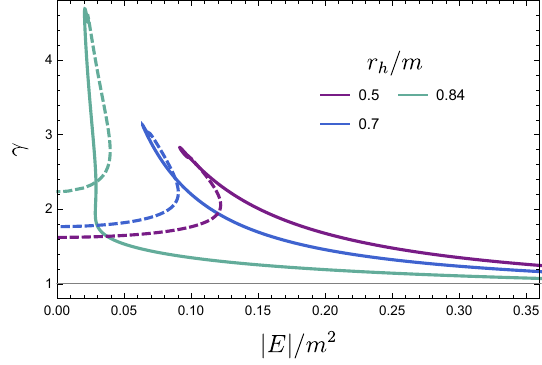}
    \end{subfigure}
    \hspace{0.02\textwidth}
    \begin{subfigure}[t]{0.48\textwidth}
        \centering
        \includegraphics[width=\textwidth]{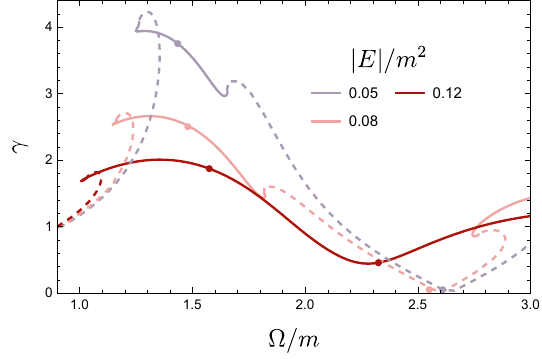}
    \end{subfigure}
    \caption{Modulus of the non-linear conductivity $\gamma = \abs{\tilde\sigma_{RC}}$ as a function of $|E|/m^2$  for fixed $\Omega/m =1.2$ at different temperatures (left plot) or as a function of $\Omega/m$ for fixed values of $|E|/m^2$ at $r_h/m=0.7$ (right plot). The dots are related to the analysis of Fig. \ref{fig:deltaBHrh0p7}. For large $|E|/m^2$ the curves asymptote to the value $\gamma = 1$.}
    \label{fig:NLconduct}
\end{figure}

Writing
\begin{equation}
    \sigma_{RC} = \gamma e^{-i\delta}, 
\end{equation}
the modulus $\gamma$ is the non-linear conductivity, whose value is plotted in Fig. \ref{fig:NLconduct}. The phase $\delta$ encodes the angle between the instantaneous vectors $\vec{J}=(J_x,J_y)$ and $\vec{E}=(E_x,E_y)$. This is why we will refer to $\delta$ as the angle, even if we use complex instead of vector notation. For a given $|E|$ this relative angle controls the Joule heating,
\begin{equation}
    q = \gamma |E|^2 \cos \delta~.
\end{equation}
Its microscopic origin is unclear although we will make an attempt to put forward a consistent picture after we have collected all the bits and pieces.

For Minkowski embeddings, $\vec{J}$ and $\vec{E}$ are perpendicular and $q=0$. This is consistent with the picture of the polarization of the meson condensate into dipoles aligned with the electric field. It leaves two possibilities for $\delta$: $\delta = \pm \pi/2$. In \cite{Kinoshita_2017}, only the positive sign was considered, as it is natural to think that the polarization and the electric field are parallel vectors. We will show here that the existence of both signs is a natural consequence of the presence of Floquet suppression points.

\begin{figure}
    \centering
    \begin{subfigure}[t]{0.48\textwidth}
        \centering
        \includegraphics[width=\textwidth]{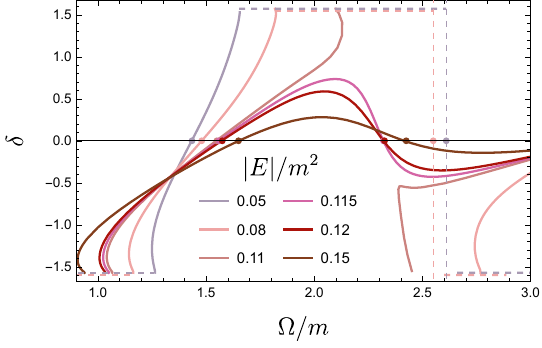}
    \end{subfigure}
    \hspace{0.02\textwidth}
    \begin{subfigure}[t]{0.48\textwidth}
        \centering
        \includegraphics[width=\textwidth]{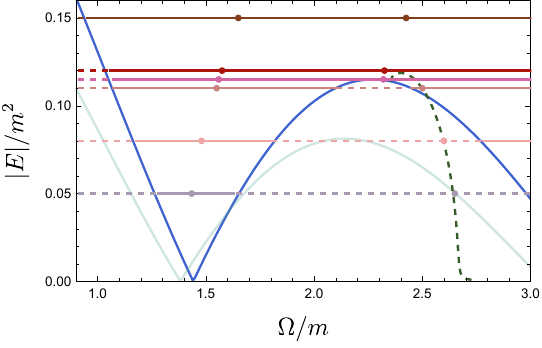}
    \end{subfigure}
    \caption{Continuous lines are black hole embeddings, whereas dashed lines are Minkowski embeddings (with a slight vertical offset for clarity). \textit{Left:} relative angle $\delta$ for different embeddings at various fixed values of the electric field and with $r_h/m=0.7$ for varying $\Omega/m$. \textit{Right:} a zoom of the first lobe region in Fig. \ref{fig:LobesEJD3D5}, where the color code for constant $|E|/m^2$ lines corresponds to the ones on the left plot. The dots indicate the frequencies where the angle $\delta$ either becomes zero in the BH segments, or flips sign in the Minkowski segments. In this later case, joining all the points gives the dashed green curve. The fact that this curve of Minkowski embeddings exits the lower lobe is related to the spiralling multivaluedness of the phase space curves in the vicinity of the critical embeddings. We have added another (dimmed green) lobed curve with higher temperature $r_h/m=0.8$ to show that the effect of rising the temperature is similar to that produced by increasing the electric field $|E|/m^2$.}
    \label{fig:deltaBHrh0p7}
\end{figure}

In Fig. \ref{fig:deltaBHrh0p7} we observe the behavior of the relative angle $\delta$ as we scan embeddings along the horizontal lines of constant $|E|$ while increasing $\Omega$, as shown in the right plot. As usual, solid (dashed) lines correspond to BH (Minkowski) embeddings. On the left plot, using the same color coding, we can see the value of the angle, $\delta$, as we move along these sets of solutions. Notice the jumps $\delta = \pi/2 \to - \pi/2$ that occur within the dashed segment, i.e. for Minkowski embeddings. They seem to reflect a discontinuous transition but this is not the case. Indeed, looking at the right plot in Fig. \ref{fig:NLconduct} we see that, precisely at those points, we find a vanishing value for the module of the polarization current $\gamma=0\to |J|=0$. The corresponding  $\Omega$ frequencies have been signalled with a dot on the right plot in Fig. \ref{fig:deltaBHrh0p7}. Joining all such Floquet suppression points yields the almost vertical green dashed curve of Fig. \ref{fig:deltaBHrh0p7} which is, precisely, the same curve represented in the lower left plot of Fig. \ref{fig:3Dj}, also in (solid) green.

In summary, the transition $\delta = \pi/2 \to - \pi/2$ occurs through a Floquet-suppressed state where the polarizability $\tilde\pi$ vanishes and transits smoothly  from positive to negative. This is remarkable as it states that, for ample intervals in the range of driving frequencies $\Omega$, the polarization of the meson condensate is \textit{antiparallel} to the applied electric field!

Looking back to the left plot in Fig. \ref{fig:deltaBHrh0p7} we notice that the opposite transition $\delta =-\pi/2\to +\pi/2$ is \textit{not} discontinuous. It occurs through a sequence of black hole embeddings that interpolate between those values along a curve that crosses smoothly the axis  $\delta = 0$ with finite slope. A look at the right plot in Fig. \ref{fig:NLconduct} reveals that, in contrast, at those points $\gamma$ stays strictly positive.

Putting  all the information together, the interpretation we find most plausible is as follows: in general, the total current will be a mixture $J = J_{con} + J_{pol}$ of conduction (dissipative) and a polarization (conservative) currents \cite{Hartnoll_2007,Karch_2008}. The precise contribution of each component is controlled by the driving frequency $\Omega$ and by $|E|$. The conduction component $J_{con}$, embodied by deconfined charged carriers, is parallel to the applied electric field. The polarizarion component $J_{pol}$, as explained above, is perpendicular. The vector sum of these two components gives $J$ and $E$ a relative phase angle $\delta$.

Changing $\Omega$ at fixed $|E|/m^2$, like on the left plot in Fig. \ref{fig:deltaBHrh0p7}, we find that $J_{pol}$ vanishes at given frequencies $\Omega$, thereby flipping $\delta = \pm \pi/2\to \mp \pi/2$. In the gapped (Minkowski phase) only this component of the current is present. In the gapless (BH)  phase, both components generically contribute. We interpret  the points where $\delta=0$ as precisely signaling that, there also, $J_{pol}=0$. Thereby the total current becomes parallel to the electric field. This is the reason why the transition in $\delta$ is continuous in the BH phase (solid segments). If this picture makes sense, the conclusion is that \textit{we also have Floquet suppression points within the BH phase}. It is just that in the BH phase this vanishing is masked by the conduction component $J = J_{con}+0$. In summary, all the dots in Fig. \ref{fig:NLconduct} and \ref{fig:deltaBHrh0p7} correspond to Floquet suppressed states. As we approach the boundaries of these segments, the conduction component disappears $J_{con}\to 0$, and the polarization component survives, making $J= J_{pol}$ and $E$ mutually perpendicular again.

For large enough $|E|/m^2$ we always stay within the phase of BH embeddings, and the (solid) curves smoothly relax down to the asymptotic regime where $\delta=0$. We interpret this as the vanishing of the $J_{pol}$ component in this limit. This is the same effect we get for large temperature $T/m\gg 1$ as both are indistinguishable from the limit of small mass $m\to 0$.

In Appendix \ref{app:masslesssol} we prove exactly this fact, $\sigma_{RC}=1$, for massless flavors. This implies that, in this case, the response is both instantaneous and linear. We make contact and fully agree here with the results  in \cite{Karch_2010}. In a sense, the claim  there is stronger as it applies to \textit{any} time dependence of a linearly polarized electric field $\mathcal{E}_x(t)$ at the boundary. Here, on one side, we go to a rotational polarization ansatz and, moreover, in Appendix \ref{app:smallmasssol} we prove this result to hold also at linearized order in a small mass $\delta m$. Linearity of the response entails that it should also extend to arbitrary two-dimensional time dependent electric fields $\vec{\mathcal{E}}(t)$ at linear order in small masses.

\subsection{Photovoltaic optical conductivities}\label{subsec:opticalcond}

The Floquet engineering of an induced Hall effect is termed usually \textit{photovoltaic Hall effect} \cite{Oka_2008}. In \cite{Hashimoto_2016}, following the proposal in \cite{Oka_2010}, the photovoltaic optical response  was obtained for massless charge carriers in the D3/D7 model and an optical Hall current was found. This study was extended in \cite{Garbayo_2020} to massive flavors in the D3/D5 model,  and observed an intricate behavior in the wedge region between the lobes in Fig. \ref{fig:LobesEJD3D5}, with multiple resonance peaks present. The physics in this wedge is presumably controlled by the vector meson Floquet condensate at zero temperature, that signals the presence of a quantum phase transition. In the present work, first, we would like to see how the presence of a temperature affects those results.

In order to study the photovoltaic current of the model, the proposal in  \cite{Oka_2010} is to analyze the response of our system to an additional linearly polarized electric field on top of the circularly driven background \eqref{eq:rotatingE}. In vector cartesian notation, the total electric field is now
\begin{equation}
    \vec{\mathcal{E}}(t)=O(t)\vec{E}+\vec{\epsilon}(t)=O(t)\vec{E}+\vec{\epsilon}\h e^{-i\omega t}~,
\end{equation}
where $\vec\epsilon$ is a constant vector such that $\abs{\vec{\epsilon}}\ll|\vec{E}|$. We want to extract the effective conductivities that arise from the effect of this perturbation on the current. In particular, we expect a change of the form $\vec{\mathcal{J}}(t)\to O(t)\vec J+\delta \vec J(t)$, giving rise to an effective conductivity defined by
\begin{equation}
    \delta\vec{J}(t)=\boldsymbol{\sigma}\cdot\vec{\epsilon}(t)~.
\end{equation}
The goal is to determine $\boldsymbol{\sigma}$.

The perturbation of the electric field will also mean a change in the gauge potential $\vec{a}+\delta \vec{a}$, so that  $c(r)$ develops a time dependent perturbation
\begin{equation}
    \vec{a}(t,r)+\delta \vec{a}(t,r)=O(t)\Big(\vec{c}(r)+\delta\vec{c}(t,r)\Big)~.
\end{equation}
Now the bulk gauge potential $\vec{a}(t,r)+\delta\vec{a}(t,r)$ has to match the full electric field at the boundary,
\begin{equation}
    \vec{a}(t,r=\infty)+\delta \vec{a}(t,r=\infty)=-\frac{1}{\Omega}O(t)\epsilon \vec{E}-\frac{i}{\omega}\vec{\epsilon}\h e^{-i\omega t}~.
    \label{eq:pert}
\end{equation}
Since the fluctuations of the gauge field will also couple to the embedding functions,  $\theta(r)\to\theta(r) + \delta \theta(t,r)$, and we will be dealing with a 3 component vector of fluctuations, $\delta \vec \xi(t,\rho) = (\delta c_x,\delta c_y,\delta \theta)$. The general formalism to study these fluctuations was developed in \cite{Garbayo_2020}, following the analysis of \cite{Hashimoto_2016} for the massless D3-D7 system. We devote Appendix \ref{app:opticalcond}  for the details of the calculation and here we only outline the main results.

\begin{figure}[H]
    \centering
    \begin{subfigure}[t]{0.48\textwidth}
        \centering
        \includegraphics[width=\textwidth]{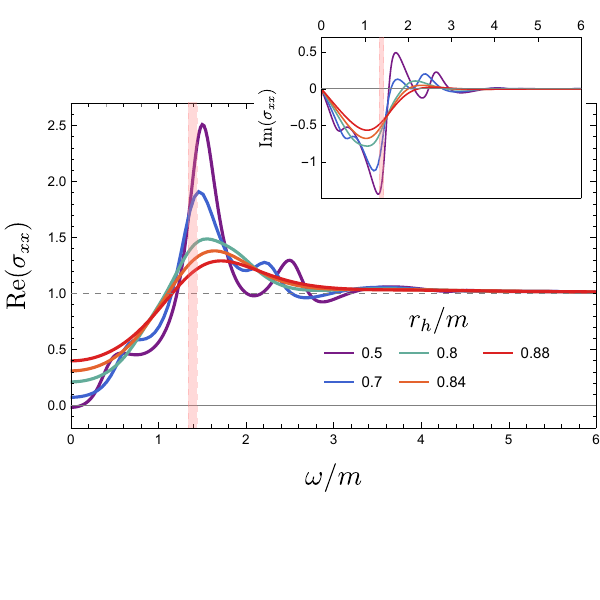}
    \end{subfigure}
    \hspace{0.02\textwidth}
    \begin{subfigure}[t]{0.48\textwidth}
        \centering
        \includegraphics[width=\textwidth]{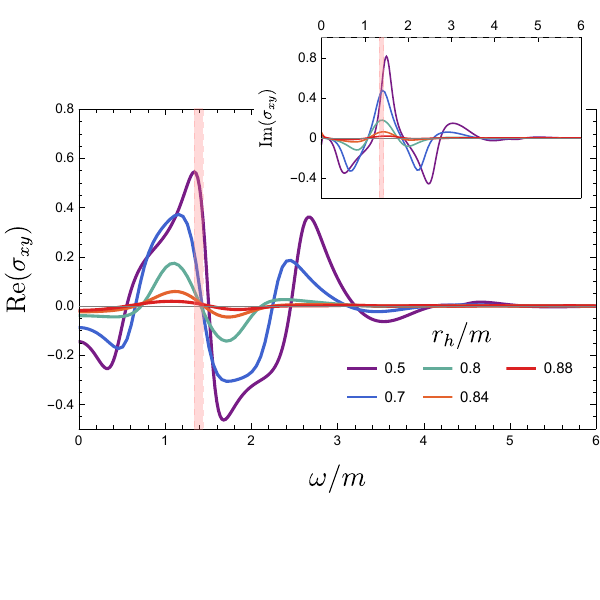}
    \end{subfigure}
    
    \centering
    \begin{subfigure}[t]{0.48\textwidth}
        \centering
        \includegraphics[width=\textwidth]{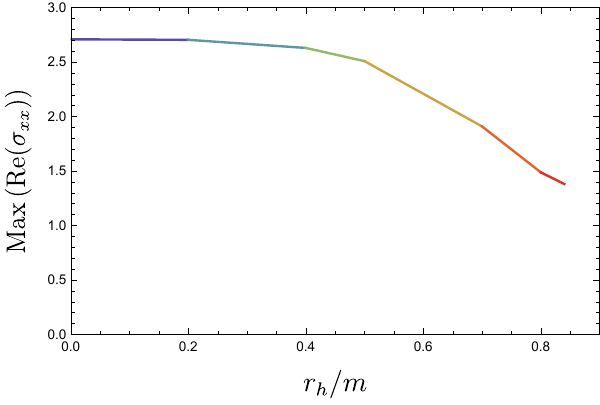}
    \end{subfigure}
    \hspace{0.02\textwidth}
    \begin{subfigure}[t]{0.48\textwidth}
        \centering
        \includegraphics[width=\textwidth]{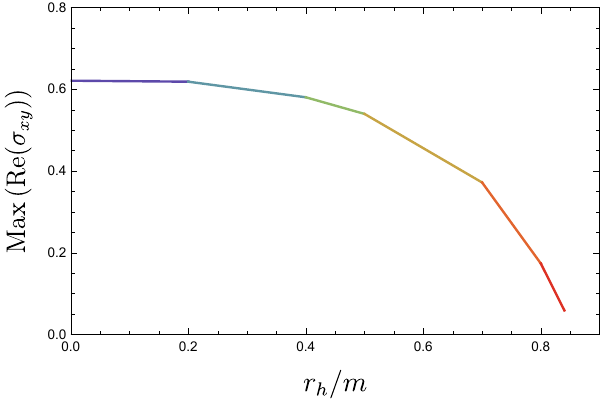}
    \end{subfigure}
    \caption{\textit{Top:} AC conductivities for four values of $r_h/m$ at their corresponding $\Omega_c(r_h)$, for $E/m^2=0.1$. The pink band marks the range where the four values of $\Omega_c(r_h)$ belong. The main peak is close to this region. \textit{Bottom:} Variation with the temperature of the maximum value of the real part of $\boldsymbol{\sigma}_{xx}$ and $\boldsymbol{\sigma}_{xy}$ for $E/m^2=0.1$.}
    \label{fig:sig_rh_Omegac}
\end{figure}

The equations for the perturbations can be solved in terms of the matrices $\mathbf{M}_\pm$, defined as
\begin{equation}
    \textbf{M}_\pm = \frac{1}{2}\begin{pmatrix} 1 & \pm i\\ \mp i & 1 \end{pmatrix}~,
\end{equation}
as well as two other matrices, $\boldsymbol{X}_{\pm}$ which, in general, have to be determined numerically. It can be shown that $\delta \vec{\mathcal{J}}$ contains three modes, which oscillate with frequencies $\omega$ and $\omega\pm2\Omega$ (the heterodyning mixing modes \cite{Oka_2016}),
\begin{equation}
    \delta \vec{\mathcal{J}}=\Big[\boldsymbol{\sigma}(\omega)e^{-i\omega t}+\boldsymbol{\sigma}^{+}(\omega)e^{-i(\omega+2\Omega)t}+\boldsymbol{\sigma}^{-}(\omega)e^{-i(\omega-2\Omega)t}\Big]\vec{\epsilon}~,
    \label{eq:sigmasgeneral}
\end{equation}
where $\boldsymbol{\sigma}(\omega)$, $\boldsymbol{\sigma^+}(\omega)$ and $\boldsymbol{\sigma}^-(\omega)$ are the conductiviy matrices corresponding to the frequencies $\omega$, $\omega+2\Omega$ and $\omega-2\Omega$, and are given by
\begin{equation}
\begin{aligned}
    \boldsymbol{\sigma}(\omega) & = -\frac{i}{\omega}\left(\textbf{M}_+\textbf{X}_+\textbf{M}_++\textbf{M}_-\textbf{X}_-\textbf{M}_-\right)~,\\
    \boldsymbol{\sigma}^+(\omega) & = -\frac{i}{\omega}\textbf{M}_-\textbf{X}_-\textbf{M}_+~,\\
    \boldsymbol{\sigma}^-(\omega) & = -\frac{i}{\omega}\textbf{M}_+\textbf{X}_+\textbf{M}_-~.
\end{aligned}
\end{equation}

See Appendix \ref{app:opticalcond} for details. The results are contained in in Fig. \ref{fig:sig_rh_Omegac}. The curves represent the absortion spectrum $\boldsymbol{\sigma}_{xx}(\omega)$ and the Hall conductivity $\boldsymbol{\sigma}_{xy}(\omega)$. The background rotating electric field has been fixed to $|E|/m^2=0.1$. Its frequency has been set to the first critical frequency $\Omega_c(r_h)$, which decreases with the temperature as seen in the lower right plot of Fig. \ref{fig:3D}. The band in which these values lie for the chosen temperatures has been signaled by a vertical band in pink. The curves for $\boldsymbol{\sigma}_{xx}$ and $\boldsymbol{\sigma}_{xy}$ show a smooth deformation of the ones in \cite{Garbayo_2020} (Fig. 10) for the same value of $|E|/m^2$. 

Succinctly stated, the temperature in the gluon bath in general destroys the AC Hall optical conductivity, and hence, also the DC Hall conductivity. Again, the effect of the temperature is similar to the one caused by an increase of the electric field. Namely, the conductivity peaks get roughened and lowered. In a sense, both agents, the temperature and the electric field, act in parallel by enhancing the amount of deconfined charge carriers in the medium.

\begin{figure}
    \centering
    \begin{subfigure}[t]{0.48\textwidth}
        \centering
        \includegraphics[width=\textwidth]{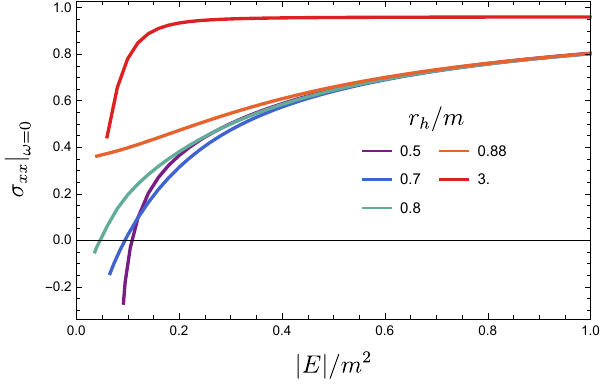}
    \end{subfigure}
    \hspace{0.02\textwidth}
    \begin{subfigure}[t]{0.48\textwidth}
        \centering
        \includegraphics[width=\textwidth]{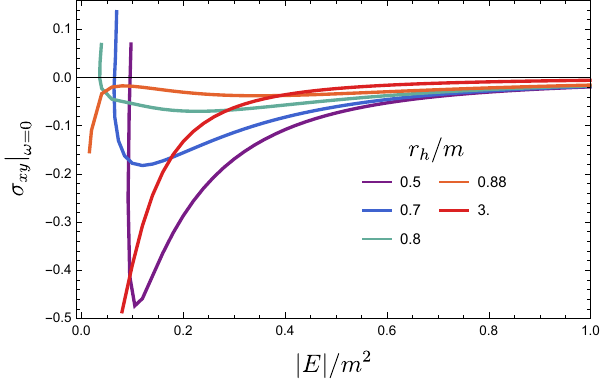}
    \end{subfigure}
    \caption{DC conductivities as functions of $|E|/m^2$ for different values of $r_h/m$ at $\Omega/m=1$. At high temperatures the conductivity tensor tends towards the identity $\boldsymbol{\sigma}_{xx} =1$ and $\boldsymbol{\sigma}_{xy} = 0$.}
    \label{fig:sigDC_rh_Omegac}
\end{figure}

The effect becomes more pronounced beyond some temperature $r_h/m \sim 0.5$, as shown in the lower plots of Fig. \ref{fig:sig_rh_Omegac}. In the large-temperature limit, $r_h \to \infty$, all conductivities, both AC and DC  tend towards $\boldsymbol{\sigma}_{xx} = 1~,~\boldsymbol{\sigma}_{xy} = 0$ (see also Fig. \ref{fig:sigDC_rh_Omegac}). This result is the same we obtained for the rotating current conductivity $\sigma_{RC}$ in the massless limit. Since in this case the electric field is linearly polarized, rather than circularly, we see this as a further evidence in favor of the fact that the response will be Ohmic and instantaneous for an arbitrary time dependence of the electric field in the plane, $J(t) = \sigma E(t)$.

A peculiar observation is that the frequencies $\omega$ of the highest peaks in the absorption spectrum $\Re(\boldsymbol{\sigma}_{xx})$ slightly deviate above the one of the driving $\omega \gtrapprox \Omega$ (within the vertical band in pink). This was also observed in \cite{Garbayo_2020} (Fig. 10), where the drift is seen to be enhanced with increasing $|E|$. We could not offer any explanation to this. Here we can see that there is a very similar shift in the driving frequencies $\Omega$ of the Floquet suppression points inside the BH phase (solid segments in Fig. \ref{fig:deltaBHrh0p7}), with increasing $|E|$ towards the right. We have tried to make sense of this qualitative coincidence but could not find an exact numerical agreement.

\subsubsection{Massless limit}

It was shown in \cite{Garbayo_2020} that, in the massless case, the perturbed system can be solved analytically. Remarkably, the same happens at finite temperature.

In the massless case, the fluctuations of the embedding function decouple from those of the gauge field $\delta \vec{c}$. The unperturbed configuration corresponds to the massless analytic solution obtained in Appendix \ref{app:masslesssol}. Surprisingly, the coupled equations for $\delta c_x$ and $\delta c_y$ can be solved analytically, obtaining the matrices
\begin{equation}
    \textbf{X}_\pm=i\omega\h \mathbf{1}~.
\end{equation}
Plugging these $\mathbf{X}_\pm$ matrices in \eqref{eq:sigmasgeneral}, we obtain the conductivity matrices in the massless case, namely
\begin{equation}
    \boldsymbol{\sigma}(\omega)  =\mathbf{1}~,\qquad\qquad  \boldsymbol{\sigma}^+(\omega)  = \boldsymbol{\sigma}^-(\omega)=0~,
\end{equation}
which is exactly the same result as the one found in \cite{Garbayo_2020} at zero temperature. As in the case of the non-linear current, after correctly normalizing the electric field and current according to \eqref{eq:dictionaryD3D5}, we obtain that the conductivity is given by
\begin{equation}
    \boldsymbol{\sigma}(\Omega)=\frac{2N_f N_c}{\pi\sqrt{\lambda}}\mathbf{1}~,
\end{equation}
where here also the result confirms the expectations put forward in \cite{Karch_2010}. The details of this analytic solution are relegated to Appendix \ref{app:masslessconduc}.

\section{Linearized Minkowski embeddings and meson spectrum}\label{sec:mesons}

We have seen that two different values of the driving frequency play an important role around each resonance. On one hand, the critical frequency $\Omega_c$, at one endpoint of the line of vector meson Floquet condensates, and $\Omega_{meson}$, at the other endpoint (see Fig. \ref{fig:3D}). We have claimed that this endpoint, $\Omega_{meson}$ corresponds to the frequencies of the vector mesons of the theory, obtained from the quadratic scalar fluctuations on the brane and written in \eqref{eq:mesonfreqs}. This observation was proven in \cite{Kinoshita_2017} for the D3/D7 model, and later proven in \cite{Garbayo_2020} for the D3/D5 model at zero temperature. In this section we confirm that this is still true at finite temperature by numerically computing the meson spectrum $\Omega_{meson}$ at different temperatures and comparing with the endpoints of the obtained vector meson condensates.

We begin by writing the DBI action in the coordinates $(\rho,w)$ for the fields $w$ and $c$:
\begin{equation}
\begin{aligned}
    S_{D5}&=-\mathcal{N}\int d\rho \frac{\rho^3}{(\rho^2+w^2)^3}\Bigg[(\rho^2+w^2)^2\left((\rho^2+w^2)^2-u_h^4\right)^2\abs{c'}^2-\Omega^2(\rho^2+w^2)^4\Re(c\bar{c}')^2+\Bigg.\\
    &\phantom{-}\Bigg.+\left((\rho^2+w^2)^2+u_h^4\right)\left(\left((\rho^2+w^2)^2-u_h^4\right)^2-\Omega^2(\rho^2+w^2)^2\abs{c}^2\right)(1+w'^2)\Bigg]^{1/2}~.
    \label{eq:SD5cartesian}
\end{aligned}
\end{equation}

Let us consider the case in which the gauge field $c(\rho)$ vanishes, and let $\tilde w=\tilde w(\rho)$ denote the Minkowski embedding function in such case. It can be obtained by solving the equation of motion of the action $\tilde{S}_{D5}=S_{D5}(c=0)$,
\begin{equation}
    \tilde{S}_{D5}=-\mathcal{N}\int d\rho\frac{\rho^2\left(u_h^4-(\rho^2+\tilde{w}^2\right))}{(\rho^2+\tilde{w}^2)^3}\sqrt{\left[(\rho^2+\tilde{w}^2)^2+u_h^4\right](1+\tilde{w}'^2)}
\end{equation}

Let us  next suppose that we perturb around the $c=0$ solution by making $c\to\delta c$ and $w\to \tilde{w}+\delta w$ in the equations of motion derived from the Lagrangian density \eqref{eq:SD5cartesian}. It is easy to see that, at first order in the variations, the equation for $\delta c$ reads
\begin{align}
\begin{split}
    \partial_\rho\Bigg(&\frac{\rho^2\left(u_h^4-(\rho^2+\tilde{w}^2)^2\right)}{(\rho^2+\tilde{w}^2)\sqrt{\left((\rho^2+\tilde{w}^2)^2+u_h^4\right)(1+\tilde{w}'^2)}} \delta c' \Bigg)\\
    & \hspace{0.25\textwidth}+ \frac{\rho^2 \Omega^2\sqrt{(\rho^2+\tilde{w}^2)^2+u_h^4}}{(\rho^2+\tilde{w}^2)\left(u_h^4-(\rho^2+\tilde{w}^2)^2\right)\sqrt{1+\tilde{w}'^2}}\delta c = 0~.
    \label{eq:eomdeltac}
\end{split}
\end{align}

\begin{figure}
    \centering
    \begin{subfigure}[t]{0.48\textwidth}
        \centering
        \includegraphics[width=\textwidth]{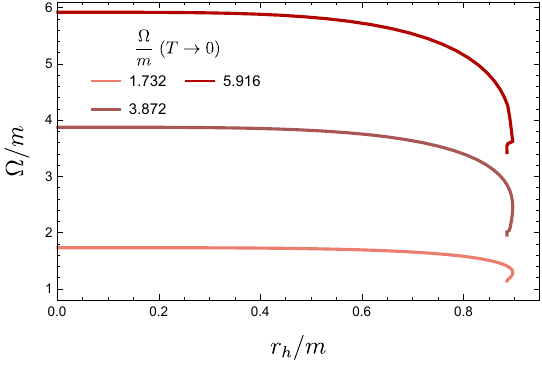}
    \end{subfigure}
    \hspace{0.02\textwidth}
    \begin{subfigure}[t]{0.48\textwidth}
        \centering
        \includegraphics[width=\textwidth]{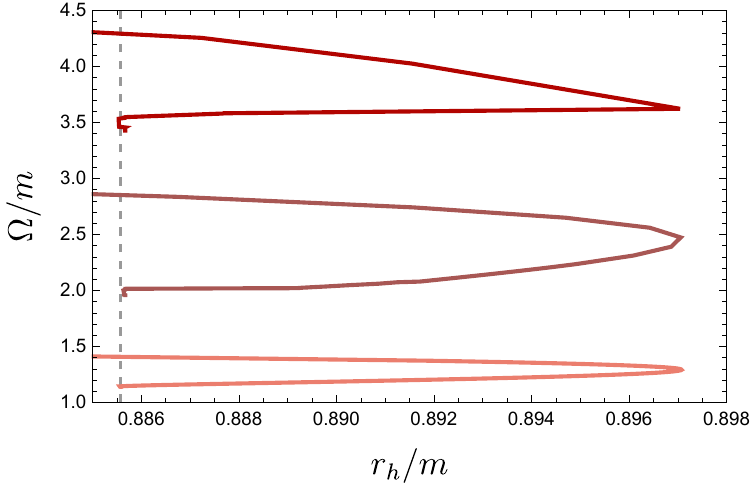}
    \end{subfigure}
    \caption{$\Omega/m$ variation with $r_h/m$. The right figure shows the endpoint of the curves. Both of them are bivaluated between $r_h/m=0.8855$ to $r_h=0.8897$, and stop for $r_h/m=0.8855$.}
    \label{fig:mesons}
\end{figure}

To obtain the resonant frequencies of the mesonic Floquet condensates we integrate \eqref{eq:eomdeltac} and find the solutions that are regular at $\rho=0$ and such that $E=0$ when we reach $J=0$ after each critical driving frequency $\Omega_c/m$ (see the lower plot in Fig. \ref{fig:3D}). This last condition is only possible when the frequency $\Omega/m$ takes values in a discrete set (which depends on the horizon radius $r_h$). For 
$r_h/m=0$, we recover the analytic results of \cite{Garbayo_2020,Arean_2006}. For $r_h/m\neq 0$, the masses of these mesonic states decrease as $r_h/m$ grows. At some value of $r_h/m$ they cease to exist (see Fig. \ref{fig:mesons}), as the mere Minkowski embedding themselves do (see Section \ref{sec:phasespace}). This behavior is dual to the meson melting phenomenon, as discussed in \cite{Mateos_2007}.

\section{Non-equilibrium thermodynamics}\label{sec:noneqthermo}

Before concluding, we briefly comment on the (unsuccessful) application of standard thermodynamic arguments in analyzing the phase structure of our system. A common method for identifying phase transitions involves computing the free energy in both competing phases. This quantity typically develops a characteristic swallow-tail structure, with a crossing point signaling a change in the dominant phase. Our goal is to apply a similar strategy to the insulator/conductor transition in the present model.

In the holographic context, the free energy is given by (minus) the on-shell Euclidean action. For our setup, this is obtained by Wick rotating the original DBI action~\eqref{eq:actionD5}, resulting in
\begin{align}
\begin{split}
    S^E_{D5}=\mathcal{N}\int du \sqrt{1-\psi^2}\Bigg[\left(u^4g^2-\Omega^2b^2\right)\left[\left(1-\psi^2\right)\left(h+b'^2\right)+u^2h\psi'^2\right]\Bigg.\\
    \Bigg.+u^4g^2b^2\left(1-\psi^2\right)\chi'^2\Bigg]^{1/2}~,
    \label{eq:actionD5Eucl}
\end{split}
\end{align}
where the functions $g$ and $h$ are defined in \eqref{eq:ghfunctions}.

To render the action finite, we introduce appropriate counterterms, which are discussed in detail in Appendix~\ref{app:holoreno}.\footnote{The overall sign differs from that in the appendix due to the Wick rotation to Euclidean signature.} These take the form
\begin{equation}
\begin{aligned}
    S^{ct}_1 & = -\frac{1}{3}\mathcal{N}\sqrt{-\gamma}~,\\
    S^{ct}_2 & = \frac{1}{2}\mathcal{N}\sqrt{-\gamma}\theta(\epsilon)^2~,
\end{aligned}
\end{equation}
where $\gamma$ denotes the determinant of the induced metric on the regulator surface, $z=\epsilon$.

\begin{figure}
    \centering
    \includegraphics[width=0.65\textwidth]{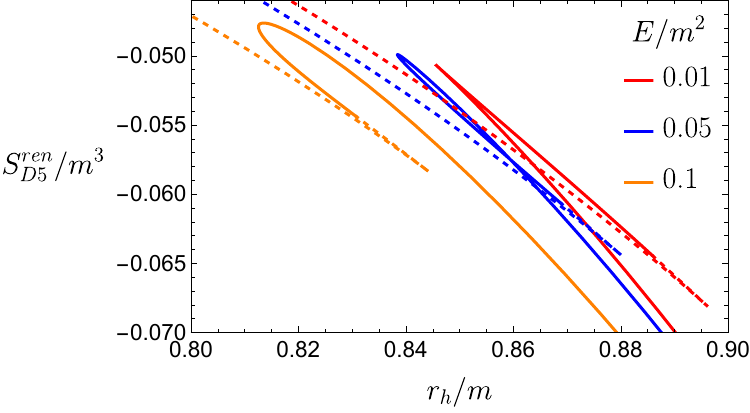}
    \caption{On-shell Euclidean action for different values of the electric field. Solid (dashed) lines correspond to BH (Minkowski) embeddings.}
    \label{fig:OnshellD5}
\end{figure}

The time-independence of the Lorentzian action suggests that the free energy can again be defined as the negative of the renormalized on-shell Euclidean action~\eqref{eq:actionD5Eucl}. Under this definition, one might expect that evaluating the free energy for both black hole and Minkowski embeddings would reveal a swallow-tail structure, characteristic of a first-order phase transition. However, as shown in Fig. \ref{fig:OnshellD5}, this expectation is not borne out.

Several comments are in order. First, the on-shell action involves an integration over the holographic radial coordinate. In the literature, two choices for the infrared cutoff are commonly considered: $u_{min} =u_h$, corresponding to the background horizon, and $u_{min} = u_c$, the critical surface of the brane embedding. This ambiguity is deeply tied to the non-equilibrium nature of the system: when the electric field is absent, the only consistent IR boundary is the background horizon.

The first choice is problematic because the phase $\chi$ diverges logarithmically as $\log (u-u_h)$. This mirrors the well-known divergence in $A_x$ on the case of a constant electric field in the $x$-direction \cite{Karch_2008}. In contrast, the second option $u_{min}=u_c$ provides the infrared regulator adopted throughout this work. This choice is widely regarded as the appropriate one~\cite{Kim_2011,Kundu_2013,Banerjee_2015}, as the singular shell at $u_c$ effectively functions as an event horizon for the open string degrees of freedom (see Section~\ref{subsec:Teff}). For this reason, we adhere to the second prescription.

We see that only for very small electric fields does the free energy curve resemble that of a conventional first-order phase transition. As $|E|/m^2$ increases, however, the behavior deviates significantly from the expected swallow-tail structure. Extending the integration domain to $u_{min} = u_h$ does not cure this problem.

The existence of the singular shell also complicates the interpretation of the on-shell action as a generator of response functions. This can be seen by considering general variations of the Euclidean action. On-shell, the variation of the bulk term~\eqref{eq:actionD5Eucl} yields only boundary contributions:
\begin{equation}
\begin{aligned}
	\delta S^E_{D5}&=\left[\frac{\partial \mathcal{L}}{\partial \psi'}\delta\psi+\frac{\partial\mathcal{L}}{\partial b'}\delta b+\frac{\partial \mathcal{L}}{\partial \chi'}\delta \chi\right]\Bigg\rvert^{z_c}_{z=\epsilon}\\
&=-\mathcal{N}\left[\frac{m\h\delta m}{\epsilon}+m\h\delta C+2C\delta m+\vec{J}\cdot\delta\vec{A}-\frac{q}{\Omega}\delta\chi\big\rvert_{z_c}\right]~,
\end{aligned}
\end{equation}
where the first four terms originate from the boundary at $z=\epsilon$, while the last term arises from the singular shell. Notably, this final contribution vanishes when the electric field is absent.

The variation of the counterterms yields
\begin{equation}
	\delta S^{ct}=\delta S^{ct}_1+\delta S^{ct}_2=\mathcal{N}\left[\frac{m\h\delta m}{\epsilon}+m\h\delta C+C\delta m\right]~.
\end{equation}

Adding up, the total variation of the renormalized action becomes
\begin{equation}
    \delta S_{D5}^{ren}=\delta S^E_{D5}+\delta S^{ct}=-\mathcal{N}\left[C\delta m+\vec{J}\cdot\delta\vec{A}-\frac{q}{\Omega}\delta\chi\big\rvert_{z_c}\right]~.
\end{equation}

The appearance of the term evaluated at the singular shell may underlie the atypical behavior of the free energy. As observed in~\cite{Ishii_2018}, this issue is related to the real-time holography prescription~\cite{Son_2002}, which instructs one to retain only boundary contributions and disregard those at the horizon. In our context, neglecting the contribution at the singular shell implies that the current and condensate defined in the main text follow this prescription. That is, they are extracted from the variation
\begin{equation}
    \delta S_{D5}^{ren}=\delta S^E_{D5}+\delta S^{ct}=-\mathcal{N}\left[C\delta m+\vec{J}\cdot\delta\vec{A}\right]~.
    \label{eq:dFnohorizon}
\end{equation}

This confirms the internal consistency of the holographic dictionary for our setup: the one-point functions $\langle\mathcal{O}_m\rangle$ and $\mathcal{J}_{\text{YM}}$ are correctly obtained from functional derivatives of the renormalized on-shell action with respect to the sources $m$ and $\vec{A}$, provided that contributions from the singular shell are excluded. 

To further verify this, we construct a family of solutions $\{\psi(u),b(u),\chi(u)\}$ parameterized by the insertion angle $\psi_0$ at the singular shell. By adjusting $b(u_c)$ and $\chi(u_c)$ so as to keep $|E|/m^2$ fixed at the boundary, we can evaluate
\begin{equation}
	S_{D5}^{ren}(\psi_0)=-\mathcal{N}\int_0^{\psi_0}\left[C(x)m'(x)+\vec{J}(x)\cdot\vec{A}'(x)\right]dx
\end{equation}
and confirm that the resulting curves coincide with the direct computation of the on-shell action shown in Fig. \ref{fig:OnshellD5}.

This implies that a naïve identification of minus the renormalized on-shell action with a free energy leads to the relation
\begin{equation}
    dF_{\text{naive}}=C\h dm+ \vec{J}\cdot d\vec{A}~,
\end{equation}
where, for simplicity, we have absorbed the overall normalization factor $\mathcal{N}$. For this expression to define a well-posed differential (\ie, for $dF_{\text{naive}}$ to be integrable), the following integrability conditions on the mixed partial derivatives must hold:
\begin{equation}
    \frac{\partial C}{\partial A_{x}}\Big\rvert_{m,A_{y}}=\frac{\partial  J_x}{\partial m}\Big\rvert_{A_{x},A_{y}}~,\qquad
    \frac{\partial C}{\partial A_{y}}\Big\rvert_{m,A_{x}}=\frac{\partial  J_y}{\partial m}\Big\rvert_{A_{x},A_{y}}~,\qquad
    \frac{\partial}{\partial \vec{A}}\times\vec{J}~\Big\rvert_m=0~,
    \label{eq:CrossDer}
\end{equation}
where $A_{i}$ denotes the boundary value of the $i$-th component of the gauge field.

\begin{figure}
    \centering
    \begin{subfigure}[t]{0.4\textwidth}
        \centering
        \includegraphics[width=\textwidth]{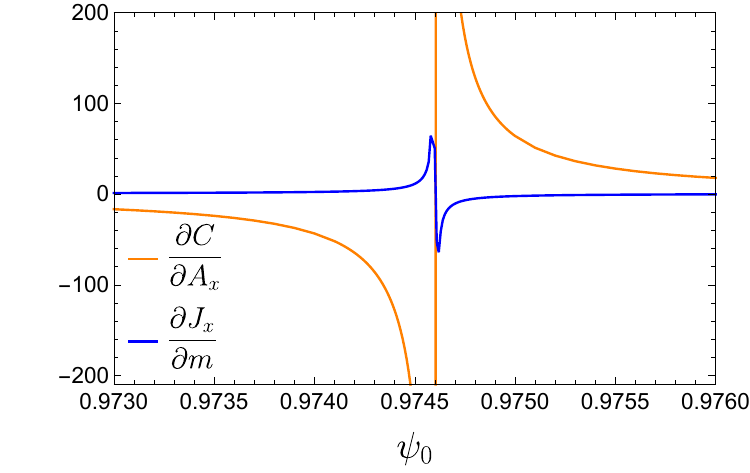}
    \end{subfigure}
    \hspace{0.02\textwidth}
    \begin{subfigure}[t]{0.4\textwidth}
        \centering
        \includegraphics[width=\textwidth]{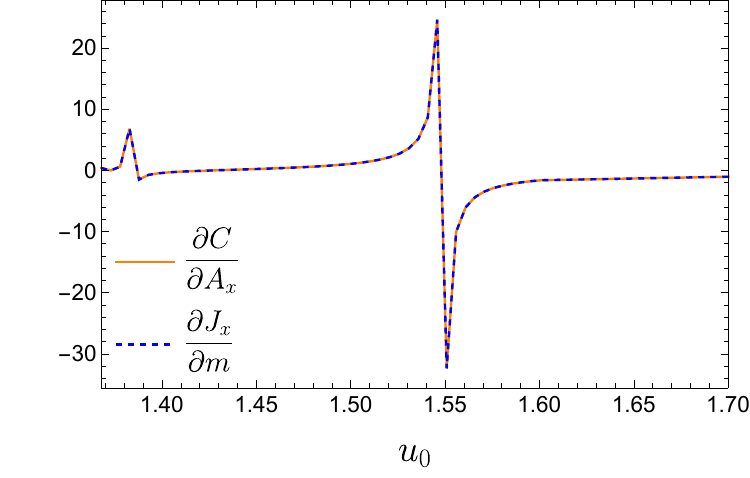}
    \end{subfigure}\\
    \centering
    \begin{subfigure}[t]{0.4\textwidth}
        \centering
        \includegraphics[width=\textwidth]{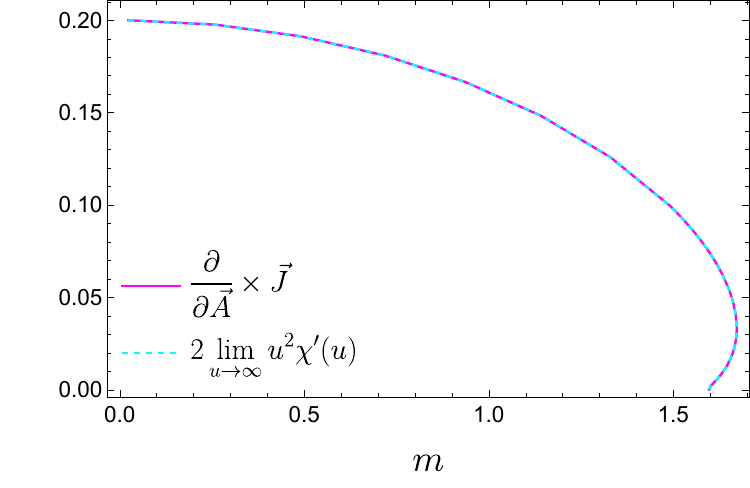}
    \end{subfigure}
    \hspace{0.02\textwidth}
    \begin{subfigure}[t]{0.4\textwidth}
        \centering
        \includegraphics[width=\textwidth]{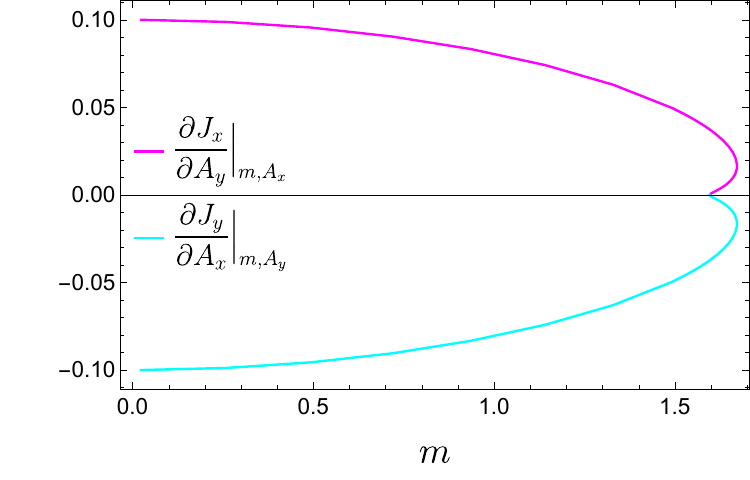}
    \end{subfigure}
    \caption{Top: First equality in Eq. \eqref{eq:CrossDer} for $\Omega=1$. On the left we show it for black hole embeddings, where it is violated, while on the right we observe that Minkowski embeddings correctly satisfy it. Bottom: on the left, numerical check of Eq. \eqref{eq:rotderlimit}. On the right, third equality in \eqref{eq:CrossDer} for $\Omega=0.1$. In the massless limit, the difference between the two derivatives approaches $2\Omega$, in agreement with the analytic result \eqref{eq:CrossDer2Omega}.}
    \label{fig:CrossDer}
\end{figure}

Evaluating the derivatives, we find that the integrability conditions are not satisfied, as illustrated in Fig. \ref{fig:CrossDer}. In fact, the violation of the last equality can be explicitly demonstrated in the massless case, where analytic solutions are known (see Appendix \ref{app:masslesssol}). Returning to the original holographic coordinate $u$, from the UV expansions \eqref{eq:UVexpansionsEJ} we identify the following expressions for the boundary values of the gauge field:
\begin{equation}
\begin{aligned}
    A_{x}&\equiv\lim_{z\rightarrow0}A_x(z)=b_0\cos\chi_0~,\\
    A_{y}&\equiv\lim_{z\rightarrow 0}A_y(z)=b_0\sin\chi_0~
\end{aligned}
\end{equation}
and also for the components of the current:
\begin{equation}
\begin{aligned}
      J_x&=b_1\cos\chi_0-b_0\chi_1\sin\chi_0~,\\
      J_y&=b_1\sin\chi_0+b_0\chi_1\cos\chi_0~,
\end{aligned}
\end{equation}
from where we immediately see that
\begin{equation}
    \frac{\partial J_x}{\partial A_{y}}=-\chi_1~,\qquad \frac{\partial J_y}{\partial A_{x}}=\chi_1\qquad \Rightarrow\qquad \frac{\partial}{\partial \vec{A}}\times\vec{J}=2\lim_{u\rightarrow\infty}u^2\chi'(u),
    \label{eq:rotderlimit}
\end{equation}
where we have used that $\displaystyle \chi_1=-\lim_{u\to\infty}u^2\chi'(u)$.

Using the analytic solution for $\chi$ in the massless case (see Appendix \eqref{app:masslesssol}), this yields
\begin{equation}
    \frac{\partial}{\partial \vec{A}}\times\vec{J}=2\Omega~,
    \label{eq:CrossDer2Omega}
\end{equation}
which matches the result reported in \cite{Ishii_2018} for a holographic superconductor under a rotating electric field. This behavior is further confirmed by the numerical results shown in Fig. \ref{fig:CrossDer}.

These results cast doubt on the interpretation of the Euclidean on-shell action as a free energy. In particular, the disappearance of the cusp would imply a negative entropy, as discussed in Appendix C of \cite{Mateos_2007}.

In conclusion, although the rotating ansatz \eqref{eq:defctu} removes the explicit time dependence from the action, the non-equilibrium character of the system remains manifest in features such as those discussed above. It is important to emphasize, however, that the extraction of response functions from the subleading terms in the near-boundary expansion of bulk fields remains valid within the framework of real-time holography \cite{Herzog_2002,Skenderis_2008_short,Skenderis_2008_long}. Nonetheless, other aspects, such as the nature and precise location of the phase transition, lie beyond the scope of equilibrium thermodynamics. This issue has been addressed in several works \cite{Sasa_2004,Albash_2007,Bergman_2008,Nakamura_2012,Nakamura_2013,Kundu_2013,Banerjee_2015,Kundu_2019}, but to our knowledge, no unambiguous resolution has been established.

\section{Conclusions}\label{sec:Conclusions}

This paper pursues, along the line of previous works \cite{Kinoshita_2017,Garbayo_2020}, the study of holographic Floquet flavor systems driven by an external rotating electric field. We focused on the D3/D5 system but we found that most of the results are qualitatively robust and shared by the D3/D7 setup. We have sharpened the findings in \cite{Garbayo_2020} in several directions.

First, we studied the effects of having the system heated at some non-zero temperature. In this  case, the dual geometry has two types of horizons: the usual event horizon of the closed string geometry in the bulk, and the effective horizon of the open string metric on the brane. Respectively, we can associate two temperatures to them: the Hawking temperature $T_H$ of the  background \eqref{eq:TBH} and the effective temperature $T_\text{eff}$ written in \eqref{eq:TeffD5} experienced by the worldsheet degrees of freedom. We have scanned throughout all our phase space and checked that $T_\text{eff}>T_H$.

The main effect of the background temperature is the addition of deconfined charged carriers to the system. Such carriers add to the ones produced by the electric field through the Schwinger mechanism of dielectric breakdown. We show that one of the main results in previous works, namely, the presence of vector meson Floquet condensates, persists at finite temperature, signaling the robustness of this non-perturbative effect. The lobbed structure of the line of critical embeddings is also found here, but it gets depleted in height and is completely washed away for high enough values of the temperature, namely, for a radius horizon $r_h/m \geq 0.8897$ in units of the quark mass (see Fig. \ref{fig:LobesEJD3D5}).

At high temperatures, some interesting effects occur when the background and effective horizons come close together. These include a multi valuedness that resembles a second-order phase transition within the conductive black hole phase (Fig. \ref{fig:JCvsE}). Also new solutions with vanishing $|E|=|J|=0$ appear within this phase (Fig. \ref{fig:JFullCurves}).

Secondly, we have remarked the relevance of the so  called {\em Floquet suppression points}. These states were missed in previous analysis but are common to both D3/D5 and D3/D7 systems both at zero and finite temperature. We have shown that the phase portrait very close to these points is strikingly similar to the one in the vicinity of the vector meson Floquet condensates, up to an exchange of $J$ with $E$. This calls for a deeper study in search for a sounder duality. From the physical point of view, these new points exhibit a dynamical suppression of the vacuum polarizability. It could be attributed to a dynamical screening of the effective dipole charge of the meson fluctuations at strong coupling and for precise frequencies. It bears resemblance to similar effects in the realm of Floquet condensed matter systems where, for example, hopping terms can be seen to vanish at fined tuned frequencies of the driving. This is the type of effects that make Floquet driving an appealing paradigm in the search for mechanisms that could help in suppressing quantum decoherence.

Thirdly, we have also pursued the analysis initiated in \cite{Garbayo_2020} concerning two different notions of conductivities in Section \ref{sec:conductivitiesD3D5}. The first one is the non-linear rotating conductivity of Section \ref{subsec:nonlinearcond}. The relative phase (angle) between $J$ and $E$ has an interesting information that we interpret in terms of a possible variable admixture of two types of currents: rotating polarization and charge flow. The polarization current is the only one present in the Minkowski phase while both are present in the BH states. The global picture that emerges from the analysis shows that the Floquet suppressed states are points where the polarizability of the vacuum switches smoothly from positive to negative. This is remarkable as it states that, for ample intervals in the range of driving frequencies, $\Omega$, the meson condensate is  polarized  {\em antiparallel} to the applied electric field! Again here, this result is amongst the class of remarkable effects that one can find in the context of Floquet engineering \cite{Giovannini_2019} of condensed matter systems. For example, it is worth citing ref. \cite{Schmidt_2019}, where paramagnetism can be turned into  diamagnetism under a strong driving in the Rabi model coupled to a heat bath.

In the limit of large electric field, and/or large temperature, we agree with the results in \cite{Karch_2010}. In this limit the polarization rotating current gets suppressed, $J_{pol}(t)\to 0$,  whereas the conduction currect satisfies an Ohmic instantaneous response for an arbitrary frequency of the rotating driving $J(t)\sim  J_{con}(t) = \sigma  E(t)$ with $\sigma$ a real constant.  Linearity suggests the possibility of this being also true for any electric field time dependence in 2+1 dimensions. We also find agreement with the predictions in \cite{Karch_2010} in the limit of small mass.

The other type of conductivities involves the optical AC and DC conductivities in the presence of a driving. The interesting pattern with peaks found in \cite{Garbayo_2020} deep inside the wedges between the lobes in phase space still exists for low to moderate temperatures, but gets dissolved as soon as the height of the lobes is depleted at high  temperature. The highest peaks shift with growing $|E|$ and stay close to the position of the (also drifting) Floquet suppression points.

Our work could be continued along several directions. One clear option would be to add chemical potential and/or magnetic components to the gauge field in order to explore the complete phase space of the D3-D5 model. To verify the universality of our results we could consider the ABJM model \cite{Aharony_2008} driven by a rotating electric field. This last model has a rich topological structure and is dual to a $(2+1)$-dimensional conformal field theory. The flavor branes in this case are D6-branes \cite{Hohenegger_2009, Gaiotto_2009} (the thermodynamics of these flavor branes has been studied in \cite{Jokela_2012}). Another direction worth pursuing is the analysis of the effects of backreaction for a large number of  flavor branes. Interestingly, backreacted D3-D5 backgrounds have been constructed in \cite{Conde_2016, Penin_2017, Jokela_2019, Garbayo_2022} using the smearing approximation reviewed in \cite{Nunez_2010}. One could also study the driving generated by moving the brane periodically in time (i.e. oscillating or rotating). This type of configurations were considered in \cite{Das_2010, Hoyos_2011}, and are the natural setup to find resonances that could be interpreted as Floquet condensates of other type of mesons, like scalar mesons.

Finally, another aspect that deserves further attention is the actual nature of the phase transition. It can be triggered by an admixture of both temperature and/or electric field. While the multi-valuedness of the state curves in Fig. \ref{fig:JCvsE} suggest an area law for the transition point, the actual location it is not consistent with the $(|J|/m^2,|E|/m^2)$ and the $(C/m^2,|E|/m^2)$ curves.

To study generic non-equilibrium configurations within holography, the most widely accepted approach is the one developed by Skenderis and van Rees \cite{Skenderis_2008_short,Skenderis_2008_long,vanRees_2009}. Their prescription requires choosing a suitable contour in the complex time plane, reflecting the field theory setup, and constructing a corresponding bulk geometry of mixed signature. This is done by gluing Euclidean segments to describe the imaginary parts of the contour and Lorentzian ones for the real-time evolution.

Despite its conceptual clarity, this prescription is technically demanding. It requires solving a coupled system of bulk equations, subject to boundary conditions that ensure the continuity of both the fields and their derivatives at the gluing surfaces between the Euclidean and Lorentzian segments. In most cases, these equations can only be tackled numerically.

An alternative approach to out-of-equilibrium holography was proposed by Glorioso, Crossley, and Liu \cite{Glorioso_2018}. Their method involves an analytic continuation of the radial coordinate in the (generally time-dependent) bulk geometry. In this construction, the gravitational dual of the Schwinger-Keldysh contour is a path that starts at the UV boundary, extends into the bulk, and encircles the point $r=r_h$ along a small circle in the complex plane.

To the best of our knowledge, no comprehensive studies have applied these formalisms to D3/D5 or D3/D7 brane intersections driven out of equilibrium. Pursuing such an analysis remains an open and compelling direction for future work.

\section*{Acknowledgements}

We are indebted for valuable mail exchanges with Takaaki Ishii, Arnab Kundu,  Carlos Hoyos, Keiju Murata, Andy O'Bannon,  Kostas Skenderis and Julian Sonner.

This work has received financial support from Xunta de Galicia (Centro singular de investigación de Galicia accreditation 2019-2022 and grant ED431C-2021/14), by European Union ERDF, and by the "María de Maeztu" Units of Excellence program MDM-2016-0692 and the Spanish Research State Agency (grant PID2020-114157GB-100). 

\appendix

\section{Coordinates}\label{app:coordinates}

We begin by considering the Schwarzschild-AdS$_5\times S^5$ geometry,
\begin{equation}
    ds^2=\frac{r^2}{L^2}\left(-f(r)dt^2+dx^2+dy^2+dz^2\right)+\frac{L^2}{r^2}\left(\frac{dr^2}{f(r)}+r^2d\Omega_5^2\right)~,
    \label{eq:AdS5xS5app}
\end{equation}
where $f(r)$ is the blackening factor,
\begin{equation}
    f(r)=1-\frac{r_h^4}{r^4}~.
\end{equation}
At finite temperature, it is convenient to introduce an "isotropic" coordinate, $u$, defined as
\begin{equation}
    u^2=\frac{r^2}{2}\left(1+\sqrt{1-\frac{r_h^4}{r^4}}\right)~.
    \label{eq:u}
\end{equation}
The metric becomes
\begin{equation}
    ds^2=\frac{u^2}{L^2}\left(-\frac{g(u)^2}{h(u)}dt^2+h(u)\left(dx^2+dy^2+dz^2\right)\right)+\frac{L^2}{u^2}\left(du^2+u^2 d\Omega_5^2\right)~,
    \label{eq:metricisotropic}
\end{equation}
where the functions $g(u)$ and $h(u)$ are given by
\begin{equation}
    g(u)=1-\frac{u_h^4}{u^4}~~,\hspace{5mm}h(u)=1+\frac{u_h^4}{u^4}~.
\end{equation}
The black hole horizon is now at $u_h=r_h/\sqrt{2}$. Notice that the effect of this change of coordinates is to remove the blackening factor from the $g_{rr}$ component of the metric, making the presence of a flat 6-plane perpendicular to the horizon manifest. At zero temperature, where $r_h=0$ and $f(r)=1$, this is already manifest, and the two sets of coordinates coincide. Moreover, near the UV boundary, $u\sim r$.

We now embed a D5-brane in this geometry. A useful parametrization of the 5-sphere is
\begin{equation}
    d\Omega_5^2=d\theta^2+\cos^2\theta d\Omega_2^2+\sin^2\theta d\tilde{\Omega}_{2}^2~.
    \label{eq:metric5spheretheta}
\end{equation}

We define also $\psi\equiv \sin\theta$, in which case the metric becomes
\begin{equation}
    ds^2=\frac{u^2}{L^2}\left(-\frac{g(u)^2}{h(u)}dt^2+h(u)d\Vec{x}_3^2\right)+\frac{L^2}{u^2}du^2+L^2\left(\frac{d\psi^2}{1-\psi^2}+(1-\psi^2)d\Omega_2^2+\psi^2 d\tilde{\Omega}_2^2\right)~.
    \label{eq:AdS5xS5psi}
\end{equation}
with $\Vec{x}_3=(x,y,z)$. These will be the coordinates that we will generically use throughout the paper, except for the analytic solutions of Appendix \ref{app:analyticsols}, as we will detail below.

We will also consider \textit{cartesian} coordinates, $(\rho,w)$, related to $(u,\theta)$ as
\begin{equation}
    \rho=u\cos\theta~,\qquad w=u\sin\theta~,
    \label{eq:relationcartesianangular}
\end{equation}
which satisfy
\begin{equation}
    d\rho^2+dw^2=du^2+u^2d\theta^2~.
\end{equation}
In these coordinates, the metric \eqref{eq:AdS5xS5psi} becomes
\begin{equation}
    ds^2=\frac{u^2}{L^2}\left(-\frac{g(u)^2}{h(u)}dt^2+h(u)d\Vec{x}_3^2\right)+\frac{L^2}{u^2}\left(d\rho^2+dw^2+\rho^2d\Omega_2^2+w^2d\tilde{\Omega}_2^2\right)~,
    \label{eq:AdS5xS5wu}
\end{equation}
where $u$ has to be understood as written in terms of $\rho$ and $w$, \ie,
\begin{equation}
    u^2=\rho^2+w^2~.
\end{equation}

Notice that, although the name \textit{cartesian} is used, $w$ denotes the radius of the $\tilde{\Omega}_2$ space in spherical coordinates.

There is one case in which we will be interested in using the original radial coordinate, $r$, at finite temperature. This is the case of the analytic solutions of Appendix \ref{app:analyticsols}. In that case, we are going to use the original metric, \eqref{eq:AdS5xS5app}, with the parametrization of the 5-sphere in terms of $\theta$, \eqref{eq:metric5spheretheta}. The metric in this case is
\begin{equation}
    ds^2=\frac{r^2}{L^2}\left(-f(r)dt^2+dx^2+dy^2+dz^2\right)+\frac{L^2}{r^2}\frac{dr^2}{f(r)}+L^2\left(d\theta^2+\cos^2\theta d\Omega_2^2+\sin^2\theta d\tilde{\Omega}_2^2\right)~.
    \label{eq:AdS5xS5formassless}
\end{equation}

\section{Analytic solutions}\label{app:analyticsols}

The equations of motion admit analytic solutions in certain limits. In particular, we can find analytic expressions for the gauge field in the massless case. By considering small masses as a deviation of this analytic solution, the equations for the deviations can also be solved analytically. This is a remarkable feature of the D3/D5 model that is not present in the D3/D7 case \cite{Hashimoto_2016,Kinoshita_2017}. There, even in the massless case, the equations have to be solved numerically. In this appendix we obtain the two analytic solutions mentioned above. The procedure follows the same steps as in \cite{Garbayo_2020}, providing a non-zero temperature generalization to the solutions found there.

In both cases, it is convenient to use the coordinates \eqref{eq:AdS5xS5formassless}, with the embedding parametrized as $\theta(r)$. The DBI action \eqref{eq:actionD5} in these coordinates is given by
\begin{equation}
\begin{aligned}
    S_{\text{D5}}=-\mathcal{N}\int dr\frac{\cos^2\theta}{\sqrt{r^4-r_h^4}}&\Big(\big[r^4-r_h^4-\Omega^2b^2\big]\big[\left(r^4-r_h^4\right)\left(b'^2+r^2\theta'^2\right)+r^4\big]\Big.\\
    &\phantom{\Big[\big[r^4-r_h^4-\Omega^2b^2\big]\big[\left(r^4-r_h^4\right)\big.\Big.}\Big.+\left(r^4-r_h^4\right)^2b^2\chi'^2\Big)^{1/2}~,
    \label{eq:SD5app}
\end{aligned}
\end{equation}
where again we have consistently chosen the fields to be time-independent, and the constant $\mathcal{N}$ is given in \eqref{eq:ND3D5}. The cyclic coordinate $\chi(r)$ is now related to $\theta(r)$ and $b(r)$ as
\begin{equation}
    \chi'(r)=\frac{q~\sqrt{r^4-r_h^4-\Omega^2 b^2}\sqrt{\left(r^4-r_h^4\right) \left(b'^2+r^2\theta'^2\right)+r^4}}{\abs{b}\left(r^4-r_h^4\right)\sqrt{\left(r^4-r_h^4\right)\Omega^2 b^2\cos^4\theta-q^2}}~.
\label{eq:chiprime}
\end{equation}
Legendre-transforming the action to eliminate $\chi'$, we obtain
\begin{equation}
\begin{aligned}
    \bar{S}_{\text{D5}}&=S_{\text{D5}}-\int dr~ \chi'\frac{\partial\mathcal{L}}{\partial \chi'}\\
    &=-\mathcal{N}\int dr \frac{\sqrt{r^4+\left(r^4-r_h^4\right)\left(b'^2+r^2\theta'^2\right)}}{\left(r^4-r_h^4\right)\Omega\abs{b}}\sqrt{\left[r^4-r_h^4-\Omega^2b^2\right]\left[\left(r^4-r_h^4\right)\Omega^2b^2\cos^4\theta-q^2\right]}~.
    \label{eq:actionLTtransftheta}
\end{aligned}
\end{equation}

The tortoise coordinates $(\tau, r_*)$ in this parameterization are defined as
\begin{align}
    d\tau &= dt-A_{\theta}(r)~dr~, & dr_* &= B_{\theta}(r)~dr~,
    \label{eq:tortoise}
\end{align}
where $A_\theta(r)$ and $B_\theta(r)$ are the functions
\begin{align}
    A_{\theta}(r) &= \frac{\Omega b^2\chi'}{r^4-r_h^4-b^2~\Omega^2}~, & B_{\theta}(r) &= \frac{\mathcal{L}}{\cos^2\theta \left(r^4-r_h^4-b^2~\Omega^2\right)}~,
\end{align}
with ${\cal L}$ being the lagrangian density in \eqref{eq:SD5app}. In terms of $(\tau, r_*)$ the $(t,r)$ part of the effective metric takes the form
\begin{equation}
    \frac{r^4-r_h^4-b^2\Omega^2}{r^2}\left(-d\tau^2+dr_*^2\right)~,
\end{equation}
which means that, indeed, these new coordinates are tortoise coordinates for the effective open string metric.

\subsection{Massless solution}\label{app:masslesssol}

Let us now consider a massless embedding with $\theta=0$. The action \eqref{eq:actionLTtransftheta} for this case takes the form
\begin{equation}
    \bar{S}_{\text{D5}}=-\mathcal{N}\int dr\frac{\sqrt{r^4+\left(r^4-r_h^4\right)b'^2}}{\left(r^4-r_h^4\right)\Omega\abs{b}}\sqrt{\left[r^4-r_h^4-\Omega^2b^2\right]\left[\left(r^4-r_h^4\right)\Omega^2b^2-q^2\right]}~.
    \label{eq:actionLTtransfmassless}
\end{equation}
Both factors in the second square root must vanish simultaneously at the singular shell, $r=r_c$, in order to keep the action real. From this condition, we get
\begin{equation}
    b_0^2=\frac{r_c^4-r_h^4}{\Omega^2}\quad, \quad\quad q^2=(r_c^4-r_h^4)\Omega^2b_0^2~.
    \label{eq:b0andqmassless}
\end{equation}
Combining both equations we obtain
\begin{equation}
    b_0^2=\frac{r_c^4-r_h^4}{\Omega^2}=\frac{q}{\Omega^2}~.
    \label{eq:b0massless}
\end{equation}
Actually, one can verify from the equation of motion of $\bar{S}_{\text{D5}}$ that a constant $b(r)=b_0$ is a solution. Moreover, plugging $b(r)=b_0$ into the right-hand side of \eqref{eq:chiprime} (with $\theta=0$), we get that the phase $\chi(r)$ in the massless case satisfies
\begin{equation}
    \chi'(r)=\Omega \frac{r^2}{r^4-r_h^4}=\frac{\Omega}{r^2f(r)}~.
    \label{eq:chi_0_prime}
\end{equation}
This equation can be integrated directly. By defining a new function $\Lambda(x)$ as
\begin{equation}
    \Lambda(x) \equiv \log\frac{x-1}{x+1}-2\arccot x~, \qquad\frac{d\Lambda}{dx} = \frac{4x^2}{x^4-1}~,
    \label{eq:Lambdadef}
\end{equation}
one has
\begin{equation}
    \chi(r)=\frac{\Omega}{4r_h}\Lambda(r/r_h)~,
    \label{eq:chimassless}
\end{equation}
where we have fixed the integration constant in such a way that $\Lambda(x)\to 0$ for $x\to\infty$ ($\Lambda(x)\approx -4/x$ for large $x$). Therefore, the complexified gauge potential $c(r)=b(r)e^{i\chi(r)}$ is given by
\begin{equation}
    c(r)=\frac{\sqrt{r_c^4-r_h^4}}{\Omega}e^{\frac{i\Omega}{4r_h}\Lambda(r/r_h)}~.
    \label{eq:c_massless}
\end{equation}

By expanding the right-hand side of \eqref{eq:c_massless} for large $r$ and comparing the result with \eqref{eq:UVexpansions}, we get the electric field $E$ and the current $J$ to be given by
\begin{equation}
    E=J=-i\sqrt{r_c^4-r_h^4}=-i\sqrt{q}~.
    \label{eq:EJmassless}
\end{equation}
Thus, $E$ and $J$ are parallel and equal, which corresponds to $\sigma_{xx}=1$ and $\sigma_{xy}=0$ in the conductivity tensor \eqref{eq:sigmaRC}. Using \eqref{eq:dictionaryD3D5} it is easy to derive the correctly normalized physical conductivity, relating the the physical electric field and the physical current, as
\begin{equation}
    \sigma(\Omega)=\frac{2N_f N_c}{\pi\sqrt{\lambda}}~,
\end{equation}
matching the results of \cite{Karch_2010}. Remarkably, the response of the system is not only linear, but instantaneous.

Finally, let's consider the effective temperature, which we found to be in general
\begin{equation}
    T_{\text{eff}}=\frac{2u_ch(u_c)-\Omega b_1}{2\pi b_0\chi_1}~.
    \label{eq:TeffApp}
\end{equation}

For the massless solution found above, $b_1=0$, and the value of $\chi_1$ can be read from \eqref{eq:chi_0_prime} by taking $r=r_c$. We obtain\footnote{The coefficients $b_1$ and $\chi_1$ in \eqref{eq:TeffApp} were defined from the series expansion in the $u$ coordinate. Therefore the change of coordinates \eqref{eq:u} needs to be used.}
\begin{equation}
    T_{\text{eff}}=\frac{r_c}{\pi}=\frac{(q+r_h^4)^{1/4}}{\pi}\ge T_H=\frac{r_h}{\pi}~.
\end{equation}

\subsection{Small-mass solution}\label{app:smallmasssol}

Let us now consider small-mass solutions, in which $b(r)$ and $\theta(r)$ are given by
\begin{equation}
    b(r)=b_0+\beta(r)~,\qquad\theta(r)= \lambda(r)~,
\end{equation}
where the functions $\beta(r)$ and $\lambda(r)$ are treated as small perturbations. At first order in these functions, they satisfy the following linear equations of motion
\begin{equation}
\begin{aligned}
    \frac{d}{dr}\left[(r^4f-q)\lambda'\right]+2r^2\lambda&=0~,\\
    \frac{d}{dr}\left[\frac{r^4f-1}{r^2}\beta'\right]+\frac{4r^2\Omega^2}{r^4f-q}\beta&=0~,
\end{aligned}
\end{equation}
which can be written as
\begin{equation}
\begin{aligned}
    (r^4-r_c^4)\lambda''+4r^3\lambda'+2r^2\lambda&=0~,\\
    r^2(r^4-r_c^4)\beta''+2r(r^4+r_c^4)\beta'+4\Omega^2\frac{r^6}{r^4-r_c^4}\beta&=0~.
    \label{eq:smallmasseqs}
\end{aligned}
\end{equation}
These equations are equivalent to those solved in Appendix B of \cite{Garbayo_2020}, with the only difference being the value of the critical radius, $r_c$. Therefore, we can just adapt their results to our case. Let us start by writing the general solution for the differential equation satisfied by $\lambda(r)$:
\begin{equation}
    \lambda(r)=c_1\h_2F_1\left(\frac{1}{4},\frac{1}{2};\frac{3}{4};\frac{r^4}{r_c^4}\right)+c_2r\h_2F_1\left(\frac{1}{2},\frac{3}{4};\frac{5}{4};\frac{r^4}{r_c^4}\right)~,
    \label{eq:lambdasol}
\end{equation}
where $_2F_1$ is the ordinary hypergeometric function, and $c_1$ and $c_2$ are constants. The hypergeometric functions in \eqref{eq:lambdasol} have a logarithmic divergence as we approach the singular shell, $r\to r_c$. Imposing the solution to be regular at $r=r_c$, we find that the integration constants $c_1$ and $c_2$ must satisfy the relation
\begin{equation}
    \frac{c_1}{c_2}=-\frac{r_c}{4}\left[\frac{\Gamma\left(\frac{1}{4}\right)}{\Gamma\left(\frac{3}{4}\right)}\right]^2
    \label{eq:c1c2}
\end{equation}
We can expand the solution \eqref{eq:lambdasol} near the boundary and, by comparing with the known expansion for $\theta(r)$,
\begin{equation}
    \theta(r)\approx \frac{m}{r}+\frac{C}{r^2}+...~, \qquad r \to \infty~,
\end{equation}
we can relate the parameters $m$ and $C$ to the integration constants. Eliminating $c_1$ using \eqref{eq:c1c2}, we find the relations
\begin{align}
    m & = r_c\h c_1 =  -\frac{r_c^2}{4}\left[\frac{\Gamma\left(\frac{1}{4}\right)}{\Gamma\left(\frac{3}{4}\right)}\right]^2c_2~,\\
    C & = r_c^3\h c_2~.
\end{align}
It follows that $C$ and $m$ are linearly related in this small-mass solutions
\begin{equation}
    C=-r_c\left[\frac{2\Gamma\left(\frac{3}{4}\right)}{\Gamma\left(\frac{1}{4}\right)}\right]^2m~,
\end{equation}
in agreement with the asymptotic behavior in the right plots of Fig. \ref{fig:JCvsE}.

In summary, in the small-mass regime, the embedding function $\theta(r)$ can be written as
\begin{equation}
    \theta(r)=\frac{m}{r_c}F\left(\frac{1}{4},\frac{1}{2};\frac{3}{4};\frac{r^4}{r_c^4}\right)+\frac{C}{r_c^3}\h r\h F\left(\frac{1}{2},\frac{3}{4};\frac{5}{4};\frac{r^4}{r_c^4}\right)~.
\end{equation}

Let us next look at the equation satisfied by the gauge field perturbation $\beta(r)$ in the system \eqref{eq:smallmasseqs}. Its general solution is
\begin{equation}
    \beta(r)=d\h\exp\left[-\frac{i\Omega}{2r_c}\Lambda(r/r_c)\right]+d^*\exp\left[\frac{i\Omega}{2r_c}\Lambda(r/r_c)\right]~,
    \label{eq:betasol}
\end{equation}
where $d$ is a complex constant and $\Lambda$ is defined in \eqref{eq:Lambdadef}. It is interesting to find the solution for the complex potential $c(r)=b(r)e^{i\chi(r)}$. Let us denote
\begin{equation}
    \delta c(r)\equiv c(r)-c_0(r)~,
\end{equation}
where $c_0(r)$ is the solution for the massless case obtained in \eqref{eq:c_massless}. $\delta c$ satisfies the following differential equation at linear order
\begin{align}
\begin{split}
    r(r^4-r_c^4) (r^4-r_h^4)^2\delta c''+ 2(r^4-r_h^4)\big[r^8+(r_c^4-r_h^4)r^3(r+i\Omega)-r_c^4 r_h^4\big]\delta c'\,+\\
    +r^5\Omega\big[\Omega~r^4+(r_c^4-2r_h^4)\Omega-4i(r_c^4-r_h^4)r\big]\delta c=0~.
\end{split}
\end{align}
Instead of trying to solve directly this equation we notice that $\delta c$, at first order, can be written as
\begin{equation}
    \delta c(r)=e^{i\chi_0(r)}\left(\beta(r)+i\h b_0\h \delta \chi(r)\right)~,
    \label{eq:deltac}
\end{equation}
where $\chi_0(r)$ and $b_0$ denote the massless solutions in \eqref{eq:b0massless} and \eqref{eq:chimassless}, respectively, and $\delta \chi(r)=\chi(r)-\chi_0(r)$. The equation satisfied by $\delta \chi(r)$ is obtained by expanding \eqref{eq:chiprime} at linear order in the perturbations, giving
\begin{equation}
    \delta \chi'=-\frac{2r^2\Omega}{b_0(r^4-r_c^4)}\beta~,
\end{equation}
which can be readily integrated:
\begin{equation}
    \delta \chi(r)=-\frac{i\h d}{b_0}\exp\left[-\frac{i\Omega}{2r_c}\Lambda(r/r_c)\right]+\frac{i\h d^*}{b_0}\exp\left[\frac{i\Omega}{2r_c}\Lambda(r/r_c)\right]+\varphi~,
    \label{eq:deltachisol}
\end{equation}
where $\varphi$ is a constant. Plugging \eqref{eq:betasol} and \eqref{eq:deltachisol} into \eqref{eq:deltac}, we obtain:
\begin{equation}
    \delta c(r)=A\exp\left[i\frac{\Omega}{4}\left(\frac{1}{r_h}\Lambda(r/r_h)-\frac{2}{r_c}\Lambda(r/r_c)\right)\right]+B\exp\left[\frac{i\Omega}{4r_h}\Lambda(r/r_h)\right]~,
    \label{eq:deltacsol}
\end{equation}
where $A=2 d$ and $B=i\frac{\varphi}{\Omega}\sqrt{r_c^4-r_h^4}$.

To proceed further we have to impose a regularity condition to the general solution 
\eqref{eq:deltacsol}. With this purpose, let us write $\delta c(r)$ in terms of the tortoise coordinates \eqref{eq:tortoise} for the open string metric. In the small-mass case it is easy to demonstrate that \eqref{eq:tortoise} can be integrated to give the following relation between the tortoise coordinates $(\tau, r_*)$ and our original coordinates $(t,r)$,
\begin{align}
    \tau & = t+\frac{1}{4r_h}\Lambda(r/r_h)-r_*~, & r_* & = \frac{1}{4r_c}\Lambda(r/r_c)~.
\end{align}
The new radial coordinate $r_*$ varies from $r_*=-\infty$ at the pseudohorizon to $r_*=0$ at the UV boundary. Actually, one can prove that in these regions it can be related to $r$ as
\begin{equation}
\begin{aligned}
    r_* & = -\frac{1}{r}+\mathcal{O}(r^{-5})~, & (r & \to \infty)~,\\
    r_* & \sim \frac{1}{4r_c}\log(r-r_c)~, & (r & \to r_c)~.
    \label{eq:tortoisesolution}
\end{aligned}
\end{equation}

Let us next consider the time-dependent gauge potential $\delta a(r,t)$, given by
\begin{equation}
    \delta a(r,t)=\delta c(r) e^{i\Omega t}~.
\end{equation}
By combining \eqref{eq:deltacsol} and \eqref{eq:tortoisesolution}) it straightforward to prove that $\delta a$ can be written in terms of $(\tau, r_*)$ simply as
\begin{equation}
    \delta a(\tau,r_*)=Ae^{i\Omega (\tau-r_*)}+Be^{i\Omega (\tau+r_*}~.
    \label{eq:deltaatortoise}
\end{equation}
We next impose an infalling boundary condition, which amounts to select the solutions with $A=0$ in \eqref{eq:deltaatortoise}. Therefore, the regular solutions we are looking for are
\begin{equation}
    \delta a(\tau,r_*)=Be^{i\Omega (\tau+r_*)}~.
\end{equation}
Equivalently, $\delta c(r)$ is given by
\begin{equation}
    \delta c(r)=B\exp\left[\frac{i\Omega}{4r_h}\Lambda(r/r_h)\right]~.
\end{equation}

We can  now read off the electric field and current from the asymptotic behavior of $\delta c$, namely
\begin{equation}
    \delta c(r) \approx B-\frac{i\h\Omega B}{r}+...~, \qquad (r  \to \infty)~.
\end{equation}

Thus, we have
\begin{equation}
    \delta E=\delta J = -i\h\Omega B~,
\end{equation}
which means that the equality of $E$ and $J$ of the massless solution is mantained at first order in the small-mass regime, in agreement with the asymptotic behavior in the left plots of Fig. \ref{fig:JCvsE}.

\section{Holographic renormalization and dictionary}\label{app:holoreno}

We begin with the metric of AdS$_5\times S^5$ written in the $\theta$-parametrization \eqref{eq:metricisotropic}-\eqref{eq:metric5spheretheta}, and we introduce the inverse radial coordinate $z=1/u$. The AdS$_5\times S^5$ metric becomes
\begin{equation}
    ds^2=\frac{1}{z^2}\left[-\frac{g(z)^2}{h(z)}dt^2+h(z)d\vec{x}_3^2+dz^2\right]+d\theta^2+\cos^2\theta d\Omega_2^2+\sin^2\theta d\tilde{\Omega}_2^2~,
\end{equation}
with
\begin{equation}
    g(z)=1-\frac{z^4}{z_h^4}~,\qquad h(z)=1+\frac{z^4}{z_h^4}~,
\end{equation}
and the horizon is located at $z_h=1/u_h$.

The ansatz for the brane embedding and the gauge field \eqref{eq:defctu} translates in these coordinates into
\begin{equation}
    (2\pi\alpha')(A_x+iA_y)=c(z)e^{i\Omega t}~,\qquad \theta=\theta(z)~.
\end{equation}

In these coordinates, the DBI action is given by
\begin{equation}
    S_{D5}=-\mathcal{N}\int_{0}^{z_0}dz\frac{\cos^2\theta}{z^4}\sqrt{z^4g^2\abs{c'}^2-z^8\Omega^2\Re(c\h c'^*)^2+h\left(g^2-z^4\Omega^2\abs{c}\right)\left(1+z^2\theta'^2\right)}~,
\end{equation}
with $\mathcal{N}=4\pi N_fT_{D5}$, and primes denote derivatives with respect to $z$. The integral goes from the boundary $z=0$ to the bulk interior, denoted by $z_0$\footnote{For black hole embeddings, $z_0$ is the location of the singular shell. For Minkowski embeddings, $z_0$ is the value of $z$ at which the brane ends.}. The asymptotic expansions \eqref{eq:UVexpansions} become
\begin{equation}
\begin{aligned}
    \theta(z) & = Mz+Cz^2+\mathcal{O}(z^4)~,\\
    c(z) & = c_\infty+Jz-\frac{\Omega^2c_\infty}{2}z^2+\mathcal{O}(z^3)~.\\
\end{aligned}
\end{equation}

Integrating the DBI action up to a UV cutoff at $z=\epsilon$, we find the asymptotic behavior
\begin{equation}
\begin{aligned}
    S_{D5}&=-\mathcal{N}\int_\epsilon dz\left[\frac{1}{z^4}-\frac{M^2}{2z^2}+\mathcal{O}(\epsilon^0)\right]\\
    &=-\mathcal{N}\left[\frac{1}{3\epsilon^3}-\frac{M^2}{2\epsilon}+\mathcal{O}(\epsilon)\right]~.
\end{aligned}
\end{equation}

The local counterterms needed to regularize the divergences were are given by \cite{Karch_2005_holorenobranes}
\begin{equation}
\begin{aligned}
    S^{ct}_1 & = \frac{1}{3}\mathcal{N}\sqrt{-\gamma}~,\\
    S^{ct}_2 & = -\frac{1}{2}\mathcal{N}\sqrt{-\gamma}\theta(\epsilon)^2~.
\end{aligned}
\end{equation}

The contribution of the counterterms is
\begin{equation}
    S^{ct}=-\mathcal{N}\left[-\frac{1}{3\epsilon^3}+\frac{M^2}{2\epsilon}+MC\right]~,
\end{equation}
where the first term is the contribution of $S^{ct}_1$, and the rest comes from $S^{ct}_2$. These contributions correctly cancel the divergences when $\epsilon\to 0$.

\subsection{Quark condensate and electric current}

We now consider the one-point functions. We begin by the quark condensate $\langle\mathcal{O}_m\rangle$,
\begin{equation}
    \langle \mathcal{O}_m\rangle =-(2\pi\alpha')\lim_{\epsilon\to 0}\epsilon\frac{\delta S_{\text{D5}}^{sub}}{\delta \theta(\epsilon)}~,
    \label{eq:condensateD3D5}
\end{equation}
with $S^{sub}_{D5}=S^{reg}_{D5}+\sum_iS^{ct}_i$. The contribution from the regularized action on-shell can be written as a boundary term,
\begin{equation}
    \delta S^{reg}_{D5}=\frac{\partial \mathcal{L}}{\partial \theta'}\delta \theta\big\rvert_\epsilon=\mathcal{N}\left[\frac{M}{\epsilon^2}+\frac{2C}{\epsilon}+\mathcal{O}(\epsilon^0)\right]\delta\theta(\epsilon)~.
\end{equation}
The counterterms contribute as
\begin{equation}
    \delta S^{ct}_2=\mathcal{N}\left[-\frac{M}{\epsilon^2}-\frac{C}{\epsilon}+\mathcal{O}(\epsilon^0)\right]\delta\theta(\epsilon)~.
\end{equation}

Adding the two contributions, we readily get from \eqref{eq:condensateD3D5}
\begin{equation}
    \langle\mathcal{O}_m\rangle=-\frac{N_fN_c}{\pi^2}C~,
\end{equation}
where we have written $\mathcal{N}$ in terms of field theory quantities using \eqref{eq:ND3D5}.

We now move on to the current,
\begin{equation}
    \mathcal{J}_\text{YM}(t) =(2\pi\alpha')\lim_{\epsilon\to 0}\left[\frac{\delta S^{sub}_{D5}}{\delta c_x(\epsilon)}+i\frac{\delta S^{sub}_{D5}}{\delta c_y(\epsilon)}\right]e^{i\Omega t}~,
    \label{eq:currentD3D5}
\end{equation}
with $c_x$ and $c_y$ defined as $c(z)=c_x(z)+ic_y(z)$.

The contribution from $S^{reg}_{D5}$ is again a boundary term,
\begin{equation}
    \delta S^{reg}_{D5}=\frac{\partial \mathcal{L}}{\partial c_x'}\delta c_x\big\rvert_\epsilon+i\frac{\partial \mathcal{L}}{\partial c_y'}\delta c_y\big\rvert_\epsilon=\mathcal{N}\left[J_x \delta c_x(\epsilon)+iJ_y\delta c_y(\epsilon)+\mathcal{O}(\epsilon)\right]~,
\end{equation}
which is the only contribution since the counterterms do not involve the gauge field\footnote{This is not the case in the D3/D7 model, where a logarithmic counterterm has to be added, leading to an ambiguity in the definition of the current \cite{Kinoshita_2017}.}.

Therefore, we readily get from \eqref{eq:currentD3D5}
\begin{equation}
    \mathcal{J}_{\text{YM}}(t) =\frac{N_fN_c}{\pi^2}Je^{i\Omega t}~,
\end{equation}
where $J=J_x+iJ_y$.

\section{Details on optical conductivities}\label{app:opticalcond}

In this appendix we provide the details concerning the photovoltaic optical conductivities of Section \ref{subsec:opticalcond}. We analyze the response of our system to an additional linearly polarized electric field on top of the circularly driven background \eqref{eq:rotatingE}. In vector cartesian notation, the total electric field is now
\begin{equation}
    \vec{\mathcal{E}}(t)=O(t)\vec{E}+\vec{\epsilon}(t)=O(t)\vec{E}+\vec{\epsilon}\h e^{-i\omega t}~,
\end{equation}
where $\vec\epsilon$ is a constant vector such that $\abs{\vec{\epsilon}}\ll|\vec{E}|$. The change in the current, $\vec{\mathcal{J}}(t)\to O(t)\vec J+\delta \vec J(t)$, defines the conductivity $\boldsymbol{\sigma}$ as $\delta\vec{J}(t)=\boldsymbol{\sigma}\cdot\vec{\epsilon}(t)$. We define the vector of fluctuations, $\delta \vec \xi(t,\rho) = (\delta c_x,\delta c_y,\delta \theta)$.

First of all, let us write the perturbed equations in terms of the tortoise coordinates $(\tau, r_*)$ defined in \eqref{eq:tortoise}. We get
\begin{equation}
    \Big( \partial_\tau^2-\partial_{r_*}^2+\textbf{A}(r)\partial_\tau+\textbf{B}(r)\partial_{r_*}+\textbf{C}(r)\Big)\delta \vec{\xi}=0~,
    \label{eq:perteq}
\end{equation}
where $\textbf{A}$, $\textbf{B}$ and $\textbf{C}$ are $3\times 3$ matrices which at the pseudohorizon $r=r_c$ satisfy
\begin{equation}
    \textbf{A}(r=r_c) =-\textbf{B}(r=r_c)\equiv \textbf{A}_c~, \qquad \textbf{C}(r=r_c)  =0~.
\end{equation}
Thus, in this limit, which corresponds to $r_*\to -\infty$, the fluctuation equation
\eqref{eq:perteq} becomes
\begin{equation}
    \left(\partial_\tau^2-\partial_{r_*}^2\right)\delta\vec{\xi}+\textbf{A}_c\left(\partial_\tau-\partial_{r_*}\right)\partial \vec{\xi}=0~,
\end{equation}
whose general solution takes the form
\begin{equation}
    \delta \vec{\xi}=\vec{f}(\tau+r_*)+e^{-\textbf{A}_cr_*}\vec{g}(\tau-r_*)~.
\end{equation}
We will impose that $\vec{g}=0$, which selects the ingoing wave boundary condition at the pseudohorizon. Let us next look at the UV boundary condition \eqref{eq:pert}. First of all, we rewrite the rotation matrix $O(t)$ as
\begin{equation}
    O(t)=\textbf{M}_+e^{i\Omega t}+\textbf{M}_-e^{-i\Omega t}~,
\end{equation}
where
\begin{equation}
    \textbf{M}_\pm = \frac{1}{2}\begin{pmatrix} 1 & \pm i\\ \mp i & 1 \end{pmatrix}~.
\end{equation}
Then, defining the frequencies $\omega_\pm=\omega\pm\Omega$, the boundary UV condition of $\delta\vec{c}$ can be written as
\begin{equation}
    \delta \vec{c}(t,r=\infty)=-\frac{i}{\omega}\Big(\textbf{M}_+e^{-i\omega_+ t}+\textbf{M}_- e^{-i\omega_- t}\Big)\vec{\epsilon}~.
    \label{eq:deltacbdry}
\end{equation}
Let us assume that $\delta\vec{c}(t,r)$ and $\delta\theta(t,r)$ oscillate with frequencies  $\omega_{\pm}$
\begin{equation}
    \delta \vec{c}(t,r)  =\vec{\beta}_+(r)e^{-i\omega_+ t}+\vec{\beta_-}(r)e^{-i\omega_- t}~, \qquad \delta \theta(t,r)  = \gamma_+(r) e^{-i\omega_+ t}+\gamma_-(r) e^{-i\omega_- t}~.
    \label{eq:deltactheta}
\end{equation}
Then, our system of equations \eqref{eq:perteq} for the fluctuations becomes
\begin{equation}
    \left[\frac{d^2}{dr_*^2}-\textbf{B}(r)\frac{d}{dr_*}+\omega_\pm^2+i\omega_\pm\textbf{A}(r)-\textbf{C}(r)\right]\vec{\xi}_{\pm}=0~.
\end{equation}
Let us write the  boundary expansions for the fields in the form
\begin{equation}
    \delta\vec{c}(t,r)  \approx \delta \vec{c}^{(0)}(t)+\frac{\delta \vec{c}^{(1}(t)}{r}+...~, \qquad \vec{\beta}_\pm(r)  = \vec{\beta}_\pm^{(0)}+\frac{\vec{\beta}_\pm^{(1)}}{r}+...~.
\end{equation}
Plugging these expansions in \eqref{eq:deltactheta} we get
\begin{equation}
    \delta \vec{c}^{(0)}(t)  = \vec{\beta}_+^{(0)}e^{-i\omega_+ t}+\vec{\beta}_-^{(0)}e^{-i\omega_- t}~, \qquad \delta \vec{c}^{(1)}(t)  = \vec{\beta}_+^{(1)}e^{-i\omega_+ t} + \vec{\beta}_-^{(1)}e^{-i\omega_- t}~.
\end{equation}
Comparing the first of these equations with \eqref{eq:deltacbdry} we conclude that
\begin{equation}
    \vec{\beta}_\pm^{(0)}=-\frac{i}{\omega}\textbf{M}_\pm\vec{\epsilon}~.
    \label{eq:betapm0}
\end{equation}
The subleading vectors $\vec{\beta}_\pm^{(1)}$ determine the variation of the current $\delta \vec{\mathcal{J}}(t)=O(t)\delta \vec{c}^{(1)}(t)$, namely
\begin{equation}
    \delta \vec{\mathcal{J}}(t)=e^{-i\omega t}\Big(\textbf{M}_+\vec{\beta}_+^{(1)}+\textbf{M}_-\vec{\beta}_-^{(1)}+\textbf{M}_+\vec{\beta}_-^{(1)}e^{2i\Omega t}+\textbf{M}_- \vec{\beta}_+^{(1)}e^{-2i\Omega t}\Big)~.
    \label{eq:deltacJ}
\end{equation}
For a regular solution the vectors $\vec{\beta}_\pm^{(1)}$ and $\vec{\beta}_\pm^{(0)}$ are related. Let us write this relation as
\begin{equation}
    \vec{\beta}_\pm^{(1)}=\mathbf{X}_\pm\vec{\beta}_\pm^{(0)}=-\frac{i}{\omega}\boldsymbol{X}_\pm\boldsymbol{M}_\pm\vec{\epsilon}~,
    \label{eq:beta1and0}
\end{equation}
where ${\textbf{X}_{\pm}}$ are $2\times 2$ matrices that, in general, must be determined numerically (see \cite{Garbayo_2020} for details). Plugging \eqref{eq:beta1and0} into \eqref{eq:deltacJ} we get a relation between the current $\delta{\vec{\cal{J}}}$ and the applied electric field $\vec{\epsilon}$
\begin{equation}
    \delta \vec{\mathcal{J}}=\Big[\boldsymbol{\sigma}(\omega)e^{-i\omega t}+\boldsymbol{\sigma}^{+}(\omega)e^{-i(\omega+2\Omega)t}+\boldsymbol{\sigma}^{-}(\omega)e^{-i(\omega-2\Omega)t}\Big]\vec{\epsilon}~,
\end{equation}
where $\boldsymbol{\sigma}(\omega)$, $\boldsymbol{\sigma^+}(\omega)$ and $\boldsymbol{\sigma}^-(\omega)$ are the conductiviy matrices corresponding to the frequencies $\omega$, $\omega+2\Omega$ and $\omega-2\Omega$, given by
\begin{equation}
\begin{aligned}
    \boldsymbol{\sigma}(\omega) & = -\frac{i}{\omega}\left(\textbf{M}_+\textbf{X}_+\textbf{M}_++\textbf{M}_-\textbf{X}_-\textbf{M}_-\right)~,\\
    \boldsymbol{\sigma}^+(\omega) & = -\frac{i}{\omega}\textbf{M}_-\textbf{X}_-\textbf{M}_+~,\\
    \boldsymbol{\sigma}^-(\omega) & = -\frac{i}{\omega}\textbf{M}_+\textbf{X}_+\textbf{M}_-~.
    \label{eq:sigmasgeneralapp}
\end{aligned}
\end{equation}

\subsection{Massless case}\label{app:masslessconduc}

In the massless case the fluctuations of the embedding function decouple from those of the gauge field $\delta \vec{c}$. Therefore, since we are interested in computing conductivities, we can concentrate in studying the equations for $\delta c_x$ and $\delta c_y$. In order to write these  equations in a more convenient form, let us define the following differential operators

\begin{equation}
\begin{aligned}
    \mathcal{O}_1 & \equiv \partial_t^2 + \frac{(r^4-r_h^4)(r_c^4-r^4)}{\rho^4(r_c^4+r^4-2r_h^4)}\partial_r^2 
    - \frac{2(r^4-r_h^4)(r_c^4-r_h^4)}{r^2(r_c^4+r^4-2r_h^4)}\partial_t\partial_r  + \frac{4r(r_c^4-r_h^4)}{(r_c^4+r^4-2r_h^4)}\partial_t
    \\
    & \hspace{0.45\textwidth}- \frac{2(r^4-r_h^4)^2(r_c^4+r^4)}{r^5(r_c^4+r^4-2r_h^4)}\partial_r~, \\
    \mathcal{O}_2 & \equiv -2\partial_t + \frac{2(r^4-r_h^4)(r_c^4-r_h^4)}{\rho^2(r_c^4+r^4-2r_h^4)}\partial_r 
    + \frac{4r(r^4-r_h^4)}{r_c^4+r^4-2r_h^4}~.
\end{aligned}
\end{equation}
Then, one can show that $\delta c_x$ and $\delta c_y$ satisfy the following system of coupled second-order differential equations
\begin{align}
    \left(\mathcal{O}_1-\Omega^2\right)\delta c_x+\Omega\h\mathcal{O}_2\h\delta c_y & =0~, & \left(\mathcal{O}_1-\Omega^2\right)\delta c_y+\Omega\h\mathcal{O}_2\h\delta c_x & =0~.
\end{align}
To decouple these equations, let us consider the following  complex combinations of $\delta c_x$ and $\delta c_y$
\begin{align}
    \eta(t,r) & \equiv \delta c_x(t,r)+i\delta c_y(t,r)~, & \tilde{\eta}(t,r) & \equiv \delta c_x(t,r)-i\delta c_y(t,r)~.
\end{align}
Notice that $\tilde{\eta}$ is not the complex conjugate of $\eta$ since $\delta c_x$ and $\delta c_y$ are not necessarily real. It is straightforward to verify that the equations for $\eta$ and $\tilde{\eta}$ are indeed decoupled and given by
\begin{align}
    \left(\mathcal{O}_1-\Omega^2\right)\eta-i\Omega\h\mathcal{O}_2\h\eta & = 0~, & \left(\mathcal{O}_1-\Omega^2\right)\tilde{\eta}-i\Omega\h\mathcal{O}_2\h\tilde{\eta} & = 0~.
\end{align}
Let us now separate variables as
\begin{align}
    \eta(t,r) & = \beta(r)\h e^{-i\omega t}~, & \tilde{\eta}(t,r) & = \tilde{\beta}(r)\h e^{-i\omega t}~,
\end{align}
for some frequency $\omega$. Then, we obtain the following differential equation for $\beta$
\begin{align}
\begin{split}
    r(r^4-r_c^4)(r^4-r_h^4)\beta''  =& -2(r^4-r_h^4)\left[r^8-r_c^4r_h^4+r^4(r_c^4-r_h^4)-ir^3(r_c^4-r_h^4)(\omega-\Omega)\right]\beta'\\
    &-r^5(\omega-\Omega)\left[4ir(r_c^4-r_h^4)+r^4(\omega-\Omega)+(r_c^4-2r_h^4)(\omega-\Omega)\right]\beta ~.
\end{split}
\end{align}
The equation for $\tilde{\beta}$ is the same, but with $(\omega+\Omega)$ instead of  $(\omega-\Omega)$. Then, remarkably, one can find the following general solutions 
\begin{equation}
\begin{aligned}
    \beta(r)&=e^{i\frac{\Omega-\omega}{4r_h}\Lambda(r/r_h)}\left[A+Be^{-i\frac{\Omega-\omega}{2r_c}\Lambda(r/r_c)}\right]~,\\
    \tilde{\beta}(r)&=e^{-i\frac{\Omega+\omega}{4r_h}\Lambda(r/r_h)}\left[\tilde{A}+\tilde{B}e^{i\frac{\Omega+\omega}{2r_c}\Lambda(r/r_c)}\right]~,
    \label{eq:betassolution}
\end{aligned}
\end{equation}
where $\Lambda$ is the function defined in \eqref{eq:Lambdadef} and $A$, $B$, $\tilde A$ and $\tilde B$ are complex constants which are determined by imposing boundary conditions both at the IR and UV. First of all, we write the solutions we found in terms of the tortoise coordinates $(\tau, r_*)$ of \eqref{eq:tortoisesolution}. Actually, by inspecting the expression of $\eta$ obtained from \eqref{eq:betassolution} one easily demonstrates that, in terms of the tortoise variables, it can be simply written as
\begin{equation}
    \eta(\tau,r_*)=e^{\frac{\Omega}{4r_h}\Lambda(r/r_h)}\left[A~e^{-i\omega (\tau+r_*)}+B~e^{-2i\Omega r_*}~e^{-i\omega(\tau-r_*)}\right]~.
\end{equation}
It is now clear that $\eta(\tau, r_*)$ is the superposition of ingoing and outgoing waves at the pseudohorizon. The infalling regularity condition requires that $B$ vanishes. Then, writing $\eta$ in our original $(t,r)$  coordinates, we have
\begin{equation}
    \eta(t,r)=A~e^{-i\frac{\omega-\Omega}{4r_h}\Lambda(r/r_h)}~e^{-i\omega t}~.
\end{equation}
We can proceed similarly with $\tilde{\eta}$ and conclude that we should require that $\tilde B=0$. Therefore,
\begin{equation}
    \tilde{\eta}(t,r)=\tilde{A}~e^{-i\frac{\omega+\Omega}{4r_h}\Lambda(r/r_h)}~e^{-i\omega t}~.
\end{equation}
Therefore, we obtain that the fluctuations $\delta c_x$ and $\delta c_y$ regular at the pseudohorizon are
\begin{equation}
\begin{aligned}
    \delta c_x(t,r)&=\frac{1}{2}\left[A~e^{-i\frac{\omega-\Omega}{4r_h}\Lambda(r/r_h)}+\tilde{A}~e^{-i\frac{\omega+\Omega}{4r_h}\Lambda(r/r_h)}\right]e^{-i\omega t}~,\\
    \delta c_y(t,r)&=\frac{1}{2i}\left[A~e^{-i\frac{\omega-\Omega}{4r_h}\Lambda(r/r_h)}-\tilde{A}~e^{-i\frac{\omega+\Omega}{4r_h}\Lambda(r/r_h)}\right]e^{-i\omega t}~.
    \label{eq:deltacxcy}
\end{aligned}
\end{equation}
Let us now impose the boundary conditions at the UV. To fulfill the UV boundary condition \eqref{eq:deltacbdry} we sum two solutions of the form \eqref{eq:deltacxcy} with frequencies $\omega_+=\omega+\Omega$ and $\omega_-=\omega-\Omega$. Let $A_{\pm}$ and $\tilde{A}_{\pm}$ denote the constants in \eqref{eq:deltacxcy} with frequency $\omega_{\pm}$. From the leading UV terms we get that, in order to satisfy the boundary condition \eqref{eq:betapm0}, the constants $A_{\pm}$ and $\tilde{A}_{\pm}$ must be related to $\epsilon_x$ and $\epsilon_y$ as
\begin{align}
    A_+ & = -\frac{i}{\omega}(\epsilon_x+i\epsilon_y)~, & \tilde{A}_- & = -\frac{i}{\omega}(\epsilon_x-i\epsilon_y)~, & \tilde{A}_+ & = A_-=0~.
\end{align}
Moreover, from the analysis of the subleading UV terms we conclude that
\begin{equation}
    \vec{\beta}_\pm^{(1)}=\textbf{M}_\pm\vec{\epsilon}~.
    \label{eq:betapm1}
\end{equation}
By comparing \eqref{eq:betapm1} and \eqref{eq:beta1and0}, we get that the matrices $\mathbf{X}_\pm$ are given by
\begin{equation}
    \textbf{X}_\pm=i\omega \mathbf{1}~.
\end{equation}
Plugging these $\mathbf{X}_\pm$ matrices in \eqref{eq:sigmasgeneralapp}, we obtain the conductivity matrices in the massless case, namely
\begin{align}
    \boldsymbol{\sigma}(\omega) & =\boldsymbol{1}~, & \boldsymbol{\sigma}^+(\omega) & = \boldsymbol{\sigma}^-(\omega)=0~,
\end{align}
which is exactly the same result as the one found in \cite{Garbayo_2020} at zero temperature. As in the case of the non-linear current, here also the result confirms the expectations put forward in \cite{Karch_2010}.

\section{Phase space structure for the D3/D7 system}\label{app:D3D7}

\begin{figure}
    \centering
    \begin{subfigure}[t]{0.48\textwidth}
        \centering
        \includegraphics[width=\textwidth]{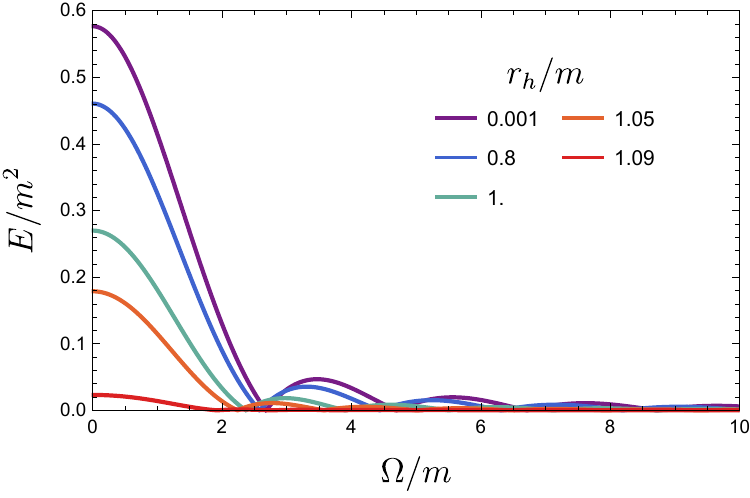}
    \end{subfigure}
    \hspace{0.02\textwidth}
    \begin{subfigure}[t]{0.48\textwidth}
        \centering
        \includegraphics[width=\textwidth]{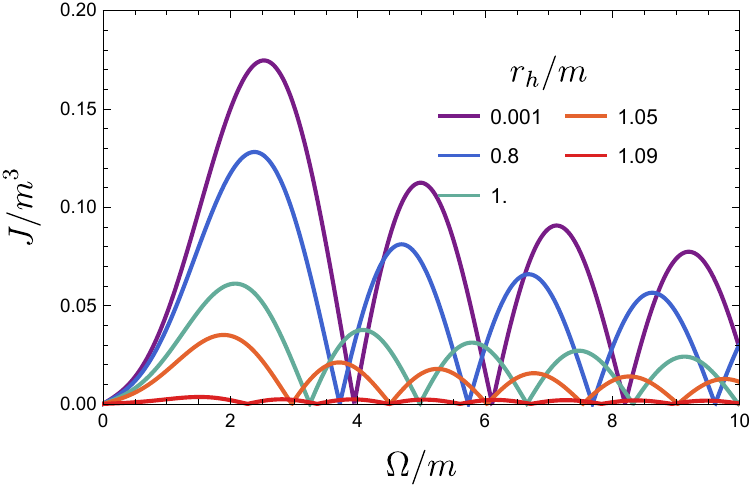}
    \end{subfigure}
    \caption{Electric field and current of the critical embeddings versus driving frequency, for different values of $r_h$ in the D3/D7 system. The structure is mainly the same as the one found for the D3/D5 case. Here the maximum temperature for the Minkowski embeddings is slightly higher, while the height of the lobes is suppressed faster as one increases the driving frequency.}
    \label{fig:LobesEJD3D7}
\end{figure}

\begin{figure}
    \centering
    \begin{subfigure}[t]{0.48\textwidth}
        \centering
        \includegraphics[width=\textwidth]{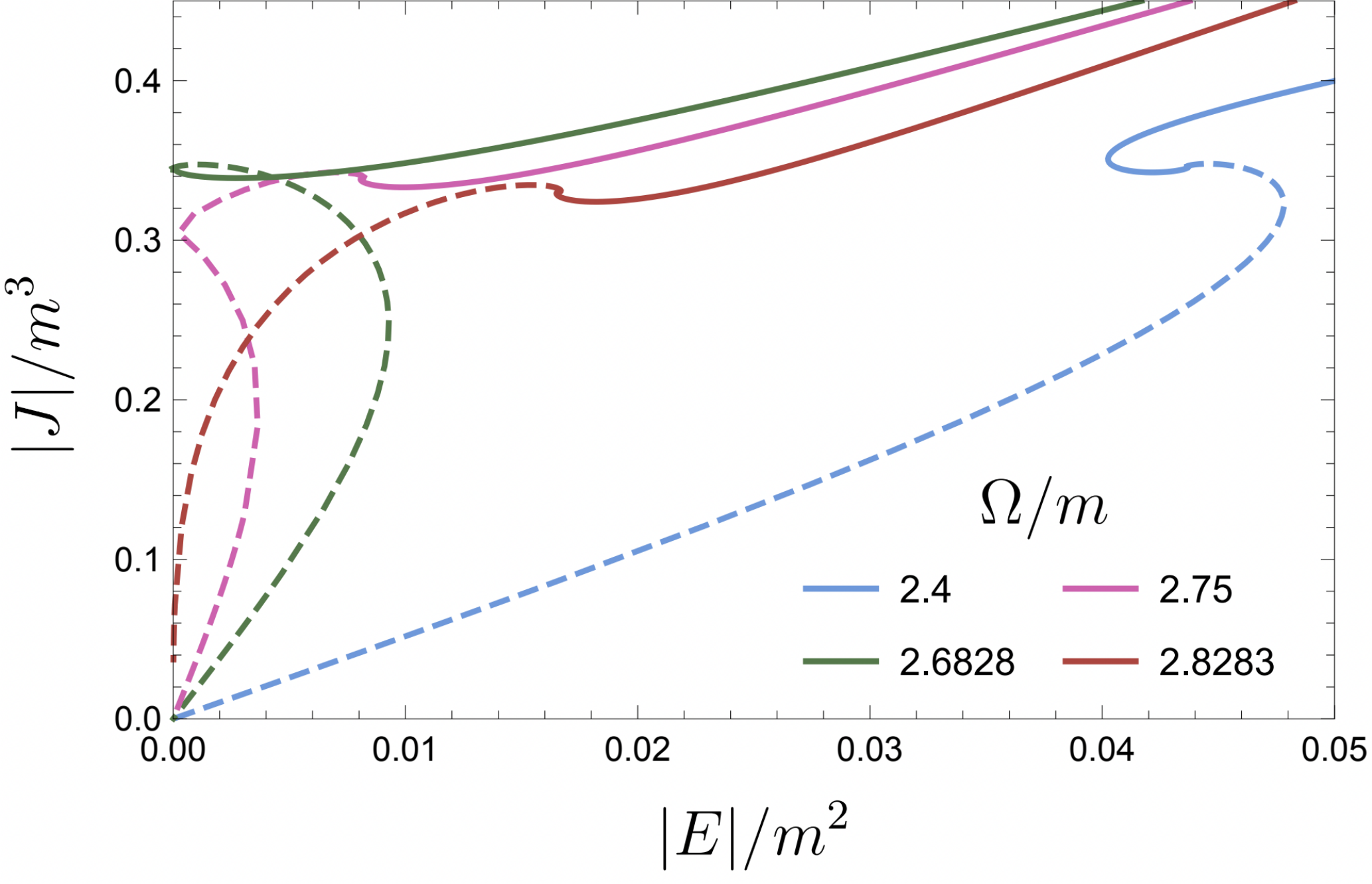}
    \end{subfigure}
    \hspace{0.02\textwidth}
    \begin{subfigure}[t]{0.48\textwidth}
        \centering
        \includegraphics[width=\textwidth]{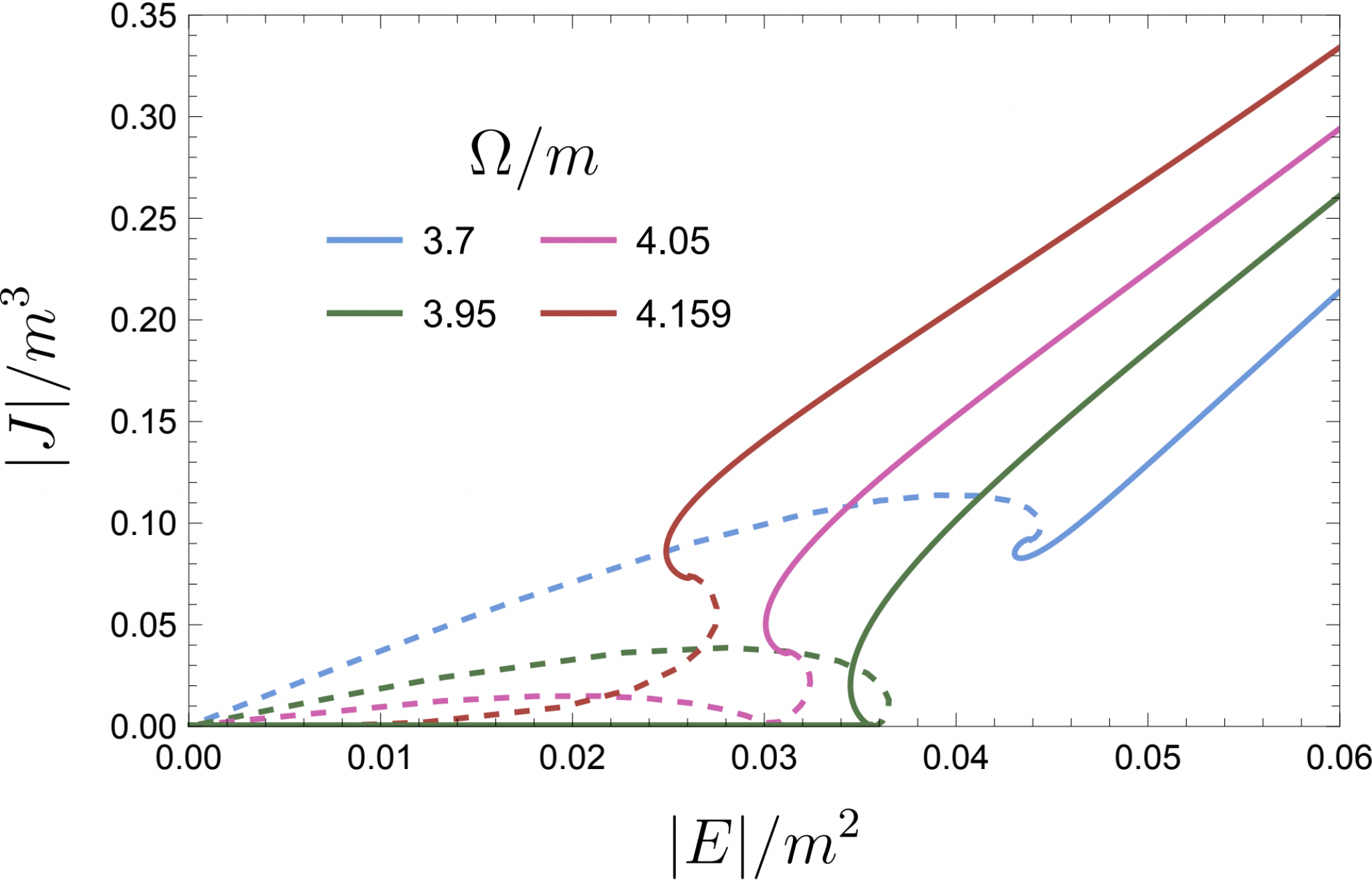}
    \end{subfigure}
    \caption{Electric current $|J|/m^3$ versus electric field $|E|/m^2$ for $r_h/m=0.001$. The solid curves represent the insulator (Minkowski) phase and the dashed curves the conductive (BH) phase.  The driving frequency is fixed to some $\Omega/m<\Omega_c/m$ (blue), $\Omega/m=\Omega_{c}/m$ (green), $\Omega_{c}/m<\Omega/m<\Omega_{meson}/m$ (pink) and $\Omega/m=\Omega_{meson}/m$ (red). The curves on the left, with $\Omega \sim 2.68$  scan the region close to the Floquet vector meson condensates and were obtained in \cite{Kinoshita_2017}. On the right, for $\Omega/m \sim 3.95$ we do the same around the first Floquet suppression point.}
    \label{fig:JvsED3D7}
\end{figure}

It is worth mentioning that the existence of these Floquet suppression points is not restricted to the D3/D5 system. In Fig. \ref{fig:LobesEJD3D7} we reproduce the lobe structure for the D3/D7 system, which is analogous to the one found in Fig. \ref{fig:LobesEJD3D5} for D3/D5. The $r_h\to0$ limit for the electric field plot coincides, of course, with the one studied in \cite{Kinoshita_2017}. 

We find again a set of points where the induced current vanishes while the electric field is close to its maximum value within the lobe. This seems to indicate the presence of a range of frequencies for which, in Minkowski embeddings, $\abs{j}=0$ with $\abs{E}\neq0$, ranging from the critical frequency of the current $\Omega_{c,j}$ ($c$ stands for critical) up to some maximum value $\Omega_{m,j}$ ($m$ stands form meson). 

The analytic computation of $\Omega_{m,j}$ in the $T=0$ limit of the D3/D7 system is more complicated than the one corresponding to the D3/D5 intersection but it can, however, be obtained numerically. We have found that the Floquet suppression points first appear for $\Omega/m\in\left(3.950,4.159\right)$, and plotted in Fig. \ref{fig:JvsED3D7} the modulus of the current and the electric field for frequencies close or within that range. As anticipated, the similarity with the structure studied in \cite{Kinoshita_2017} is quite obvious, and thus we conclude that a 3D diagram as that of Fig. \ref{fig:3Dj} is also found when studying D3/D7 branes.

\bibliographystyle{JHEP}
\bibliography{main}

\end{document}